\documentclass[printer]{aa}
\usepackage{amssymb}
\usepackage{amsmath}
\usepackage{txfonts}
\usepackage{graphicx}
\usepackage{natbib}
\usepackage{rotating}
\usepackage{morefloats}
\usepackage{soul}
\usepackage{url}
\usepackage{epsfig}
\usepackage{longtable}

\usepackage{color}
\usepackage[utf8]{inputenc}
\usepackage{setspace}

\begin{document}

\title{The XMM Cluster Outskirts Project (X-COP):\\
 Physical conditions to the virial radius of Abell 2142}
\author{C. Tchernin\inst{1,2}, D. Eckert\inst{2,3}, S. Ettori\inst{4,5}, E. Pointecouteau\inst{6,11}, S. Paltani\inst{2}, S. Molendi\inst{3}, G. Hurier\inst{7,12}, F. Gastaldello\inst{3,8}, E. T. Lau\inst{9}, D. Nagai\inst{9}, M. Roncarelli\inst{4}, M. Rossetti\inst{10,3}}
\institute{
Center for Astronomy, Institute for Theoretical Astrophysics, Heidelberg University, Philosophenweg 12, 69120 Heidelberg, Germany,
\email{Tchernin@uni-heidelberg.de}
\and
Department of Astronomy, University of Geneva, ch. d'Ecogia 16, 1290 Versoix, Switzerland
\and
INAF - IASF-Milano, Via E. Bassini 15, 20133 Milano, Italy
\and
INAF - Osservatorio Astronomico di Bologna, Via Ranzani 1, 40127 Bologna, Italy
\and
INFN, Sezione di Bologna, viale Berti Pichat 6/2, 40127 Bologna, Italy
\and
CNRS; IRAP; 9 Av. colonel Roche, BP 44346, F-31028 Toulouse cedex 4, France
\and
Centro de Estudios de Fisica del Cosmos de Aragon, Plaza San Juan 1, Planta-2, 44001, Teruel, Spain 
\and
Department of Physics and Astronomy, University of California at Irvine, 4129 Frederick Reines Hall, Irvine, CA 92697-4575, USA
\and
Department of Physics, Yale University, New Haven, CT 06520, USA
\and
Universit\` a degli studi di Milano, Dip. di Fisica, via Celoria 16, 20133 Milano, Italy
\and
Universit\'e de Toulouse; UPS-OMP; IRAP; Toulouse, France
\and
Institut d'Astrophysique Spatiale, CNRS (UMR8617) Universit\'e Paris-Sud 11, Batiment 121, Orsay, France 
}
\authorrunning{C. Tchernin et al. }
\titlerunning{Joint X-ray/SZ analysis of Abell 2142}

\abstract
{Galaxy clusters are continuously growing through the accretion of matter in their outskirts. This process induces inhomogeneities in the gas density distribution (clumping) which need to be taken into account to recover the physical properties of the intracluster medium (ICM) at large radii. }{We studied the thermodynamic properties in the outskirts ($R>R_{500}$) of the massive galaxy cluster Abell 2142 by combining the Sunyaev Zel'dovich (SZ) effect with the X-ray signal.} {We combined the SZ pressure profile measured by \emph{Planck} with the \emph{XMM-Newton} gas density profile to recover radial profiles of temperature, entropy and hydrostatic mass out to $2R_{500}$. We used a method that is insensitive to clumping to recover the gas density, and we compared the results with traditional X-ray measurement techniques.} {When taking clumping into account, our joint SZ/X-ray entropy profile is consistent with the predictions from pure gravitational collapse, whereas a significant entropy flattening is found when the effect of clumping is neglected. The hydrostatic mass profile recovered using joint X-ray/SZ data agrees with that obtained from spectroscopic X-ray measurements and with mass reconstructions obtained through weak lensing and galaxy kinematics.} {We found that clumping can explain the entropy flattening observed by \textit{Suzaku} in the outskirts of several clusters. When using a method insensitive to clumping for the reconstruction of the gas density, the thermodynamic properties of Abell 2142 are compatible with the assumption that the thermal gas pressure sustains gravity and that the entropy is injected at accretion shocks, with no need to evoke more exotic physics. Our results highlight the need for X-ray observations with sufficient spatial resolution, and large collecting area,  to understand the processes at work in cluster outer regions.}

\maketitle

%%%%%%%%%%%%%%%%%%%%%%%%%%%
\section {Introduction}
%%%%%%%%%%%%%%%%%%%%%%%%%%%
Galaxy clusters are the largest gravitationally bound systems in the Universe. According to the concordance cosmological model, they are the latest structures to be formed. For this reason, they are expected to continue growing at the present epoch through the accretion of matter in their outskirts. Thus, information on the processes governing structure formation can be obtained through the study of galaxy cluster outskirts \citep[see ][ for a review]{reiprich13}. The distribution of matter in the outer regions of galaxy clusters is expected to become clumpy \citep{nagai11,vazza13} and asymmetric \citep{vazza11}, and the impact of non-thermal energy in the form of turbulence, bulk motions, cosmic rays and magnetic fields is expected to be significant, even if still poorly constrained by theory and simulations.

Spectroscopic X-ray measurements of cluster outskirts became possible recently with the \textit{Suzaku} experiment thanks  to its low particle background \citep{mitsuda07}. With the help of  \textit{Suzaku}, many bright galaxy clusters have been observed out to the viral radius \citep[$\sim R_{200}$, e.g.][]{reiprich09,hoshino10,akamatsu11,simionescu11,walker12,walker13,urban14,okabe14}. These works studied the radial profiles of the density, temperature and entropy out to $R_{200}$. In several cases, the authors observed a flattening of the entropy profile beyond $R_{500}$ compared to the expectation of the self-similar accretion model \citep{voit05}. Several studies also observed a decrease in the hydrostatic mass profile in the same radial range, which might suggest that the medium is out of  hydrostatic equilibrium. This could be caused by a significant non-thermal pressure in the form of turbulence, bulk motions or cosmic rays \citep[e.g.][]{vazza09,lau09,battaglia13}, non-equilibration between electrons and ions \citep{hoshino10,avestruz15} or weakening of the accretion shocks \citep{lapi10,fusco14}. 

Alternatively, \citet{simionescu11} proposed that the measured gas density is overestimated because of gas clumping, which would lead to an underestimated entropy \citep[see also][]{eckert13b,walker13,urban14,morandi13,morandi14}. Recently, several numerical studies have focused on quantifying the effect of gas inhomogeneities on X-ray observations \citep[see e.g., ][]{nagai11,zhu13,eckertclumping,vazza13,roncarelli13}. For instance, using hydrodynamical simulations \citet{zhu13} showed that the density distribution inside a shell at a given distance from the cluster center can be described by a log normal distribution plus a high density tail \citep[see also][]{rasia14,khedekar13}. While the log-normal distribution contains information about the bulk of the ICM, the high density tail is due to the presence of infalling gas clumps. The authors showed that the median of the distribution coincides with the mode of the log-normal distribution, whereas the mean is biased high by the presence of clumps. This result has been confirmed observationally by \citet{eckertclumping}, where the authors reproduced this result using \emph{ROSAT} and \emph{XMM-Newton} data. In this paper, the authors computed the surface brightness distribution in an annulus at $\sim1.2R_{500}$ from the cluster center and showed that the median of the distribution corresponds to the mode of the log-normal distribution, while the mean is shifted toward higher surface brightness values. They concluded that the azimuthal median method allows us to recover the true gas density profile even in the presence of inhomogeneities. 

In addition, the recent years have seen great progress in the study of the ICM through the Sunyaev-Zel'dovich (SZ) effect \citep{sunyaev72}. The SZ effect arises when photons of the cosmic microwave background photons (CMB) interact with the electrons of the ICM through inverse Compton scattering. The observed distortion of the CMB spectrum is proportional to the thermal electron pressure integrated along the line of sight (the Compton $y$ parameter). Therefore, the SZ signal decreases less sharply with radius than the X-ray emissivity. Furthermore, it is less sensitive to density inhomogeneities. Indeed, it has been shown that at $R_{200}$ , variations in the X-ray signal are $\sim$3 times larger than in the SZ flux  \citep[see for instance Fig.~6 of][]{roncarelli13}, since the fluctuations are nearly isobaric  \citep{khedekar13}. This makes the SZ signal highly complementary to X-ray observations. Recent SZ experiments (e.g. \emph{Planck}, \emph{Bolocam}, \emph{SPT}) enabled us to extend the measurements of the SZ signal well beyond $R_{500}$ \citep{planck13,sayers13}. These breakthroughs opened the possibility of combining the SZ signal with X-ray observations to study the thermodynamical properties of the gas, bypassing the use of X-ray spectroscopic data \citep{ameglio07,nord09,basu10,eckert13a}. Joint X-ray/SZ imaging studies can also lead to a reconstruction of the cluster mass profile through the hydrostatic equilibrium assumption \citep{ameglio09,eckert13b}. 

In this paper, we combine SZ and X-ray observations from the \textit{Planck} and \textit{XMM-Newton} satellites to study the outskirts of Abell 2142. This cluster belongs to the sample selected in the framework of the \emph{XMM} Cluster Outskirts Project (X-COP), a Very Large Program on \emph{XMM-Newton} which aims at studying the outskirts of an SZ-selected sample of 13 massive, nearby clusters. Abell 2142 is a massive cluster \citep[$M_{200}\sim1.3\times10^{15} M_\odot$,][]{munari14} at a redshift of 0.09 \citep{owers11}. The cluster hosts a moderate cool core \citep[$K_0=68$ keV cm$^2$, ][]{cavagnolo09} and exhibits multiple concentric cold fronts in its central regions \citep{markevitch00,rossetti13}, which are indicative of ongoing sloshing activity extending out to 1~Mpc from the cluster center \citep{rossetti13}. The sloshing activity may have triggered the formation of an unusual radio halo \citep{farnsworth13}. \citet{owers11} studied the 3D galaxy distribution out to $\sim2$~Mpc from the cluster core and identified several substructures associated with minor mergers. \citet{eckert1a2142} discovered an infalling galaxy group located $\sim1.5$~Mpc north-east (NE) of the main cluster. This subcluster is in the process of being stripped from its hot gas by the ram pressure applied by the main cluster, forming a spectacular X-ray tail. On the larger scales, A2142 is located in the core of a collapsing supercluster \citep{einasto15,gramann15}. Together, these studies reveal that A2142 is a dynamically active cluster located at a node of the cosmic web. 

The paper is organized as follows. In Sect.~\ref{sec:spectralfit} we describe the analysis of the X-ray data to obtain radial profiles of the surface brightness, temperature and metal abundance of A2142. In Sect.~\ref{sec:deproj}, we perform a deprojection of the profiles of X-ray surface brightness and SZ $y$ parameter from \emph{Planck} assuming spherical symmetry to recover the three-dimensional gas density and pressure profiles. In Sect.~\ref{sec:combRXSZ}, we combine the resulting SZ pressure profile with the X-ray density profile to obtain radial profiles of entropy, temperature, hydrostatic mass, and gas fraction. We also estimate the effects of gas clumping by comparing the results obtained with the azimuthal median method with the ones obtained using the traditional approach. Our results are discussed in Sect.~\ref{sec:discussion}. 

Throughout the paper, we assume a $\Lambda$CDM cosmology with $\Omega_\Lambda=0.7$, $\Omega_m=0.3$ and $H_o=70$~km/s/Mpc. At the redshift of A2142, 1 arcmin corresponds to 102 kpc. Uncertainties throughout the paper are provided at the 1-$\sigma$ confidence level.
We use as reference values for $R_{200}$ and $R_{500}$,  $R_{200}=2160$ kpc and $R_{500}=1408$ kpc, which are the results of a joint analysis performed in \citet{munari14} based on kinematics, X-ray and gravitational lensing observations of A2142. 

%%%%%%%%%%%%%%%%%%%%%%%%%%%
\section{X-ray spectral analysis}\label{sec:spectralfit}
%%%%%%%%%%%%%%%%%%%%%%%%%%%

\subsection{Description of the XMM data}

Abell 2142 was mapped by \textit{XMM-Newton} through five pointings: a central one (50 ks) and four 25 ks offset pointings obtained in extended full frame mode for pn and full frame for MOS. The data were processed with the \textit{XMM-Newton} Scientific Analysis System (XMMSAS) v13.0. using the Extended Source Analysis Software package \citep[ESAS][]{snowden08}. We filtered out the time periods affected by soft proton flares using the tasks \texttt{MOS-filter} and \texttt{pn-filter} to obtain clean event files. In Table~\ref{tab:obsid} we provide the OBSID and the clean exposure time of the pointings. Point sources were detected and masked down to a fixed flux threshold ($10^{-14}$erg/cm$^{-2}$s$^{-1}$) using the ESAS task \texttt{cheese}.
The presence of anomalous MOS CCDs was also taken into account.

In Fig.~\ref{fig:annulus_region} we show the combined EPIC mosaic of the cluster in the energy band [0.7-1.2] keV corrected for the exposure time for the three instruments \citep{eckert1a2142}. Circular black areas in this image represent masked point sources. 

\begin{table*}
\caption{OBSID, clean exposure time and IN/OUT ratio for the five observations used in this paper.}
\begin{center}
\begin{tabular}{ccccccc}\hline
observation&OBSID& Total [ks] & pn [ks]&MOS1  [ks]&MOS2 [ks] &IN/OUT ratio\\ \hline\hline
center&0674560201 & 59.1 &48.8 & 52.3 &53.8&1.074\\
NW&0694440101  & 24.5 & 12.5&19.6&18.2&1.260\\
SE&0694440501 &34.6 & 29.8&33.1&32.5 &1.139\\
SW&0694440601&38.6 & 24.1&30.0&31.7 &1.154\\
NE&0694440201 & 34.6 & 29.7&32.9&33.2&1.060\\\hline
\end{tabular}
\label{tab:obsid}
\end{center}
\end{table*}

\subsection{Spectral analysis}
We performed a spectral analysis in the [0.5-12] keV energy band of the cluster in the concentric regions shown in Fig.~\ref{fig:annulus_region} and estimated the spectra of the local sky background components from the regions in the four red sectors in the same energy band. These regions are located at a distance of 28 arcmin from the cluster center, where we see no evidence for cluster emission. Spectra and response files were extracted using the ESAS tasks \texttt{MOS-spectra} and \texttt{pn-spectra}. For each of the annuli for which it was possible, we combined the different observations (center, NE, SE, SW, NW, see Fig.~\ref{fig:annulus_region}) in the spectral analysis. 

The modeling of the background and of the source is described in Sect.~\ref{sec:background} and ~\ref{sec:source_spec}, respectively. The fitting procedure was performed using XSPEC v12.7.1.

 \subsubsection{Background modeling}\label{sec:background}
As shown in a number of recent studies \citep[e.g.][]{leccardi08,ettori11}, the modeling of the background is critical to obtain reliable measurements of the properties of the ICM in cluster outskirts. The total background is made of two main components: the sky background and the non X-ray background (NXB). The procedure adopted here to model these components follows \citet{eckert1a2142} and is described in the following.
 
 \begin{itemize}
\item{}The sky background can be modeled with three components: the cosmic X-ray background (CXB), the Galactic halo and the local hot bubble. The emission of the CXB can be described by a power law with a photon index fixed to 1.46 \citep{deluca04}. This component is absorbed by the Galactic column density along the line of sight. We used an hydrogen column density of $3.8\cdot10^{20}$cm$^{-2}$ in this analysis as measured from the LAB HI Galactic survey \citep{kalberla05}. The emission of the Galactic halo can be represented by a thermal component at a temperature of 0.22 keV \citep{mccammon02}. We modeled this thermal emission with the thin plasma model \textit{APEC} \citep{smith01}, with Solar abundance. This emission is also affected by absorption along the line of sight. The local hot bubble is modeled with an unabsorbed thermal component at 0.11 keV. We used the \textit{APEC} model to represent this thermal component, again with Solar abundance. The normalization of these components was fit to each background region independently and then rescaled to the area of the region used for the spectral extraction. For these three background components, only the normalization was allowed to vary.

\item{}The second source of background, the NXB, is induced by charged particles interacting within the detector. It is dominated by cosmic rays, i.e. relativistic charged particles that hit and excite the detector. Fluorescence emission lines are then emitted once the atoms of the detector de-excite. The spectrum of this background contribution can be estimated from the spectrum obtained during closed-filter observations using the method outlined in \citet{snowden08}. 
A model of the NXB for all three EPIC detectors (pn, MOS1 and MOS2) was extracted from filter-wheel-closed data using the ESAS procedures \texttt{MOS-spectra} and \texttt{pn-spectra}.  For each observation, we rescaled these spectra by comparing the count rates measured in the unexposed corners of the detectors with the mean count rates of the closed-filter observations. The resulting closed-filter spectra can be characterized by a flat continuum with several fluorescence emission lines. For each considered region, we modeled these spectra with a phenomenological model consisting in a broken power law and several Gaussians and used the resulting fit as an appropriate model of the NXB. 

A priori, soft protons could also affect the data. In Table~\ref{tab:obsid}, we show  the IN/OUT ratio \citep{deluca04,leccardi08}, which is the ratio between the surface brightness in the FOV and the surface brightness in the unexposed corners (out the FOV) in the hard energy band. Since soft protons are focused by the telescope mirrors, while cosmic rays can induce X-ray emission over the all detector, this ratio is used as an indicator of the contamination of the soft protons to the NXB background. We find that the contamination by soft protons is very low, except for the NW observation, where it reaches a level of 25\%. \citet{leccardi08} estimated the effect of soft protons on the spectral fitting analysis and found that for regions that are bright enough (R$ < R_{500}$) this contribution is subdominant. Given that we stop the spectral extraction at $R_{500}$, we decided to neglect the contribution of the soft protons in the spectral fitting procedure.

The Solar Wind Charge Exchange \citep[SWCX,][]{carter08,carter11} is also  a potential source of background in X-ray observations of cluster outskirts. 
Given that SWCX emission is time variable, the consistency of the spectral fits of the sky components at different times in the four regions considered in our analysis (see Table~\ref{table:bgd_fit4obs}) argues against an significant contamination by this component. Furthermore, the level of the solar proton flux detected with the Alpha Monitor (SWEPAM) instrument on board of the Advanced Composition Explorer (ACE) satellite is below $4\cdot10^{8}$ protons/(s cm$^2$), which is typical of the quiescent Sun, and is not sufficient to trigger SWCX. Therefore, we neglect this background component in our spectral analysis.

\end{itemize}

We estimated the normalizations of the various components of the local sky background (CXB, Galactic halo and local bubble) and of the NXB from the combined analysis of the four regions delimited in red in Fig.\ref{fig:annulus_region}. The results of the fits of the sky background components for each of the four observations separately and for the combined fit are provided in Table~\ref{table:bgd_fit4obs}. To fit the spectrum of these regions, the intensity of the NXB fluorescence lines in the energy range 1.2-1.9 keV (and also in the 7-9 keV energy range for the pn instrument), as well as that of the continuum were left free to vary. The normalization of the CXB was fitted locally to take into account cosmic variance, which is expected to be of the level of 15\% \citep{moretti03}.

By the comparison of the result of the fit in the four regions, we note that the normalization of the local hot bubble is not well constrained by these measurements. This is easily explained by the low temperature of this component (0.11 keV), which is below the energy range covered by \textit{XMM-Newton}/EPIC (0.5-12 keV). This renders the overall model largely insensitive to this component, such that uncertainties in the local hot bubble normalization are not expected to affect the result of the present study.

\begin{table*}
\caption{Fit of the local sky background components (local hot bubble (LHB), Galactic halo (GH) and cosmic X-ray background (CXB)) using the model: $\textit{constant(apec + wabs(apec + powerlaw))}$ on the four regions delimited by the red sectors in Fig.~\ref{fig:annulus_region}. The name of the observation represents the considered region. ALL means that the four regions have been fitted simultaneously. Only the normalization parameters were allowed to vary in this model.  Units are: [$10^{-6}$/cm$^{-5}$] for the normalization of the local bubble and of the Galactic halo and [$10^{-6}$photons/(keVcm$^2$s)] for the normalization of the CXB.}
\begin{center}
\begin{tabular}{ccc|cc|cc}\hline
&LHB norm &LHB $\Delta$norm &GH norm &GH  $\Delta$norm &CXB norm &CXB $\Delta$norm \\ \hline \hline
NE&1.69&[1.29;2.04]&1.20&[1.07;1.30]&0.938&[0.878;0.988]\\
NW& 3.39&[2.71;3.90]&1.11&[0.978;1.30]&0.949&[0.832;1.00]\\
SE& 0.430&[0.181;0.921]&1.36&[1.20;1.42]&0.844&[0.770;0.912]\\
SW &1.64&[1.13;1.93]&1.34&[1.25;1.48]&0.925&[0.832;0.971]\\
ALL&1.53&[1.35;1.81]&1.25&[1.21;1.32]&0.896&[0.866;0.931]\\\hline
\end{tabular}
\label{table:bgd_fit4obs}
\end{center}
\end{table*}

\subsubsection{The source emission}\label{sec:source_spec}
\begin{figure}\begin{center}
  \includegraphics[height=0.8\columnwidth,angle=0]{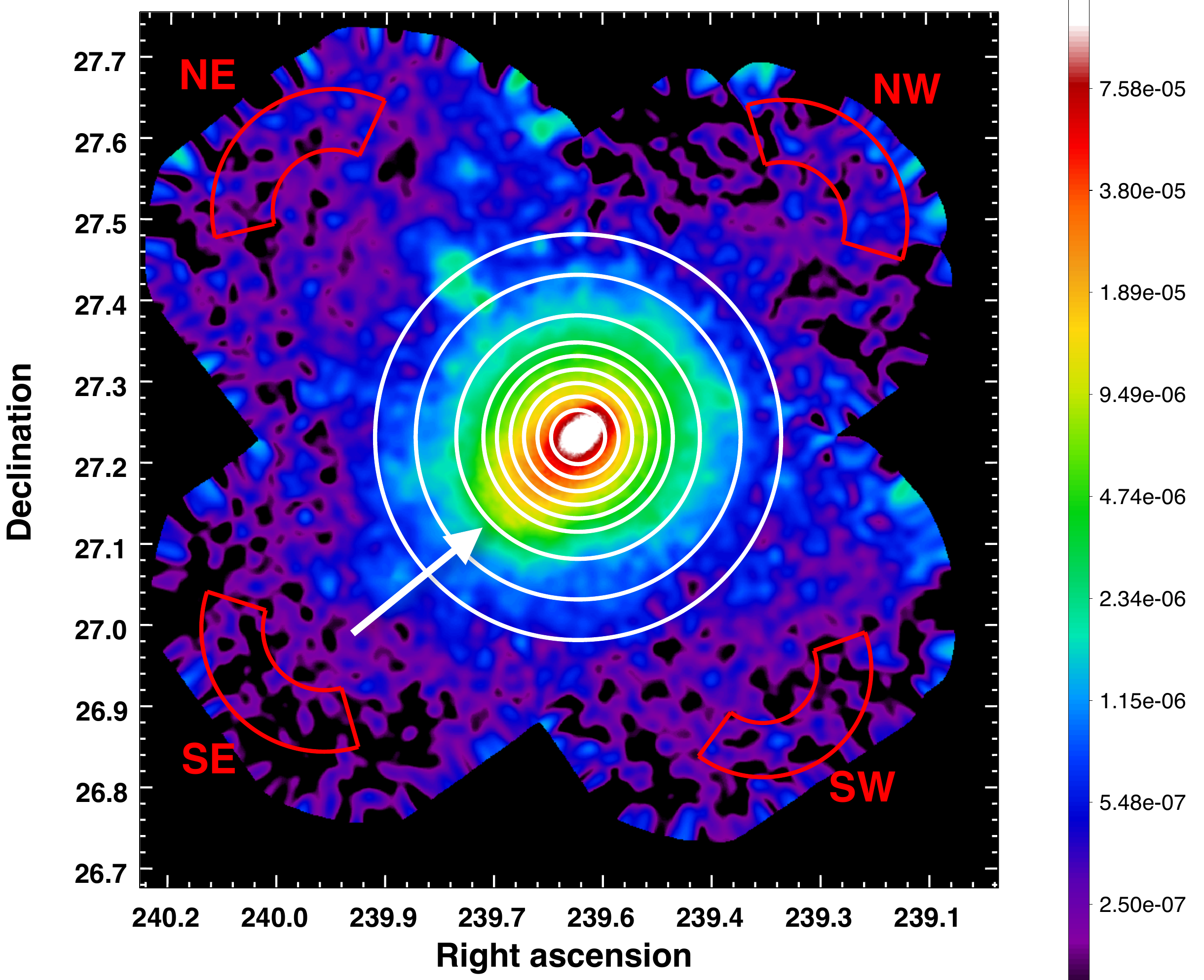}
\caption{Combined XMM-Newton mosaic in the energy band [0.7-1.2] keV corrected for the different exposure times and for the NXB. The units in the color bar are MOS count/s. The concentric white circles show the regions chosen for the source spectral extraction. The outermost circle has a radius of 15 arcmin (corresponding to $\sim$1530 kpc). The four regions delimited by the red curves have been used to estimate the local sky background components. The labels represent the four regions used in the analysis: the north east (NE), north west (NW), south east (SE) and south west (SW) observations. The white arrow indicates the location of the outermost cold front \citep[][see Sect.~\ref{sec:sloshing}]{rossetti13}. The tip of the accreting substructure observed in the region NE, as reported in \citet{eckert1a2142}, is located at approximatively (R.A.=239.72$^{\circ}$, Dec=27.40$^{\circ}$). }
\label{fig:annulus_region}\end{center}
\end{figure}

\begin{figure}
\begin{center}
  \includegraphics[height=0.7\columnwidth,angle=0]{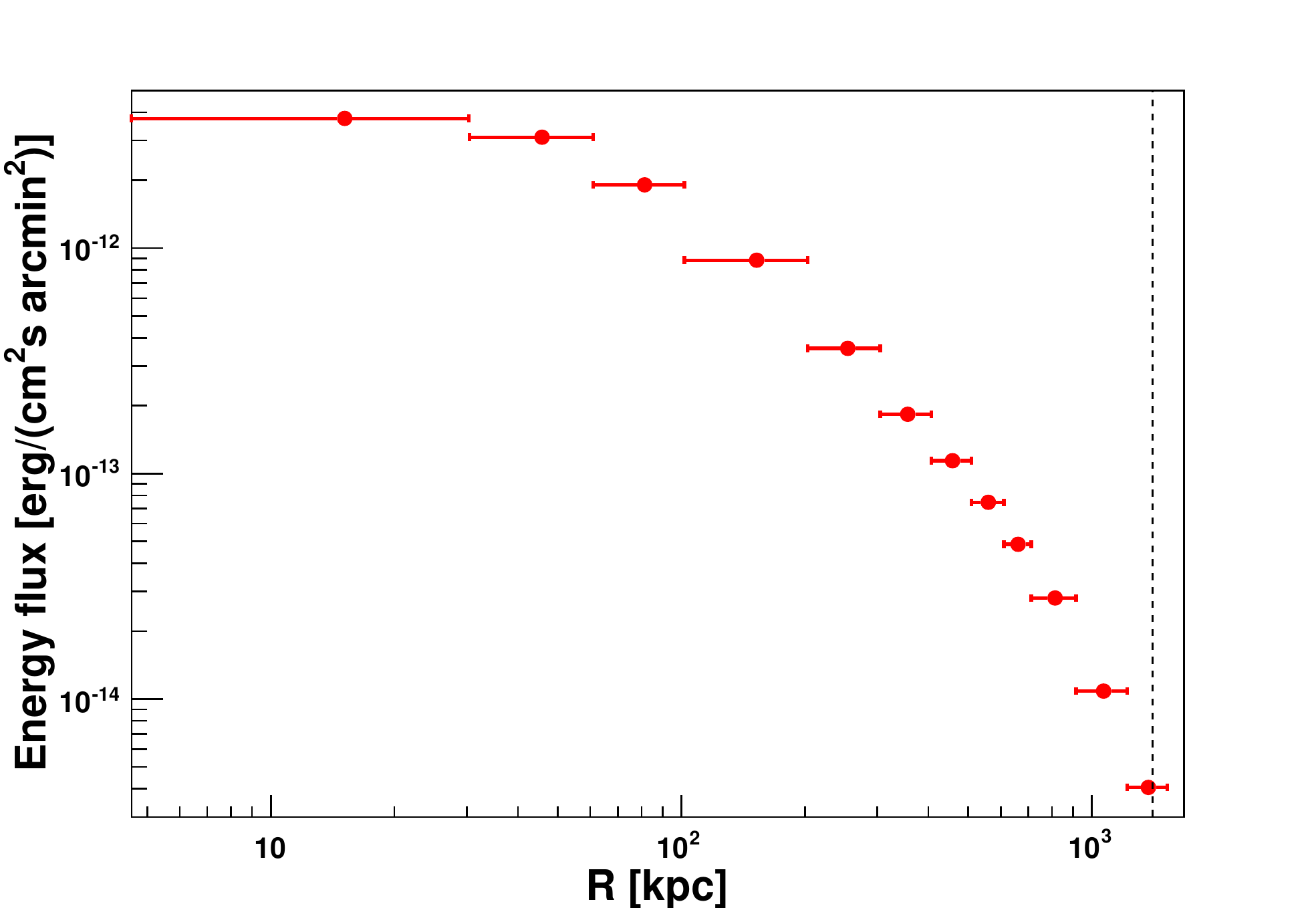}
\caption{Surface-brightness profile obtained by spectral fitting of the regions shown in Fig.~\ref{fig:annulus_region}. The dashed line represents the location of $R_{500}$. }
\label{fig:norm2D}
\end{center}
\end{figure}

\begin{figure}
\begin{center}
  \includegraphics[height=0.7\columnwidth,angle=0]{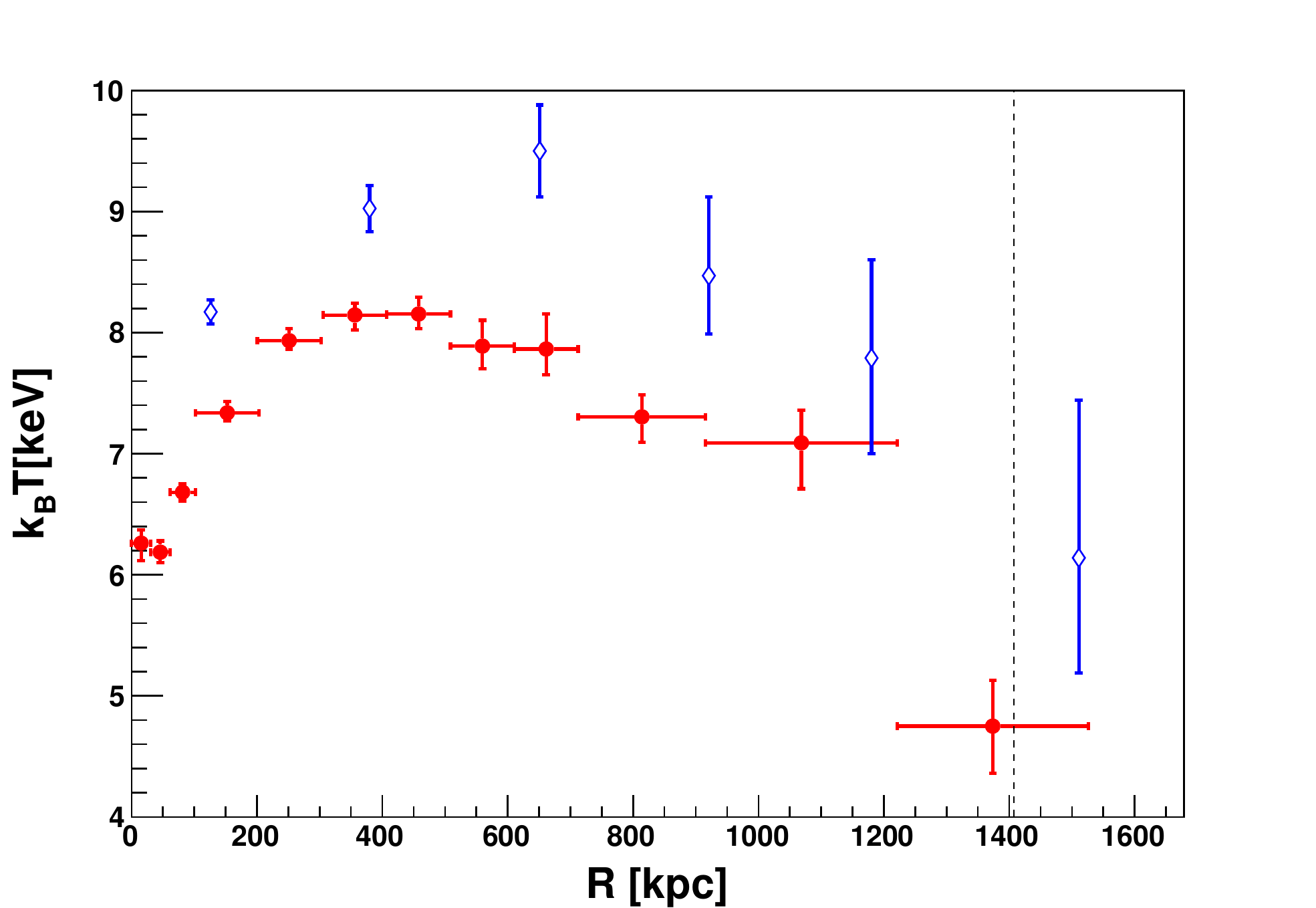}
\caption{Temperature profile. Red: \emph{XMM-Newton} measurements for the regions shown in Fig.~\ref{fig:annulus_region}. Blue: \textit{Suzaku} results \citep[from][]{akamatsu11}. The dashed line represents the value of $R_{500}$.}
\label{fig:T2D}
\end{center}
\end{figure}

\begin{figure}
\begin{center}
  \includegraphics[height=0.7\columnwidth,angle=0]{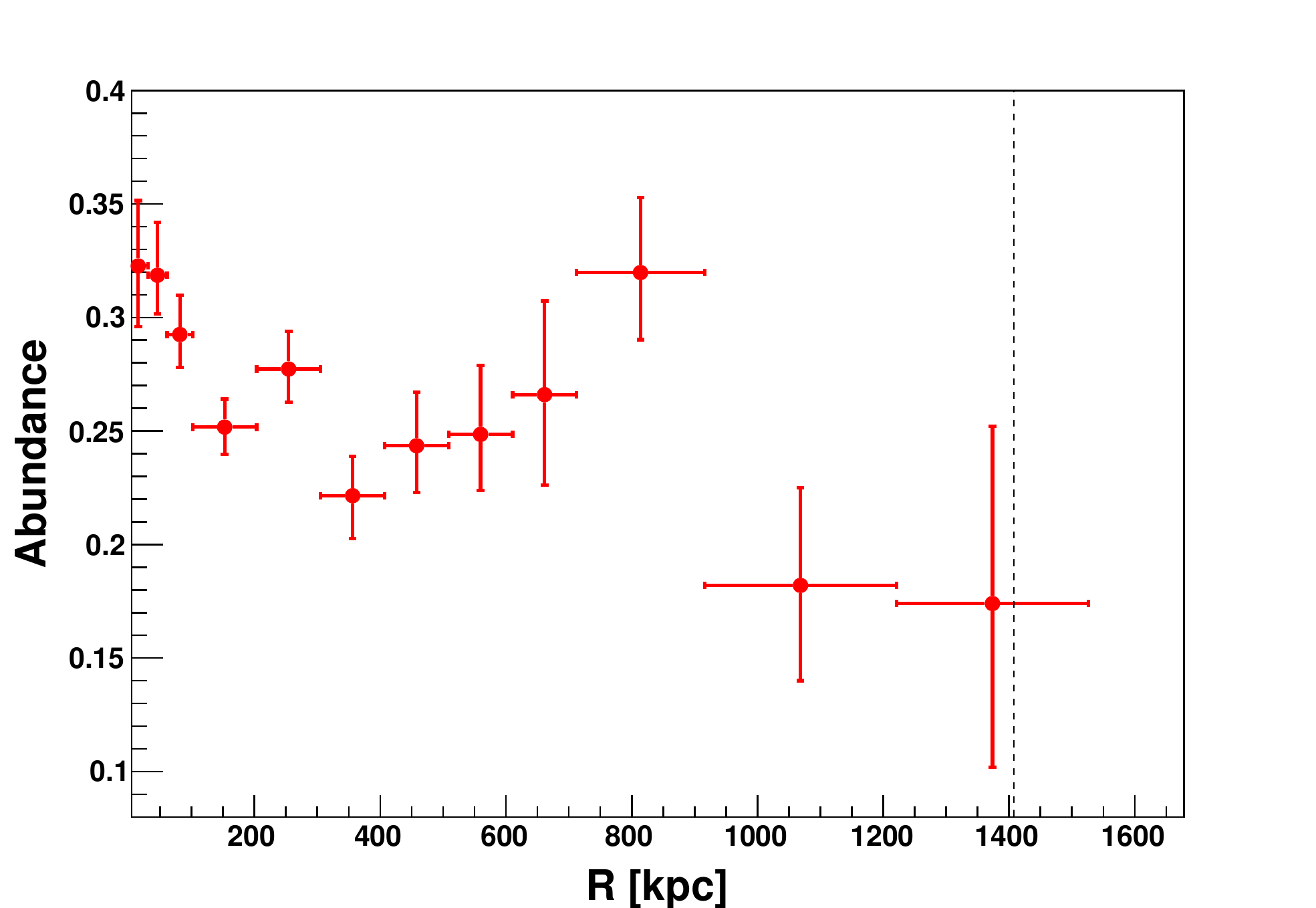}
\caption{Metal abundance profile for the regions shown in Fig.~\ref{fig:annulus_region}. The dashed line indicates the location of $R_{500}$.}
\label{fig:abundance2D}
\end{center}
\end{figure}

We extracted spectra from concentric annuli centered on the cluster (R.A.=239.58$^{\circ}$, Dec=27.23$^{\circ}$) as depicted in Fig.~\ref{fig:annulus_region} by the white circles. In each annulus, we fitted the resulting spectra with the thin plasma emission code \textit{APEC} and we derived projected radial profiles of emission measure, temperature and metal abundance. The surface brightness profile obtained from the normalization of the \textit{APEC} thermal model and the temperature profile are shown in Figs.~\ref{fig:norm2D} and \ref{fig:T2D}, respectively. The excellent data quality allowed us to extract the abundance profile up to 15 arcmin ($\sim R_{500}$). The resulting abundance profile is shown in Fig.~\ref{fig:abundance2D}, and exhibits a slightly decreasing behavior from $Z=0.35Z_\odot$ in the core down to $\sim0.15Z_\odot$ at $R_{500}$, where $Z_\odot$ represents the Solar abundance \citep{anders89}. We note that the best-fit temperature and normalization are relatively unaffected by the metal abundance even in the outermost bins. Indeed, fixing the metal abundance to $0.25Z_\odot$ instead of leaving it free to vary does not change the output parameters. For completeness, we also show the two-dimensional spectroscopic temperature profile obtained by \textit{Suzaku} \citep{akamatsu11} in Fig.~\ref{fig:T2D}. We note that the temperatures measured by \textit{Suzaku} significantly exceed the ones obtained in our analysis in the central regions. However, the temperature profile derived by \citet{akamatsu11} was extracted only along the NW, while the one obtained in our analysis is azimuthally averaged. The spectroscopic temperature profile extracted with \textit{XMM-Newton} in the NW direction agrees with the one of \citet{akamatsu11} (see Appendix~\ref{T_NW}, Fig.~\ref{TemperatureAPEC_NW}), which shows that the difference is caused by genuinely higher temperatures in the NW direction rather than by systematic differences between the two instruments. This conclusion is reinforced by independent observations performed with Chandra \citep[see the temperature map in the work of ][]{owers09}, which also indicate an increase of the temperature in the NW direction.

The best-fit values for the parameters are listed in Table~\ref{table:source_fit}. For the outermost three annuli (7-9, 9-12, and 12-15 arcmin), we performed a combined fit of the different observations (NE, NW, SW, SE and center, as defined in Fig.~\ref{fig:annulus_region}).  To do so, we fixed the sky background components of each observation to the values obtained in the previous section (see Table~\ref{table:bgd_fit4obs}) and rescaled them by the ratio of the source area to the background area.
  
 The parameter values shown in Table~\ref{table:source_fit} are the results of the combined fit. The results of the fit performed in each region separately are listed in Tables ~\ref{table:source_indivfit8}, \ref{table:source_indivfit9}, and \ref{table:source_indivfit10}. We note that we did not include neither the NE region in the combined fit of the annulus 7-9 arcmin, nor the central region in the combined fit of the annulus 12-15 arcmin, because the areas of overlap were too small to be included in the analysis.

\begin{table*}
\caption{Fit of the \emph{XMM-Newton} spectra with an \textit{APEC} model in the regions delimited by the concentric green circles in Fig.~\ref{fig:annulus_region}. The free parameters of the model are the temperature (in keV), the norm (in $10^{-3}$cm$^{-5}$) and the abundance (in solar metallicity). For the outermost three radial bins, all the available observations were combined (for details, see text and Tables ~\ref{table:source_indivfit8} -\ref{table:source_indivfit10} for the results of the individual fits).}
\begin{center}
\begin{tabular}{lllllll}\hline
R(arcmin) & T & $\Delta $T & norm & $\Delta$norm & Z & $\Delta$ Z\\\hline \hline
0-0.3 &6.26&[6.12, 6.37]&7.67&[7.60, 7.71]&0.321&[0.295, 0.350]\\
0.3-0.6&6.19&[6.10, 6.28]&6.38&[6.33, 6.41]&0.319&[0.301, 0.342]\\
0.6-1&6.68&[6.61, 6.75]&3.92&[3.90, 3.93]&0.293&[0.278, 0.310]\\
1-2&7.34&[7.27, 7.43]&1.82&[1.81, 1.82]&0.252&[0.240, 0.264]\\
2-3&7.93&[7.86, 8.03]&7.38$\cdot10^{-1}$&[7.34$\cdot10^{-1}$, 7.40$\cdot10^{-1}$]&0.277&[0.263, 0.294]\\
3-4&8.15&[8.02, 8.24]&3.77$\cdot10^{-1}$&[3.75$\cdot10^{-1}$, 3.78$\cdot10^{-1}$]&0.221&[0.203, 0.239]\\
4-5&8.15&[8.03, 8.29]&2.34$\cdot10^{-1}$&[2.33$\cdot10^{-1}$, 2.36$\cdot10^{-1}$]&0.244&[0.223, 0.267]\\
5-6&7.89&[7.70, 8.10]&1.53$\cdot10^{-1}$&[1.52$\cdot10^{-1}$, 1.54$\cdot10^{-1}$]&0.248&[0.224, 0.279]\\
6-7&7.86&[7.65, 8.15]&9.98$\cdot10^{-2}$&[9.90$\cdot10^{-2}$, 1.01$\cdot10^{-1}$]&0.266&[0.226, 0.307]\\
7-9&7.30&[7.10, 7.49]&5.77$\cdot10^{-2}$&[5.72$\cdot10^{-2}$, 5.81$\cdot10^{-2}$]&0.320&[0.290, 0.352]\\
9-12&7.09&[6.71, 7.36]&2.23$\cdot10^{-2}$&[2.21$\cdot10^{-2}$, 2.25$\cdot10^{-2}$]&0.182&[0.140, 0.225]\\
12-15&4.75&[4.36, 5.13]&8.34$\cdot10^{-3}$&[8.12$\cdot10^{-3}$, 8.57$\cdot10^{-3}$]&0.174&[0.102, 0.252]\\ \hline
\label{table:source_fit}
\end{tabular}
\end{center}
\end{table*}

%%%%%%%%%%%%%%%%
%%%%%%%%%%%%%%%%

%%%%%%%%%%%%%%%%
%%%%%%%%%%%%%%%%

%%%%%%%%%%%%%%%%
%%%%%%%%%%%%%%%%

%%%%%%%%%%%%%%%%
%%%%%%%%%%%%%%%%

%%%%%%%%%%%%%%%%%%%%%%%%%%%%
\section{X-ray and SZ imaging analysis}\label{sec:deproj}
%%%%%%%%%%%%%%%%%%%%%%%%%%%

\subsection{X-ray surface brightness profile}
Because of the faint cluster emission and of the relatively high background of \emph{XMM-Newton}, spectroscopic measurements beyond $\sim R_{500}$ are affected by systematic uncertainties \citep{leccardi08,ettori11}. For this reason, we adopted a different approach that allows us to extract surface brightness profiles with much poorer signal-to-noise than spectroscopic profiles and therefore out to $R_{200}$ and beyond (see Appendix \ref{app:blank-sky}). We refer to this surface brightness profile as "photometric" in the following, to be distinguished from the spatially limited "spectroscopic" surface brightness profile.

To obtain the photometric surface brightness profile we extracted photon images in the energy band [0.7-1.2] keV  and created exposure maps for each instrument using the SAS task \texttt{eexpmap} and the \texttt{PROFFIT} v1.2 software \citep{eckert11}. The choice of the [0.7-1.2] keV band is motivated by the fact that this particular band maximizes the signal-to-noise ratio \citep{ettori10,ettori11} and avoids the bright and variable Al K$\alpha$ and Si K$\alpha$ fluorescence lines, without affecting too much the statistics. Surface-brightness profiles were accumulated in concentric annuli starting from the surface-brightness peak (R.A.=239.58$^{\circ}$, Dec=27.23$^{\circ}$), taking vignetting effects into account. NXB profiles were accumulated in the same regions from the NXB maps taking both the contribution of the quiescent particle background and the soft protons into account. To model the contamination from residual soft protons, we extracted the spectra of the entire observations and fitted the high-energy part of the spectra (7.5-12 keV) using a broken power-law model \citep[see][]{leccardi08}. A 2D model for the contamination of residual soft protons was created using the ESAS task \texttt{proton} following \citet{kuntz08}. The details of the soft-proton modeling technique are provided in Appendix \ref{app:sp}, and a careful validation using blank-sky pointings is presented in Appendix \ref{app:blank-sky} together with an assessment of systematic uncertainties. 

In addition, we also derived the azimuthal median surface brightness profile following the method described in \citet{eckertclumping}. Namely, Voronoi tessellation was applied on the count image to create an adaptively binned surface-brightness map with a minimum of 20 counts per bin. The median surface brightness was then estimated in each annulus by weighting the surface brightness of each bin by its respective surface. To estimate the uncertainty in the median, we performed $10^4$ bootstrap resampling of the surface-brightness distributions binned uniformly using Voronoi tessellation. The standard deviation of the bootstrap realizations was then adopted as the error on the median. In the [0.7-1.2] keV energy band, the systematic uncertainty in the subtraction of the background amounts to 5\% of the sky background component, as shown by an analysis of a set of 22 blank-sky pointings (see Appendix \ref{app:blank-sky}). This uncertainty was added in quadrature to the surface-brightness profiles. To convert the resulting surface brightness profiles into emission measure, we folded the \textit{APEC} model through the \textit{XMM-Newton} response and computed the conversion between count rate and emission measure. We note that the conversion factor is roughly independent of the temperature in the energy band [0.7-1.2] keV, provided that the temperature does not fall below $\sim1.5$ keV. 

\begin{figure}
\begin{center}
  \includegraphics[height=0.7\columnwidth,angle=0]{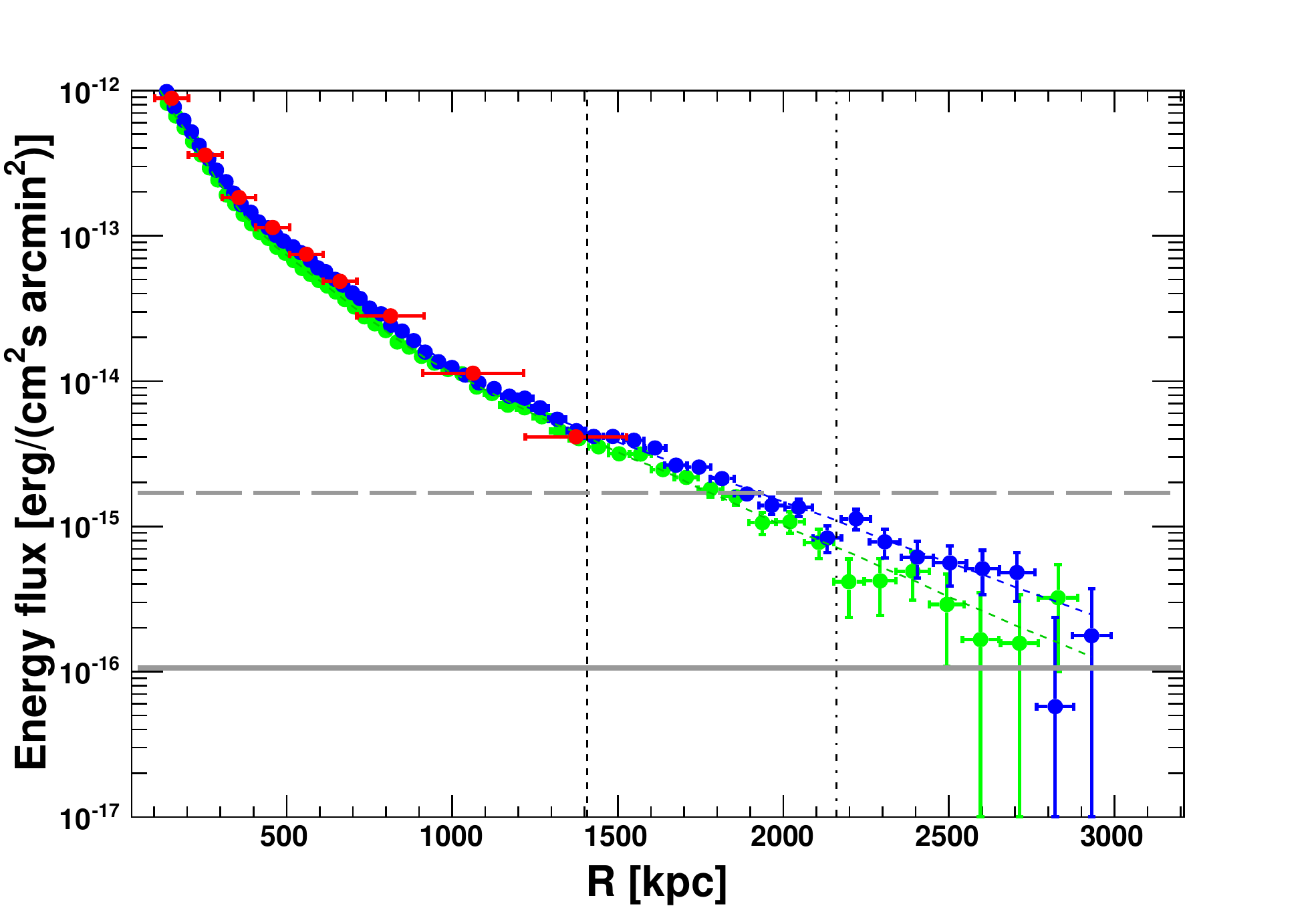}
\caption{Surface-brightness profiles of A2142 obtained with different methods. The data points show the spectroscopic measurements (red), the azimuthally averaged profile (blue), and the azimuthal median (green). The green and blue thin dashed lines show the corresponding best-fits obtained with the multiscale deprojection method. The solid and dashed horizontal lines correspond to the total background level, and to the uncertainty on the background, respectively. The dashed and dashed-dotted vertical lines represent $R_{500}$ and $R_{200}$, respectively.}
\label{fig:compaSB}
\end{center}
\end{figure}

In Fig.~\ref{fig:compaSB} we show the comparison between the spectroscopic and the photometric surface-brightness profiles. An excellent agreement is found between the profiles obtained with the two methods. We can see that \emph{XMM-Newton} detects a significant emission out to almost 3~Mpc from the cluster core, which corresponds to the virial radius $R_{100}\sim2R_{500}$.

%%%%%%%%%%%%%%%%
%%%%%%%%%%%%%%%% 

\subsection{\textit{XMM-Newton} deprojected electron density profile}\label{sec:density}

The normalization of the \textit{APEC} model, in units of cm$^{-5}$, is related to the electron density ($n_e$) by
\begin{equation}\label{eq:norm}
\mbox{Norm}=\frac{10^{-14}}{((1+z)\cdot D_a)^2}\int n_en_pd^3r,
\end{equation}
where $D_a\sim 349.8$~Mpc is the angular distance to the cluster in cm, $z=0.09$ is its redshift and $n_p$ is the proton density (in cm$^{-3}$), which is related to the electron density by $n_e=1.21n_p$, assuming that the plasma is fully ionized. 

After having converted the surface-brightness profile into emission measure, we deprojected the resulting profiles to estimate the 3D electron density profile. For the deprojection, we compared the output of two different methods: the multiscale method described in \citet{XXL} and an onion-peeling method \citep{ettori10}. Both methods assume spherical symmetry. The differences between the output of the two procedures thus gives us a handle of the uncertainties associated with the deprojection.

In the multiscale deprojection method, the projected profile is decomposed into a sum of multiscale basis functions. Each component can then be easily deprojected to reconstruct the 3D profile. Following \citet{XXL}, we decompose the projected profile into a sum of King profiles, with $s$ the projected radius, related to the line-of-sight distance and the 3D radius $r$, by: $r^2=s^2+\ell^2$

\begin{equation}
EM(s)=\sum_{i=1}^N N_{i}\left[1+\left(\frac{s}{r_{c,i}}\right)^2\right]^{-3\beta_i/2},
\end{equation}

where  $i$ represents the $i^{th}$ basis function and $s$ the projected radius. The parameters of this fit are the normalization ($N_i$), the core radii ($r_{c,i}$) and the slopes ($\beta_i$). The number of components and the core radii used for the fit of the projected profile are determined adaptively from the total number of data points, with the condition that one basis function is used for each block of 4 data points. The relation between the projected and 3D profiles can then be computed analytically \citep[see Appendix A of][for details]{XXL}. This method provides an adequate representation of the observed profile and of the underlying density 3D profile, although the derived parameters have no actual physical meaning. The confidence intervals are derived using the Markov Chain Monte Carlo (MCMC) code \texttt{emcee} \citep{foreman13}. In the following, all chains have a burn-in length of 5,000 steps, and contain 10,000 steps. Chains are started from the best-fit parameters. All reported errors correspond to 68\% confidence interval around the median of the MCMC distribution.

As an alternative, we used a direct non-parametric geometrical deprojection method based on the method of \citet{fabian81}  \citep[see also][]{kriss83,mclaughlin99,buote00}.  
The observed surface brightness profile is considered as the sum along the line of sight of the gas emissivity weighted by the fraction of the shell volume sampled in the given annular ring. From the outermost radial bin, and moving inward with the ``onion-peeling'' method, the gas emissivity (and density) is recovered in each shell. To avoid un-physical solutions induced by sharp fluctuations in the surface brightness profile, the radial points that deviate more than 2-$\sigma$ from the median-smoothed profile are replaced by the latter values.
In the present case, only 2 (out of 64) data points have been replaced. The error bars are estimated from the distribution of the deprojected values of the 100 MonteCarlo realizations of the X-ray surface brightness profile.

\begin{figure}
\begin{center}
  \includegraphics[height=0.7\columnwidth,angle=0]{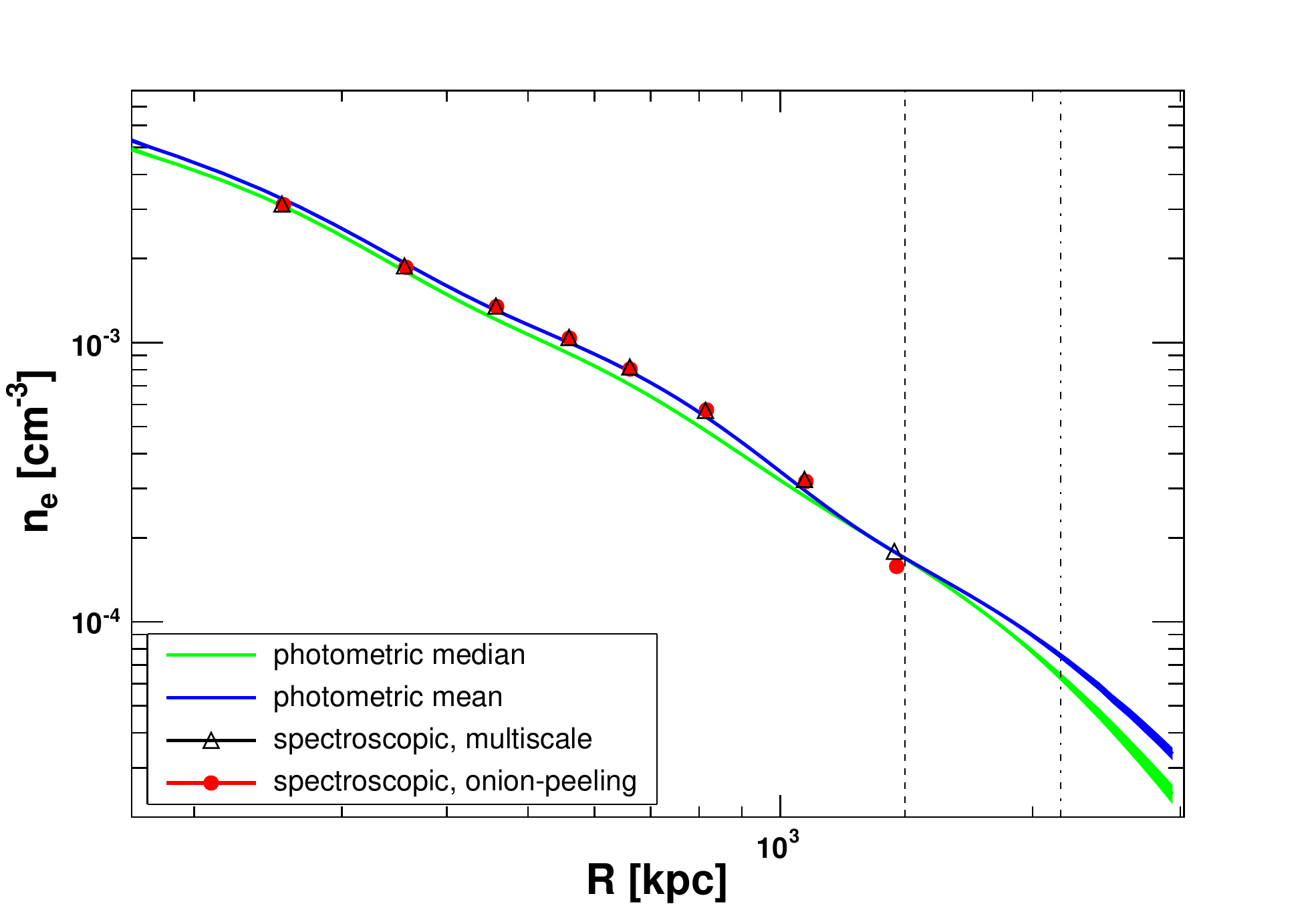}
\caption{Electron density profile. The data points show the deprojected spectroscopic data (from Table \ref{table:source_fit}) using the multiscale method \citep[black triangle,][]{XXL} and the onion peeling method \citep[red dot,][]{ettori10}. The green and blue data curves show the density profiles recovered using the azimuthal median and azimuthal mean photometric surface brightness profiles, respectively. Both profiles were deprojected using the multiscale method. The dashed and dashed-dotted vertical lines represent $R_{500}$ and $R_{200}$, respectively.}
\label{fig:A2142_n}
\end{center}
\end{figure}

The density profiles obtained with these two methods for the spectroscopic surface brightness are shown in Fig.~\ref{fig:A2142_n}. They are in excellent agreement. 
In this figure, we also compared these spectroscopic profiles to the azimuthal mean and azimuthal median density profiles obtained from the photometric analysis (Sect. 3.1). Both photometric profiles have been calculated using the multiscale deprojection method. Hereafter, we adopt the multiscale deprojection method to provide the gas density profiles of reference. As expected, the spectroscopic data points are following the trend of the azimuthal mean density profile, and overestimate the density in each shell compared to the azimuthal median. The systematic difference is even more visible than in the \textit{APEC} norm profiles (Fig.~\ref{fig:compaSB}). This illustrates the potential bias induced by gas clumping when using the spectral fitting procedure.

%%%%%%%%%%%%%%%%
%%%%%%%%%%%%%%%%

%%%%%%%%%%%%%%%%
\subsection{Clumping factor profile}\label{sec:clumping}
In X-rays, the measured emissivity provides information on $\langle n_e^2\rangle$, where $\langle \cdot\rangle$ represents the mean inside spherical shells. The level of inhomogeneities in the ICM can be estimated by the clumping factor $C$, as $C=\langle n_e^2\rangle/\langle n_e\rangle^2$ \citep{mathiesen99}. 

This definition of the clumping factor reflects variations of the gas density in a given volume. Such variations are expected to be accompanied by variations of others thermodynamic properties. In the following, we exploit the fact that the X-ray volume emissivity in the energy range addressed in this work ([0.7-1.2] keV) is essentially independent of the gas temperature, which allows us to use the analysis based on the X-ray photometry as a direct proxy of the clumping factor.

Given that the density distribution inside a shell can be described by a log-normal distribution skewed with denser outliers, the median of the density distribution is robust against the presence of outliers, whereas the mean of the distribution overestimates the density inside the considered region \citep{zhu13,eckertclumping}. Thus the ratio between the azimuthal mean and the azimuthal median density profiles (see Fig.~\ref{fig:A2142_n}) can be used as an estimator of the square root of the clumping factor profile.

\begin{figure}
\begin{center}
  \includegraphics[height=0.7\columnwidth,angle=0]{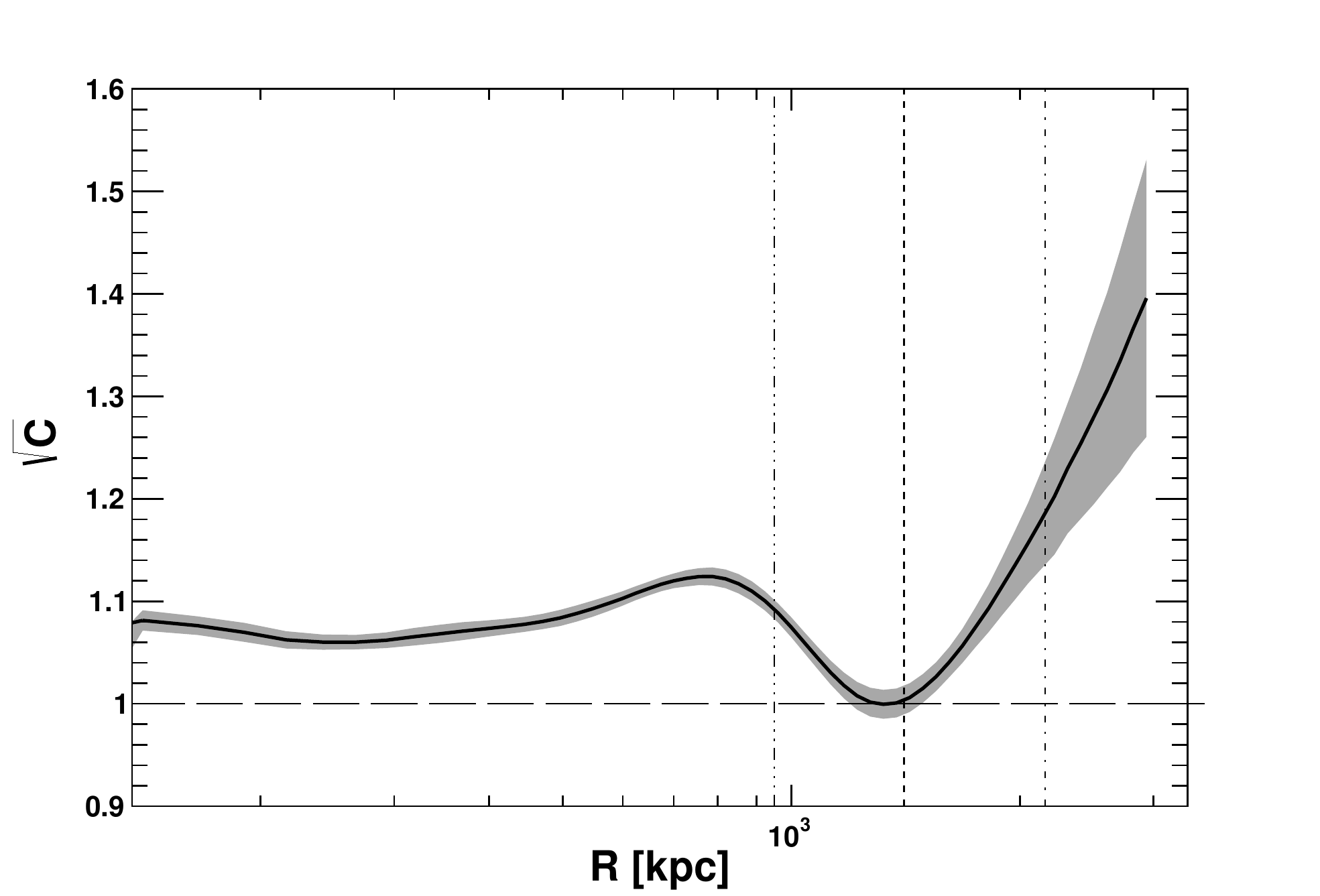}
\caption{Clumping factor profile. Solid line: median of the MCMC simulation; shaded area: 68\% confidence interval around the median. The dashed and dashed-dotted vertical lines represent $R_{500}$ and $R_{200}$, respectively. The triple dot-dashed line shows the approximative position of maximal radius of the sloshing region reported in \citet{rossetti13}.}
\label{fig:A2142_clumping}
\end{center}
\end{figure}

The resulting clumping factor is shown in Fig.~\ref{fig:A2142_clumping}. Beyond $R_{500}$, we observe that the clumping factor increases with the distance to the cluster center, while at smaller radii, the clumping factor is roughly constant at the value $\sqrt{C}=1.1$, followed by a decrease at about 1~Mpc from the cluster core. This behavior will be discussed in detail in Sect. \ref{sec:clumpingdisc}.

%%%%%%%%%%%%%%%%
%%%%%%%%%%%%%%%%

\subsection{Planck deprojected electron pressure profile}\label{sec:pressure}

To derive the electron pressure profile ($P_e$), we first need to estimate the thermal SZ signal from A2142. The SZ effect provides a measurement of the thermal pressure integrated along the line-of-sight (through the dimensionless $y$ parameter),
\begin{equation}
y(s)=\frac{\sigma_T}{m_ec^2}\int P_e(\ell)d\ell,
\end{equation}
where $\ell$ is the distance along the line of sight, $\sigma_T$ the Thomson cross section, $m_e$ the mass of the electron, and $c$ the speed of light. 

\begin{figure}
\resizebox{\hsize}{!}{\includegraphics{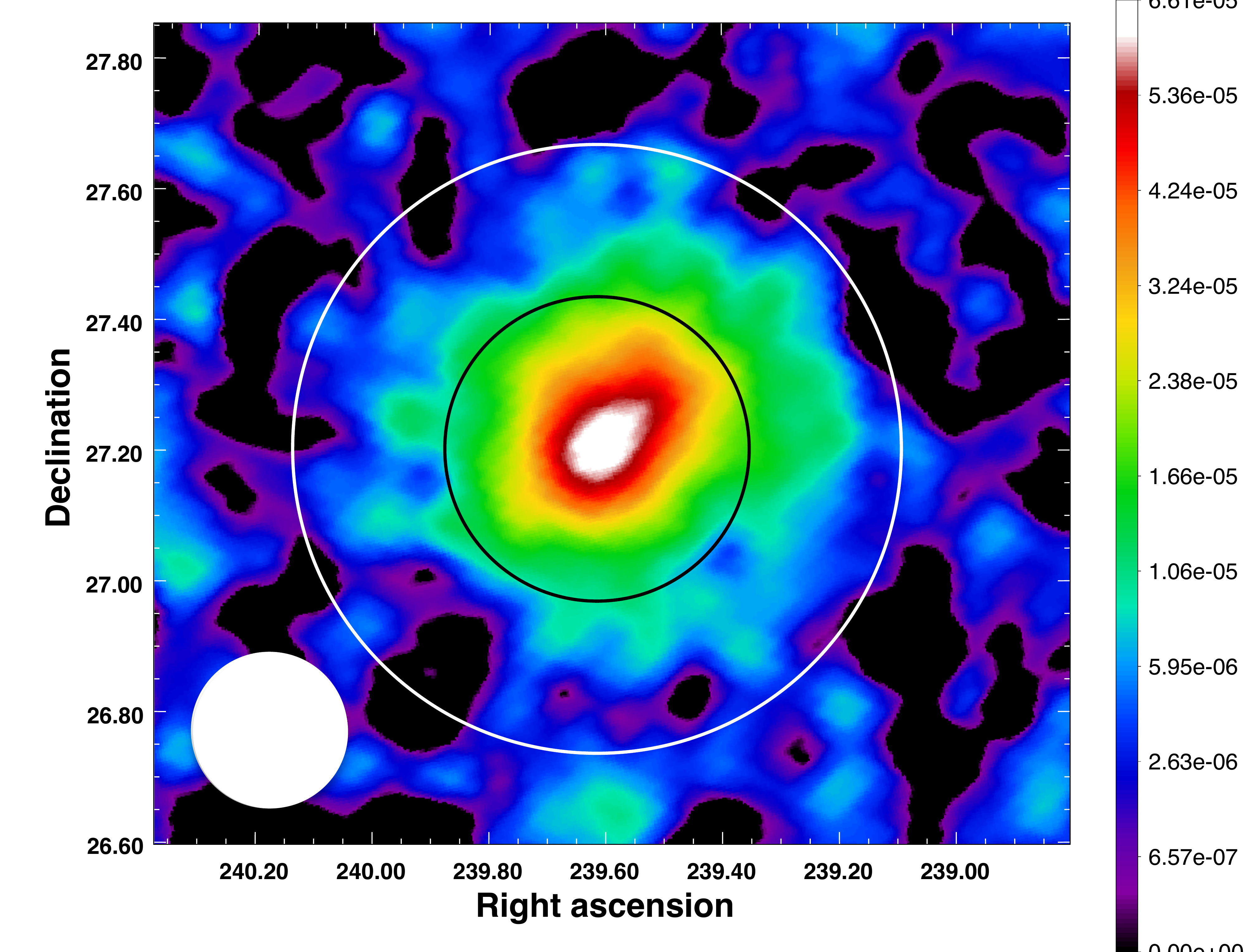}}
\caption{\emph{Planck} map of the Comptonisation parameter, $y$,  for A2142. The black and white circles indicate the approximate location of $R_{500}$ and $R_{100}\sim2R_{500}$, respectively. The white circle in the bottom left corner indicates the size of a 7 arcmin beam FWHM. }
\label{fig:ymap}
\end{figure}

We make use of the all-sky survey from the \emph{Planck} mission  \citep{tauber10, planckdr2015}, and more specifically from the full survey data from the six frequency bands of the high frequency instrument \citep{lamarre10,planckHFI}. The SZ signal map was reconstructed over a patch map of $1024\times1024$ pixels$^2$ centered at the location of A2142 and with a size of $20\times R_{500}$ ( i.e., 4.6~degrees). We applied the Modified Internal Linear Combination Algorithm \citep[MILCA,][]{hurier13}  to produce a map of the Comptonisation parameter, $y$, in a tangential Galactic coordinates referential. This algorithm was also used to produce the full sky $y$ map delivered by the Planck Collaboration to the community  \citep{planck14Ymap,planck15Ymap}. MILCA offers the possibility to perform the SZ signal reconstruction in multiple bins of angular scales. As a consequence, we have been able to produce a SZ map of A2142 at 7 arcmin FWHM angular resolution. Our A2142 SZ-map has therefore a significantly better resolution than the public full sky SZ-map at 10 arcmin FWHM. Thus, our SZ map uses the information from the 100 GHz Planck channel (roughly 10 arcmin FWHM) only for large angular scales \citep[see][for a more detailed description of the procedure]{hurier13}.
The resulting $y$-map for A2142  is shown in Fig.~\ref{fig:ymap}. 
A2142 was very well detected as an SZ source in the \emph{Planck} survey, with an overall signal-to-noise ratio of 28.4 \citep{planckSZ2013, planckSZ2015}. Due to its extension over the sky A2142 is among the clusters spatially resolved in the \emph{Planck} survey through its SZ signal which clearly extends well beyond $R_{500}$ out to $R_{100}\sim 2\times R_{500}$ (as shown in Fig.~\ref{fig:ymap}).

\begin{figure}
\begin{center}
  \includegraphics[height=0.9\columnwidth,angle=0]{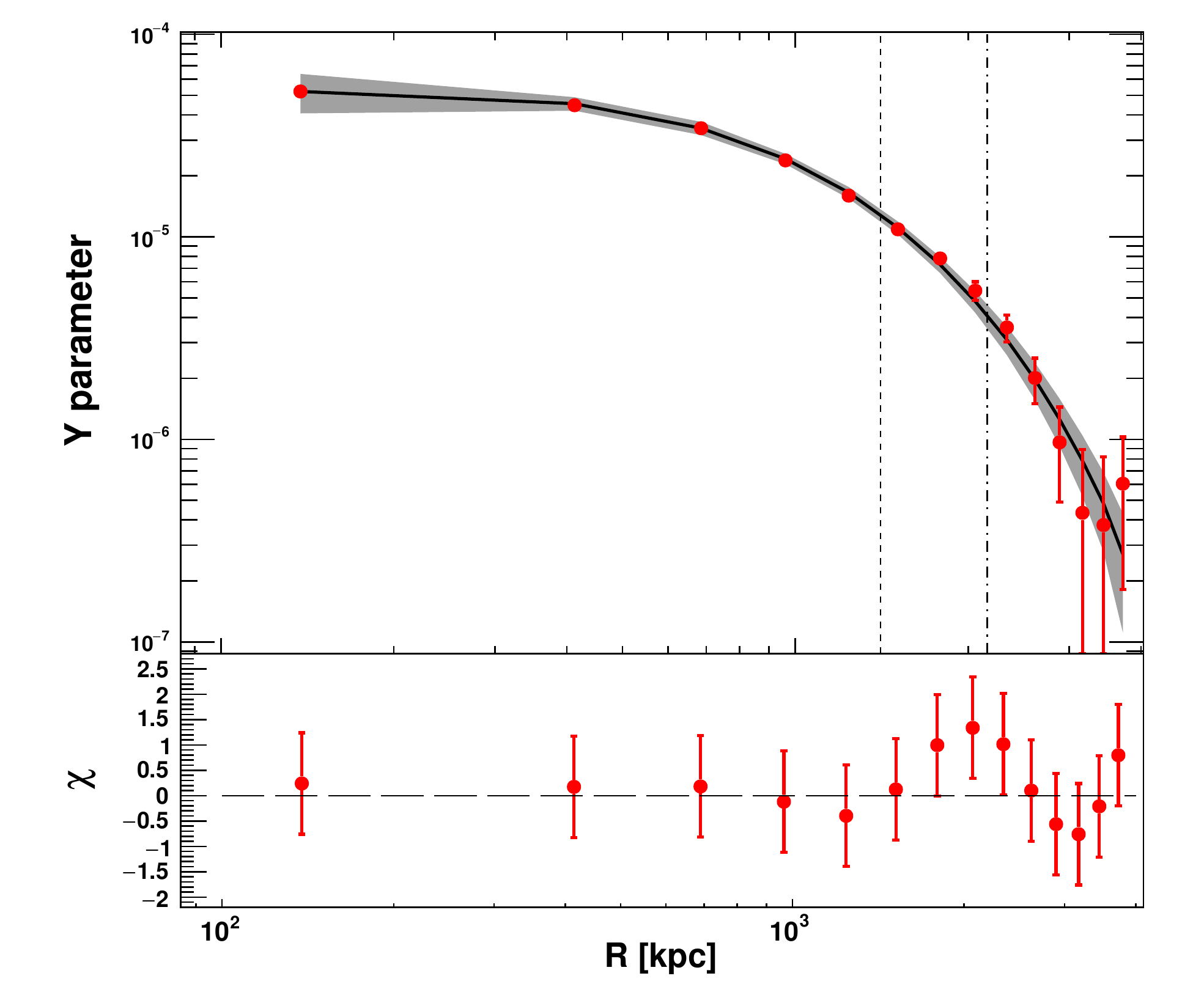}
\caption{Top panel: Compton $y$ parameter profile from \textit{Planck} data (red points). The grey solid line and shaded area show the best-fit profile convolved with the instrument PSF. The data points are correlated and the associated errors correspond to the square root of the diagonal elements of the covariance matrix. Bottom panel: residual of the fit to the \textit{Planck} data. The dashed and dashed-dotted vertical lines represent $R_{500}$ and $R_{200}$, respectively.}
\label{fig:Yparam}
\end{center}
\end{figure}

We further proceeded in extracting the $y$-parameter profile of A2142 from our MILCA $y$-map following the exact same method developed by \citet{planck13}. We recall that the $y$ profile is extracted on  a regular grid with bins of width $\Delta\theta/\theta_{500}=0.2$. The local background offset is estimated from the area surrounding the cluster beyond $5\times \theta_{500}=69$~arcmin. The resulting profile is shown in Fig.~\ref{fig:Yparam} together with a fit to the data obtained with the multi-scale method. For the fitting procedure, we take into account the covariance between the data points, which conveys the statistical properties of the noise of each \emph{Planck} frequency band used to compute the $y$-map and  the oversampling factor of our patch with respect to the 1.71~arcmin resolution element in the \emph{Planck} \verb|HEALPIX| map \citep{gorski05}. In addition, the model was convolved with the PSF of the instrument, which we approximated as a Gaussian with a full width at half maximum (FWHM) of 7~arcmin. The residual between the best fit convolved with the PSF and the $y$-parameter data is shown in the bottom panel of Fig.~\ref{fig:Yparam}. 

The best fit $y$-parameter profile (shown in Fig.~\ref{fig:Yparam}) was then converted into a 3D electron pressure profile using the multiscale deprojection method. For completeness, we also deprojected the y-parameter data using the same methodology as \citet{planck13}. In the latter case, the underlying pressure profile was  obtained from a real space deconvolution and deprojection regularisation method adapted from \citet{croston06} assuming spherical symmetry for the cluster. The correlated errors were propagated from the covariance matrix of the $y$ profile with a Monte Carlo procedure and led to the estimation of the covariance matrix of the pressure profile $P_e(r)$.

To compare the resulting SZ pressure profiles with the one obtained from purely X-ray analysis, we estimated the 3-dimensional pressure profile from the spectroscopic X-ray measurements using the method outlined in \citet{vikhlinin06}. In this method, the 3D temperature profile is assumed to be represented by a parametric form with a large number of free parameters,
\begin{equation}\label{eq:Tspec}
T(r)=T_0\frac{(r/r_{\rm cool})^{a_{\rm cool}}+T_{\min}/T_0}{(r/r_{\rm cool})^{a_{\rm cool}}+1}\frac{(r/r_t)^{-a}}{(1+(r/r_t)^b)^c/b)}.
\end{equation}

This functional form was projected along the line of sight weighted by the 3D emissivity profile, and subsequently fit to the observed temperature profile described in Sect.~\ref{sec:source_spec}. We then ran an MCMC to sample the parameter space and draw the 3D temperature profile. The 3D X-ray pressure profile was computed by combining the deprojected temperature with the electron density profile obtained from the spectral X-ray analysis (black points in Fig. \ref{fig:A2142_n}). 

\begin{figure}
\begin{center}
  \includegraphics[height=0.7\columnwidth,angle=0]{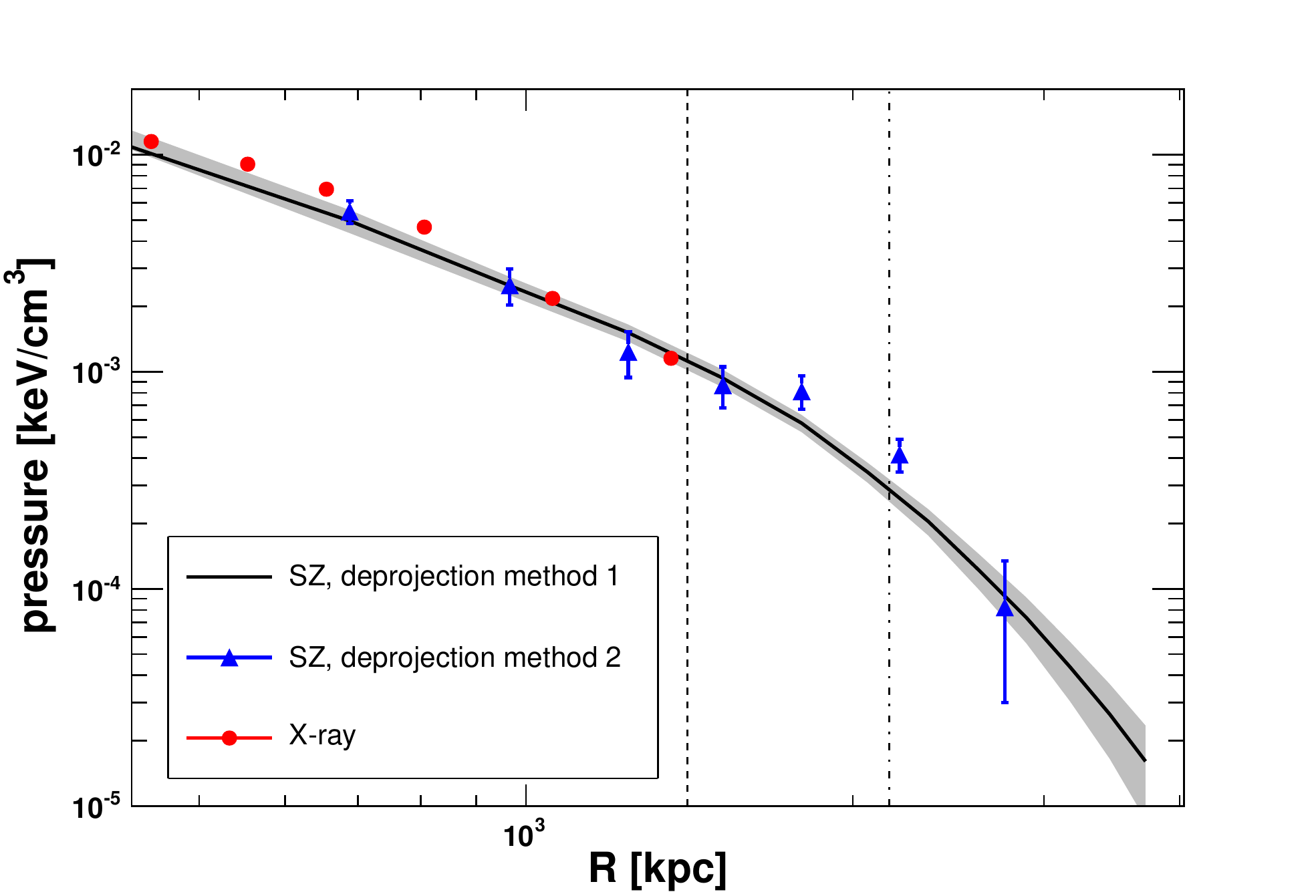}
\caption{Electron pressure profile.  The grey solid line and shaded area show the best-fit pressure profile obtained by deprojecting the SZ data using the multiscale method (deprojection method 1).  The red points show the spectroscopic X-ray data deprojected with the method of \citet{vikhlinin06} and the blue triangles show the result of the deprojection of the $y$-parameter data using the same methodology as \citet{planck13} (deprojection method 2). The blue data points are correlated and the associated errors correspond to the square root of the diagonal elements of the covariance matrix. The dashed and dashed-dotted vertical lines represent $R_{500}$ and $R_{200}$, respectively.}
\label{fig:A2142_p}
\end{center}
\end{figure}

In Fig.~\ref{fig:A2142_p} we show all three pressure profiles: the two SZ pressure profiles obtained by the two different deprojection methods described above (method 1: multiscale method; method 2: same methodology as  \citet{planck13}) and the spectroscopic X-ray pressure profile. All three pressure profiles are consistent, although we note a slight excess of the X-ray pressure profile compared to the SZ pressure profile at a distance of 500 kpc from the cluster center. Given that the thermal SZ signal is less affected by clumping \citep[e.g. ][]{roncarelli13}, this observed difference can be explained by fluctuations in the X-ray signal (see the value of the clumping factor $\sqrt{C}\sim1.1$ in this radial range in Fig.~\ref{fig:A2142_clumping}). We note also that around $R_{200}$ the two pressure profiles recovered from SZ observations are slightly different. This may be due to the fact that the multiscale deprojection method smoothes the fluctuations to fit the data with a superposition of King profiles. In addition to that, the points on the deconvolve/deprojected profile obtained using the same method as in \citet{planck13} are correlated. However, the errors shown on this figure are only the square root of the diagonal of the covariance matrix. This likely bias a direct visual comparison. The observed discrepancy around $R_{200}$ may therefore not be a physical effect. We will come back to the excess in the purely X-ray pressure profile compared to the SZ profile in the discussion section.

Due to the moderate resolution of the Planck satellite (translating into 7 arcmin on our y-map), we can not recover constraints on the SZ pressure profile from the y parameter measurements close to the cluster center. Therefore, we will  consider in the following only the radial range beyond 400 kpc ($\sim$ 4 arcmin), radius beyond which the constraints on the pressure profile from the SZ data are less impacted by the PSF blurring and therefore more reliable. 

%%%%%%%%%%%%%%%%
%%%%%%%%%%%%%%%%

%%%%%%%%%%%%%%%%
%%%%%%%%%%%%%%%%

%%%%%%%%%%%%%%%%%%%%%%%%%%%%
\section{Joint X-ray/SZ analysis}\label{sec:combRXSZ}
%%%%%%%%%%%%%%%%%%%%%%%%%%%

The combination of the X-ray and SZ signal can be used to recover the thermodynamical quantities that characterize the ICM. In this section, we combine the three-dimensional SZ pressure profile with the X-ray gas density profile to recover the radial distribution of temperature, entropy, hydrostatic mass, and gas fraction. Moreover, we can recover these quantities largely corrected for the effect of the clumped gas. This is obtained by comparing the X-ray surface brightness measured using the mean of the azimuthal photon counts distribution with and the one estimated with the median of the distribution \citep[as detailed in][]{eckertclumping}.

\subsection{Temperature profile}\label{sec:T}
Assuming that the ICM is an ideal gas, the joint X-ray/SZ 3D temperature profile can be recovered  by combining the X-ray density profiles obtained in Sect.~\ref{sec:density}  (Fig.\ref{fig:A2142_n}) with the SZ pressure profile derived in Sect.~\ref{sec:pressure} (Fig.~\ref{fig:A2142_p}). Using the equation $k_BT=P_e/n_e$, we derived the 3D temperature profile for both the azimuthal mean and the azimuthal median density profiles. We remind that while the density profile obtained using the azimuthal median is corrected for the presence of clumps, the one obtained using the azimuthal mean is not.  

\begin{figure}
\begin{center}
  \includegraphics[height=0.7\columnwidth,angle=0]{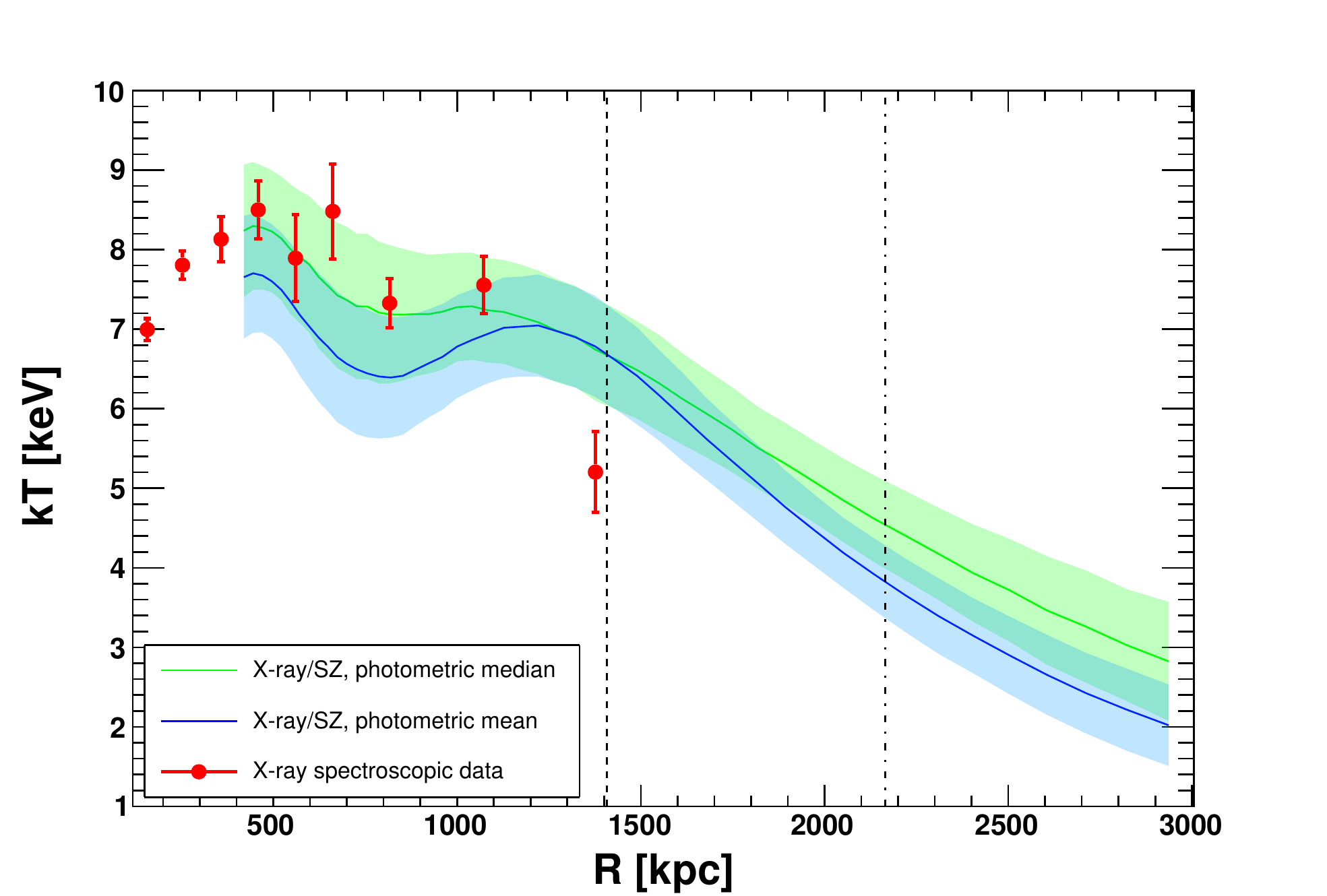}
\caption{Deprojected temperature profile.  Red points: Spectroscopic data (from Fig.~\ref{fig:T2D}) deprojected using the method of \citet{ettori10}; green and blue: combined X-ray/SZ profile using the azimuthal median and azimuthal mean density profiles, respectively. The dashed and dashed-dotted vertical lines represent $R_{500}$ and $R_{200}$, respectively.}
\label{fig:A2142_T}
\end{center}
\end{figure}

The resulting joint X-ray/SZ 3D temperature profiles are shown in Fig.~\ref{fig:A2142_T}. The uncertainties in the temperature profile were estimated by combining the MCMC runs for the pressure and density. At each radius, the temperature and its uncertainty were drawn from the distribution of output temperature values.  In this figure we also show the deprojected spectroscopic temperature profile obtained with the method of \citet{ettori10}. %Eq. (\ref{eq:Tspec}). 

As expected, we observe different behaviors of the temperature profile depending on if gas clumping is taken into account or not. Indeed, the increase in the clumping factor towards the outskirts (see Fig. \ref{fig:A2142_clumping}) causes the temperature profile obtained from the azimuthal mean to steepen with cluster-centric distance compared to the profile estimated using the azimuthal median technique. We also note that the spectroscopic X-ray profile closely follows the X-ray/SZ profile obtained using the azimuthal median (except for the very last data point, but this is an artefact of the deprojection method). We will come back to this point in the discussion section. 

The effect of the overestimate the density has a clear signature in the temperature profile. Similar effects are also expected in the other thermodynamic quantities derived from the X-ray analysis.

%%%%%%%%%%%%%%%%
%%%%%%%%%%%%%%%%
%%%%%%%%%%%%%%%%
\subsection{Entropy profile}\label{sec:S}
%%%%%%%%%%%%%%%%

Assuming that entropy is only generated by spherical virialisation shocks driven by hierarchical structure formation, we expect an entropy profile that follows the gravitational collapse model. In such a case, the gas with low entropy sinks into the cluster center, while the high-entropy gas expands to the cluster outskirts  \citep{voit05,pratt10} and the resulting entropy profile has a power-law shape given by,
\begin{equation}\label{eq:voit}
K(R)=K_{500}\cdot1.42\left(\frac{R}{R_{500}}\right)^{1.1}\text{ keV cm}^2.
\end{equation}
The quantity $K_{500}$ is defined as \citep{pratt10},
\begin{equation}
K_{500}=106\cdot \left(\frac{M_{500}}{10^{14}M_{\sun}}\right)^{2/3}\left(\frac{1}{f_b}\right)^{2/3}h(z)^{-2/3} \text{ keV cm}^2,
\end{equation}
where $M_{500}$=$8.66\cdot10^{14}M_{\odot}$  is the cluster mass at $R_{500}=1408$ kpc \citep[values derived from][]{munari14}, $f_b=\Omega_b/\Omega_m=0.15$ is the cosmic baryon fraction,  with $\Omega_b$ the baryon density, $\Omega_m$ the matter density, and $h(z)=\sqrt{\Omega_m(1+z)^3+\Omega_\Lambda}$ the ratio of the Hubble constant at redshift $z$ to its present value.

\begin{figure}
  \includegraphics[height=0.9\columnwidth,angle=0]{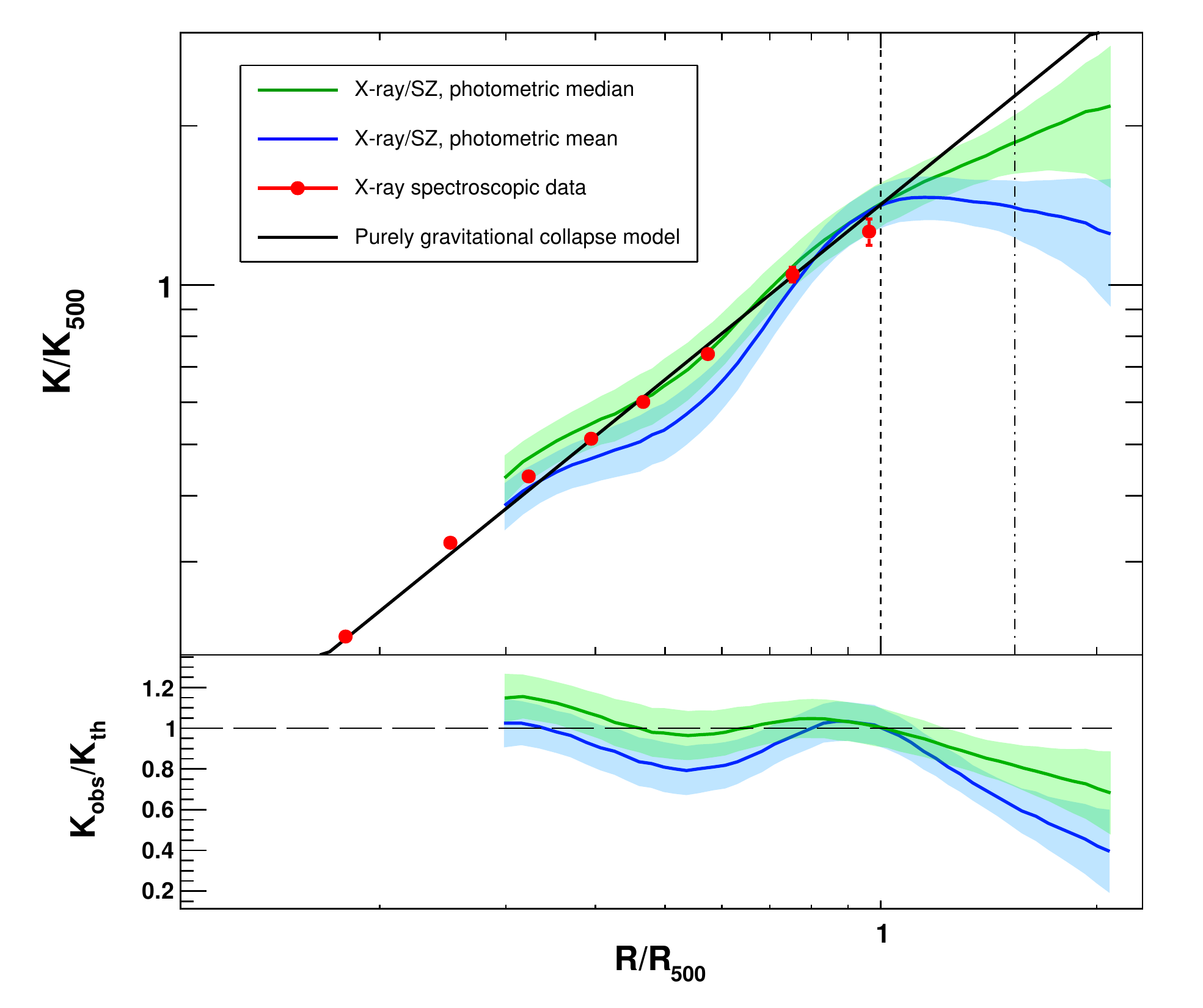}
\caption{Deprojected entropy profile. Top panel: the red dots show the X-ray spectroscopic data, while the green and blue curves represent the X-ray/SZ profiles obtained using the azimuthal median and azimuthal mean density profiles, respectively. The black line shows the expectation from purely gravitational collapse \citep[Eq.~(\ref{eq:voit}),][]{voit05}. The dashed and dashed-dotted vertical lines represent $R_{500}$ and $R_{200}$, respectively. Bottom panel: ratio of the X-ray/SZ entropy profile ($K_{obs}$) over the entropy profile expected from the purely gravitational collapse model ($K_{th}$):  for the azimuthal median (in blue) and azimuthal mean (in green) density profiles. The horizontal dashed line represents the expectation for $K_{obs}=K_{th}$.}
\label{fig:A2142_K}
\end{figure}

We derived the combined SZ and X-ray 3D entropy profile by using the equation $K=P_e/n_e^{5/3}$ with the X-ray density profiles obtained in Sect.~\ref{sec:density}  (Fig.~\ref{fig:A2142_n}) and the SZ pressure profile derived in Sect.~\ref{sec:pressure} (Fig.~\ref{fig:A2142_p}).
In Fig.~\ref{fig:A2142_K} we show the entropy profiles obtained using the azimuthal mean and median density profiles. For comparison, we also show the entropy profile obtained with the spectroscopic X-ray information ($K=k_BT/n_e^{2/3}$) from our deprojected temperature and gas density spectroscopic profiles. All profiles are rescaled by $K_{500}$ and compared to the expectations of the self-similar model \citep{voit05}. 
\\\\Excellent agreement is found between the X/SZ and spectroscopic X-ray profiles out to $R_{500}$.  

At larger radii, the use of a method sensitive to outliers leads to an entropy profile that deviates from the self-similar prediction ($\propto R^{1.1}$) in the outskirts and produces a feature which resembles an entropy flattening.

On the contrary, we can see that the X/SZ profile obtained using the azimuthal median method rises steadily with radius out to the maximum radius accessible in this study ($3000$ kpc $\approx R_{100}$).  
Therefore, if the presence of clumps is taken into account in the X-ray data the deviation observed with the blue curve almost completely disappears and at $R_{200}$ the entropy falls within just 1-$\sigma$ of the self-similar expectation.

We stress that the SZ effect is nearly insensitive to clumping \citep[e.g.][]{battaglia15}, the difference between the two profiles is caused only by our treatment of gas clumping in the X-ray data. This shows the importance of taking into account the effects due to the presence of clumps in the derivation of the thermodynamics quantities.

%%%%%%%%%%%%%%%%
%%%%%%%%%%%%%%%%

\subsection{Hydrostatic mass}\label{sec:Mtot}
Assuming that the ICM is in hydrostatic equilibrium within the gravitational potential of the cluster, the total enclosed mass at the distance $r$ from the cluster center can be estimated as
\begin{equation}\label{eq:Mtot}
\frac{dP_g(r)}{dr}=-\rho_g(r)\frac{\text{GM}_{\text{tot}}(<r)}{r^2},
\end{equation}
where $P_g=P_e+P_p$ is the gas pressure profile, $\rho_g=(n_e+n_p)\cdot m_p\mu$ is the gas mass density, with $m_p$  the mass of the proton, $\mu=0.6$ the mean molecular weight, and $G$ the universal gravitational constant.

Following \citet{ameglio09}, we combined the \textit{Planck} electron pressure profile (Sect.~\ref{sec:pressure}) with the \emph{XMM-Newton} electron density profile (Sect.~\ref{sec:density}) to derive the hydrostatic mass profile. As above, we investigated the effect of clumping on the hydrostatic mass by comparing the results obtained with the azimuthal mean and median density profiles. 

\begin{figure}
\begin{center}
  \includegraphics[height=0.78\columnwidth,angle=0]{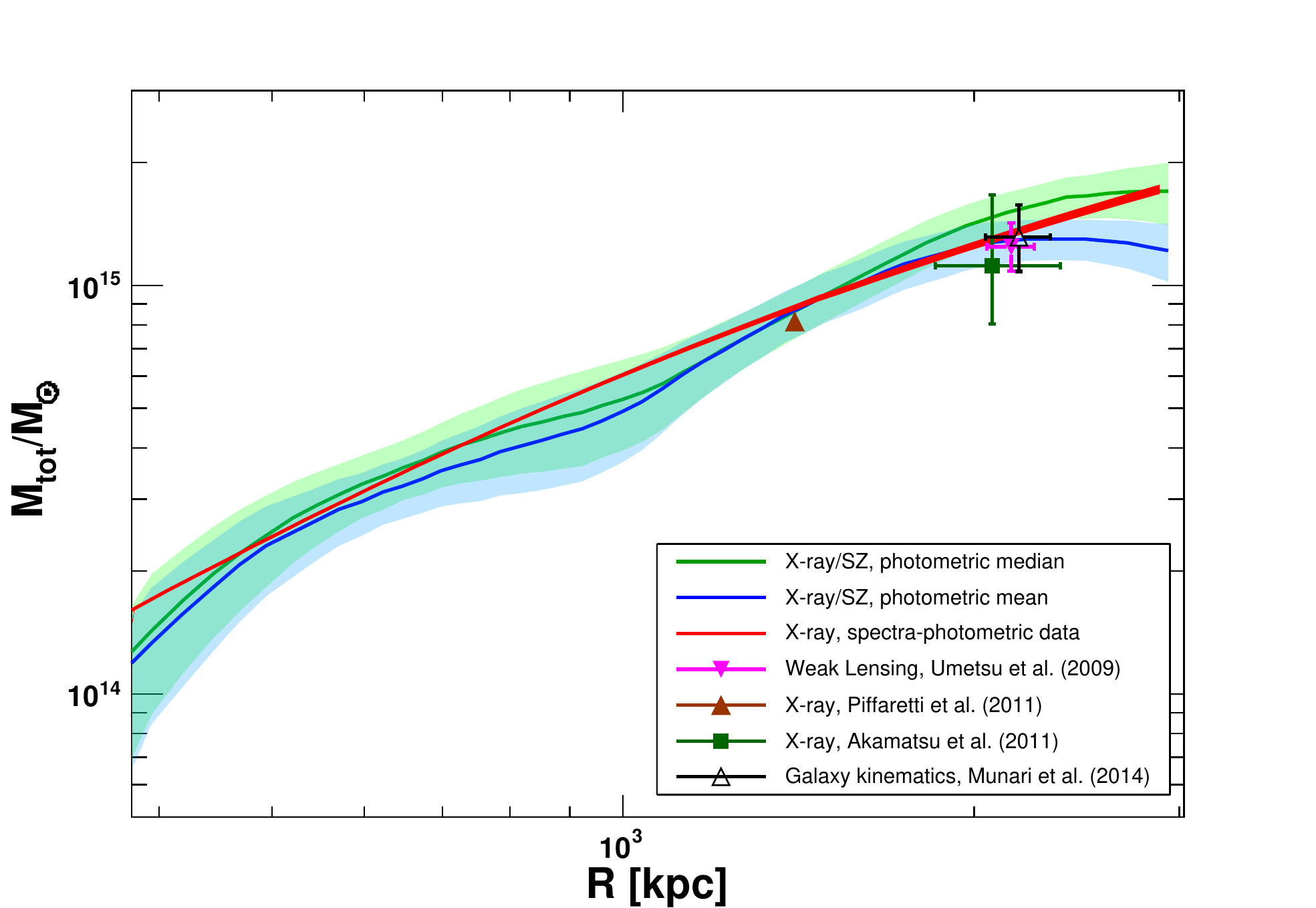}
\caption{Mass profile of A2142. Green: X-ray/SZ combined profile using the azimuthal median density profile. Blue: X-ray/SZ combined profile using the azimuthal mean density profile. Red: NFW fit to the spectroscopic X-ray data using the method of \citet{ettori10}. For comparison, we also plot the mass measurements reported in the literature. Brown triangle: $M_{500}$ from $L_X-M$ relation \citep{mcxc}; pink reversed triangle: $M_{200}$ from \emph{Subaru} weak lensing \citep{umetsu09}; dark green square: $M_{200}$ from \textit{Suzaku} X-ray \citep{akamatsu11}; black empty triangle: $M_{200}$ from galaxy kinematics \citep{munari14}.}
\label{fig:A2142_Mtot}
\end{center}
\end{figure}

In Fig.~\ref{fig:A2142_Mtot} we show the combined X-ray/SZ hydrostatic mass profiles obtained for the different input density profiles. The mass profile obtained using the azimuthal median increases steadily, while the one obtained with the azimuthal mean density profile shows an unphysical turnover at $R>R_{200}$. Such turnovers have been reported in the literature and interpreted as evidence for a significant non-thermal pressure contribution to sustain gravity \citep[e.g.][]{kawa10,bonamente13,fusco14}.

We also derived the hydrostatic mass profile using X-ray-only information with the method described in \citet{ettori10}. This method assumes a Navarro-Frenk-White (NFW) form for the underlying mass profile and uses the deprojected gas density profile to reproduce the observed temperature profile estimated with the spectral analysis by inversion of the hydrostatic equilibrium equation applied on a spherically symmetric object. The best-fit on the 2 parameters describing the NFW mass model (i.e. the concentration and $R_{200}$ in the present analysis) is then obtained using a $\chi^2$ minimization technique. Applying this method to the photometric median density profile, we measure a concentration $c= 3.00\pm 0.06$ and $R_{200}=2249 \pm 16$ kpc. Hereafter, all references to the method of \citet{ettori10} will be applied to the photometric median density profile.

In Fig. ~\ref{fig:A2142_Mtot} we compare the resulting mass profile with that obtained using the X/SZ method. We can see that the two methods lead to consistent results. Good agreement is found in particular between the X-ray-only and the median X-ray/SZ profiles. 
We also show the comparison of several mass measurements from the literature: \citet{akamatsu11} (\textit{Suzaku}, assuming hydrostatic equilibrium); \citet{umetsu09} (\textit{Subaru}, weak gravitational lensing); \citet{munari14} (optical spectroscopy, galaxy dynamics) and \citet{mcxc} (\emph{ROSAT}, $L_X-M$ relation). These measurements are summarized in Table~\ref{tab:mass}. All our mass measurements are consistent within the error bars with the mass measurements made in \citet{akamatsu11}, \citet{umetsu09}, \citet{munari14} and \citet{mcxc}.

\begin{table*}
\caption{\label{tab:mass}M$_{200}$ and r$_{200}$ corresponding to the three hydrostatic mass profiles shown in Fig.~\ref{fig:A2142_Mtot} resulting from the combined X-ray/SZ or spectroscopic X-ray study. For comparison, we also list the values of M$_{200}$ and $R_{200}$ obtained from \emph{Subaru} weak lensing \citep{umetsu09}, \emph{Suzaku} X-ray \citep{akamatsu11} and galaxy kinematics \citep{munari14}.}
\begin{center}
\begin{tabular}{lll}\hline
 & $M_{200} $[10$^{14}M_{\odot}$]&$R_{200}$[kpc] \\ \hline\hline
X/SZ, median &$16.1\pm 2.6$& $2347 \pm 154$ \\
X/SZ, mean &$12.9\pm 1.8$& $2179 \pm 129$ \\
X-ray, spectroscopic+median &$14.1\pm 0.3$ & $2249 \pm 16$ \\
X-ray, \emph{Suzaku} & $11.1^{+5.5}_{-3.1}$&$2080^{+300}_{-220}$ \\
Weak Lensing &$12.4^{+1.8}_{-1.6}$& $2160\pm 100$ \\
Kinematics &$13.1^{+2.6}_{-2.3}$ & $2190\pm 140$ \\\hline
\end{tabular}
\label{tab:M200}
\end{center}
\end{table*}

%%%%%%%%%%%%%%%%

%%%%%%%%%%%%%%%%
%%%%%%%%%%%%%%%%
\subsection{Gas fraction profile}\label{sec:fgas}
%%%%%%%%%%%%%%%%
Because of their deep gravitational well, massive clusters are expected to retain the matter collected since their formation. Thus, the relative amount of baryonic and dark matter should be close to the Universal value. Recent \emph{Planck} observations of the power spectrum of CMB anisotropies indicate a Universal baryon fraction $\Omega_{b}/\Omega_m=0.153\pm0.003$  \citep{planck15cosmo}. Corrected for the stellar fraction, which accounts for 10-20\% of the total amount of baryons in galaxy clusters \citep[e.g.,][]{gonzales07}, we expect a gas fraction of 13-14\%. 

\begin{figure}
\begin{center}
  \includegraphics[height=0.7\columnwidth,angle=0]{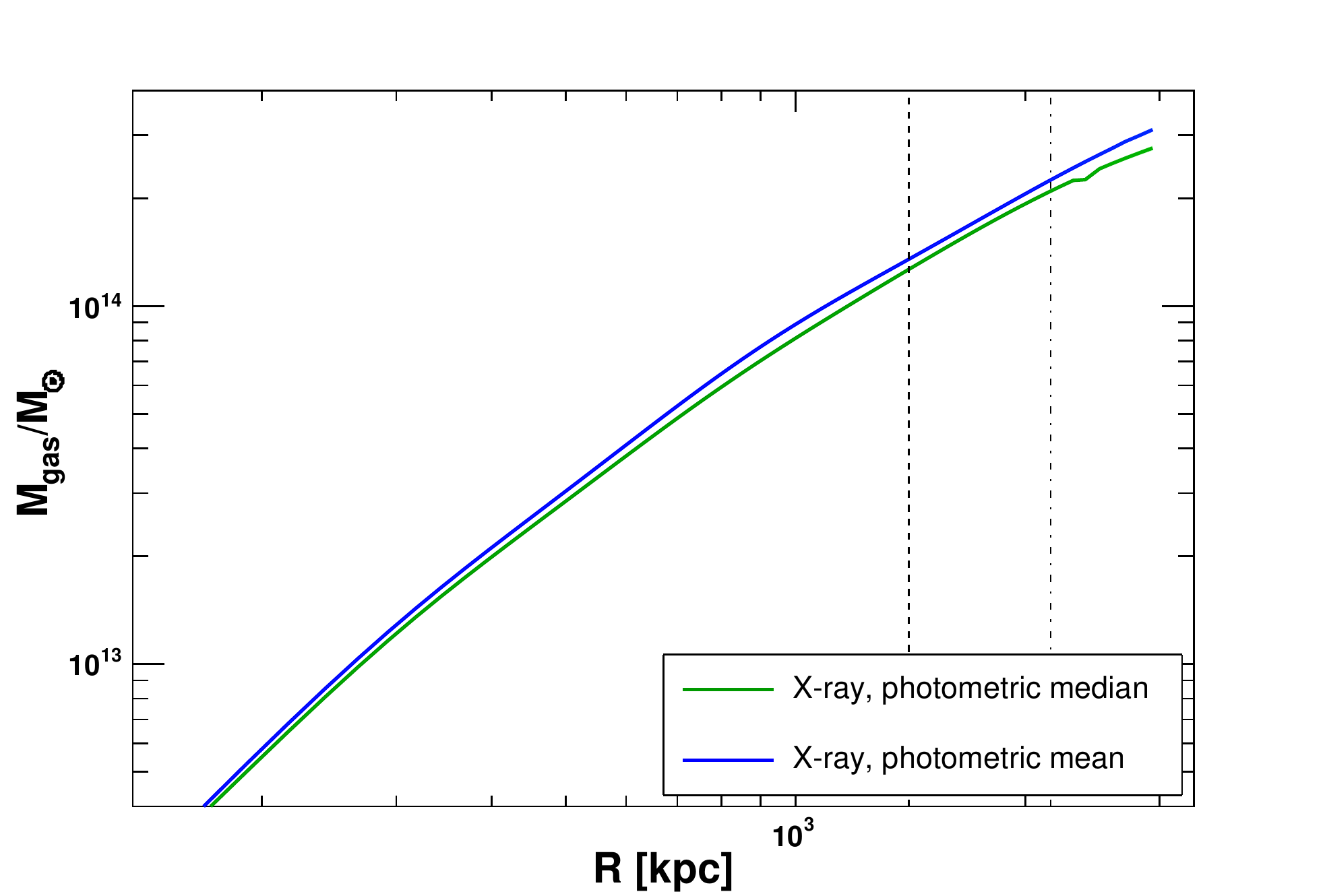}
\caption{Gas mass profiles obtained with Eq.~(\ref{eq:Mgas}). Green: using the azimuthal median density profile (green curve in Fig.~\ref{fig:A2142_n}). Blue: using the azimuthal mean density profile (blue curve in Fig.~\ref{fig:A2142_n}). The dashed and dashed-dotted vertical lines represent $R_{500}$ and $R_{200}$, respectively.}
\label{fig:A2142_Mg}
\end{center}
\end{figure}

\paragraph{}
Assuming spherical symmetry, the total gas mass is given by the integral of the gas density over the cluster volume,
\begin{equation}\label{eq:Mgas}
M_{\rm gas}(<r)=4\pi\int_0^{r}\rho_{g}(r')r'^2dr',
\end{equation}
where $\rho_g$ is defined as in Eq.~(\ref{eq:Mtot}). In Fig.~\ref{fig:A2142_Mg} we show the resulting $M_{\rm gas}$ profiles obtained for the azimuthal mean and the azimuthal median density profiles (see Fig.~\ref{fig:A2142_n}). Consistent results within few per cent are obtained using the method of  \citet{ettori10}. As expected,  the $M_{\rm gas}$ profile resulting from the use of the azimuthal median density profile lies slightly below the profile obtained from the azimuthal mean. At $R_{200}$, the difference between the azimuthal median and the azimuthal mean profiles is of the order of 6\%.

\begin{figure}
\begin{center}
  \includegraphics[height=0.7\columnwidth,angle=0]{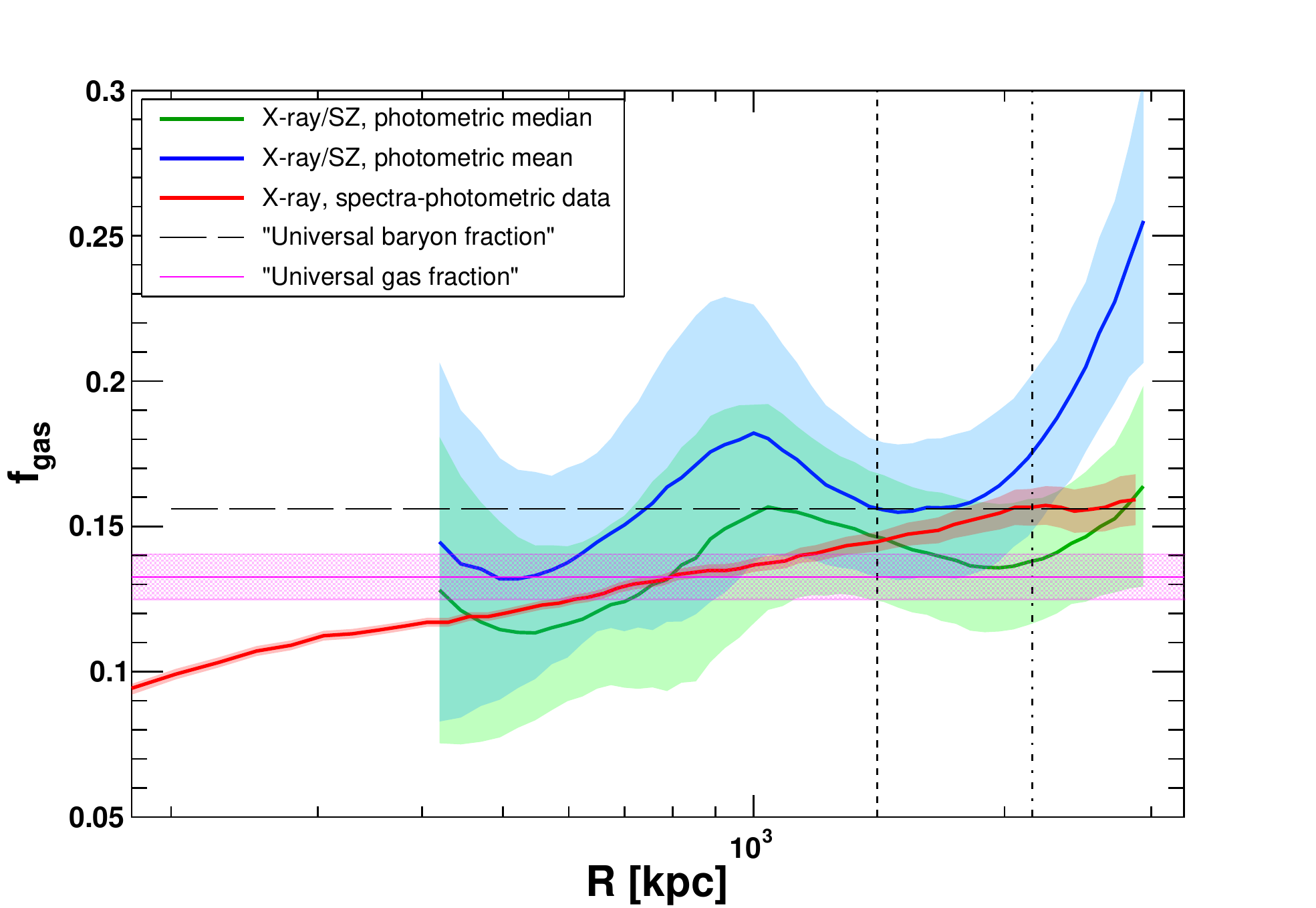}
\caption{Gas fraction profile.  Green: X-ray/SZ combined profile using the azimuthal median density profile. Blue: X-ray/SZ combined profile using the azimuthal mean density profile. Red: NFW fit to the spectroscopic X-ray data using the method of \citet{ettori10}. The dashed and dashed-dotted vertical lines represent $R_{500}$ and $R_{200}$, respectively. The dashed horizontal line represents the Universal baryon fraction from \emph{Planck} \citep{planck15cosmo}, whereas the hatched pink area shows the expected gas fraction corrected for the fraction of baryons in the form of stars \citep[][]{gonzales07}.}
\label{fig:A2142_fgas}
\end{center}
\end{figure}

We derived the gas fraction as a function of radius by combining the gas mass profiles with the corresponding hydrostatic mass profiles (see Sect. \ref{sec:Mtot}). In Fig.\ref{fig:A2142_fgas} we compare the resulting gas fraction profiles with the expected Universal baryon fraction from \emph{Planck}, corrected for the baryon fraction in the form of stars \citep{gonzales07}. Interestingly, we can see that while the gas fraction profile derived from the combination of the azimuthal median density profile with the SZ pressure profile is almost constant and close to the expected value ($\sim$ 13-14\%), the gas fraction profile derived using the azimuthal mean density profile and the SZ pressure profile increases with radius and exceeds the cosmic baryon fraction. We also note that the gas fraction profile obtained from purely X-ray information is slightly above the expected value. We will discuss these points further in Sect. \ref{sec:fgasdisc}.

%%%%%%%%%%%%%%%%

%%%%%%%%%%%%%%%%
%%%%%%%%%%%%%%%%

\section{Discussion}\label{sec:discussion}

The combination between deep X-ray and SZ data presented in this work allowed us to extend the measurements of the thermodynamic properties of the ICM out to $2R_{500}\sim R_{100}$, which corresponds roughly to the cluster's virial radius. This is the first study in which we are able to estimate self-consistently the effects of gas clumping and non-thermal energy. Here we discuss our main results and their implications.

\subsection{Gas clumping}
\label{sec:clumpingdisc}
\subsubsection{The clumping factor beyond $R_{500}$}

We estimated the effects of gas clumping on the cluster's gas density profile by applying the azimuthal median method presented in \citet{eckertclumping}. This method allows us to resolve all the clumps whose sizes exceed the size of the Voronoi bins, which given the depth of our \emph{XMM-Newton} observation corresponds to scales of $\sim20$ kpc around $R_{200}$, and to remove them from our analysis. On the other hand, in the spectroscopic analysis, the total number of detected photons is used, which leads to an overestimate of the surface brightness in the presence of inhomogeneities. This is illustrated in Fig. \ref{fig:A2142_n}, where we can see that the gas density estimated using the spectroscopic analysis closely follows the results obtained with the azimuthal mean, but overestimate the azimuthal median. This shows that the azimuthal median is a more reliable estimator of the mean gas density, especially when studying the outskirts of galaxy clusters. 

At $R_{200}$, we measured $\sqrt{C}=1.18\pm0.06$, which is consistent with the value $\sqrt{C}=1.25^{+0.31}_{-0.21}$ estimated by \citet{eckertclumping} using lower-resolution \emph{ROSAT}/PSPC data. Note however that because of projection effects, the method used here is only expected to provide an accurate measurement of the clumping factor when averaging over a sufficiently large number of clusters. 

The relatively mild clumping factor estimated here is somewhat lower (albeit consistent) with the value $\sqrt{C}\sim1.5$ measured by \citet{morandi13} from the dispersion of the surface-brightness distribution and with the value $\sqrt{C}\sim1.4$ expected by \citet{urban14} to reconcile the measured entropy profile of the Perseus cluster with the expectation from pure gravitational collapse. Numerical simulations \citep[see e.g., ][]{nagai11,zhu13,vazza13} consistently predict a mean value $\sqrt{C}\sim1.4$ around $R_{200}$, albeit with a rather large cluster-to-cluster scatter. Moreover, the exact value of the clumping factor was found to depend significantly on the adopted baryonic physics \citep{nagai11,roncarelli13}. Indeed, gas cooling removes the most structured phase of the gas from X-ray-emitting temperatures, which results in a smoother gas distribution and a lower clumping factor $\sqrt{C}\sim1.2$ \citep{nagai11}. Our measurements are therefore in better agreement with simulations including additional physics.

\subsubsection{Origin of the clumping}

An important question to ask is whether the mild, but significant level of clumping observed in this study originates mainly from a population of compact infalling clumps or from large-scale asymmetries in the gas distribution, e.g. coinciding with intergalactic filaments. In a recent paper, \citet{roncarelli13} divided the clumping effect into the contribution of individual infalling clumps and that of smooth, large-scale accretion patterns. While the former component strongly depends on the adopted baryonic setup, the latter (called the \emph{residual clumping $C_R$}) is robust against the implementation of additional physical effects. To investigate whether the clumping observed here is caused mainly by a large population of small accreting clumps or by smooth, large-scale accretion patterns, following \citet{roncarelli13} we estimated the level of residual clumping $C_R$ in our observing. To this aim, we computed the surface-brightness profile in 12 sectors of constant opening angle and computed the scatter of the surface-brightness values as a function of cluster-centric radius \citep{vazza11}. The azimuthal scatter can then be related to the residual clumping using the recipe described in \citet{roncarelli13}, allowing us to assess the level of clumping induced, on average, by large-scale asymmetries. 

\begin{figure}
\resizebox{\hsize}{!}{\includegraphics{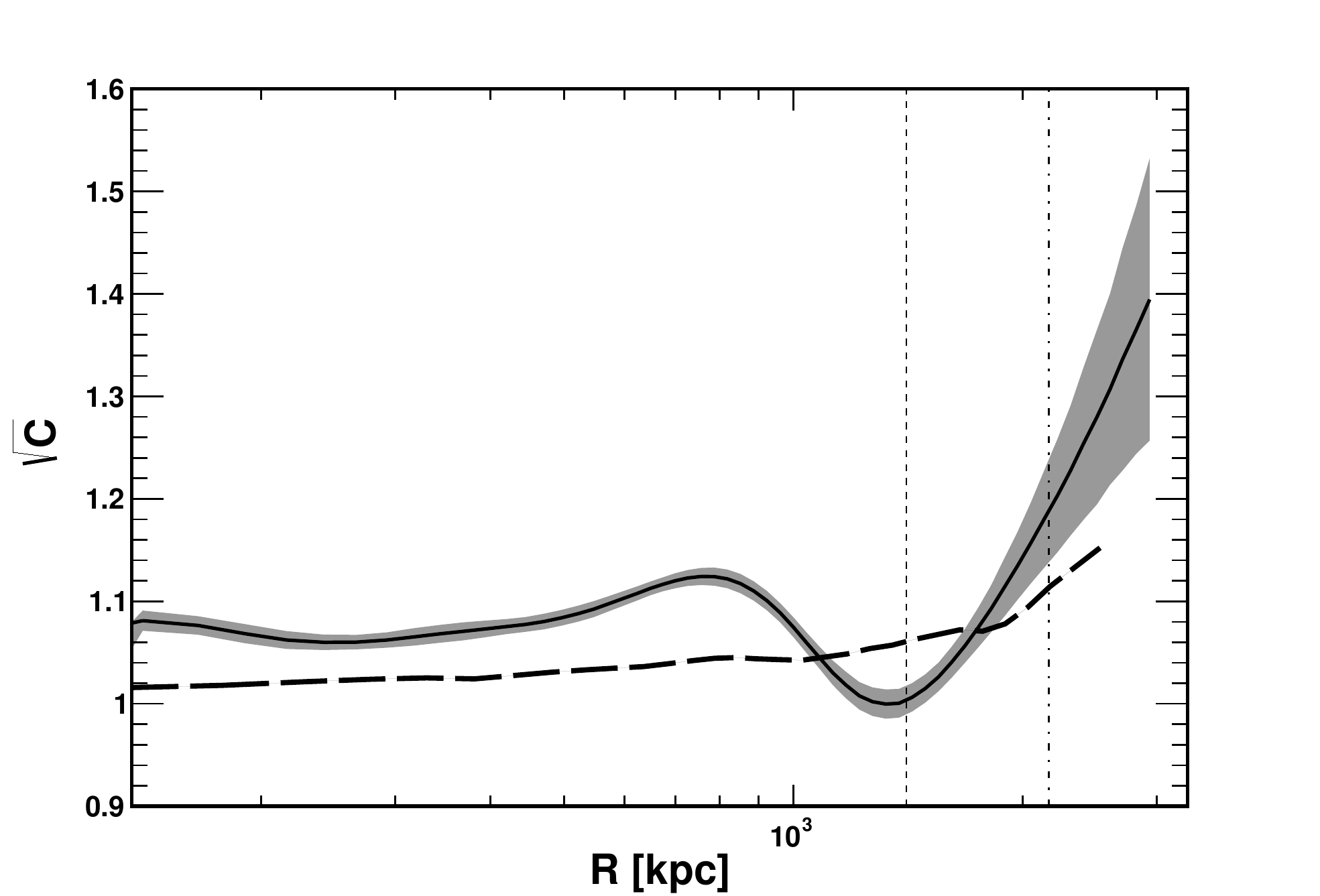}}
\caption{Clumping factor estimated using the azimuthal median method (same as Fig. \ref{fig:A2142_clumping}, solid line and shaded area) compared to the \emph{residual clumping} \citep[see text for details, dashed line,][]{roncarelli13}. }
\label{fig:CR}
\end{figure}

In Fig. \ref{fig:CR} we compare the total clumping from Fig.~\ref{fig:A2142_clumping} with the residual clumping as a function of radius. We can see that our measurement of the clumping factor consistently exceeds the residual clumping, with the exception of a small region around $R_{500}$. This shows that large-scale asymmetries account for a part, but not the entirety of the measured effect. At $R_{200}$, the residual clumping is roughly half of the total clumping, which reveals the presence of a population of small-scale clumps in the outskirts of the clusters \citep[such as the accreting group,][]{eckert1a2142}. Note however that the residual clumping is obtained `on average' as representative of the simulated dataset.

\subsubsection{The sloshing region}
\label{sec:sloshing}

In addition to the increase in the clumping factor beyond $R_{500}$, we also noticed a slight excess in the clumping factor in the inner regions followed by a decrease at about 1~Mpc. This can be explained by the large-scale sloshing phenomenon taking place in A2142 \citep{rossetti13}. Indeed, as pointed out in \citet{rossetti13}, A2142 exhibits three concentric cold fronts distributed along the cluster's main axis (NW-SE), the largest of which being located nearly 1~Mpc SE of the cluster. The presence of this substructure is indicated in Fig.~\ref{fig:annulus_region}. The sloshing phenomenon induces alternating surface-brightness excesses which, in turn, bias the mean surface-brightness high. On the other hand, the azimuthal median technique allows us to filter out the regions where the excess surface brightness is observed, leading to a lower estimate of the gas density. This example nicely illustrates the effectiveness of the azimuthal median technique in returning the surface brightness associated with the bulk of the ICM.

\subsection{Thermodynamic properties}

As discussed above, the main advantage of our analysis compared to previous works is that we are able to disentangle non-gravitational and clumping effects. Here we discuss the thermodynamic properties out to the virial radius measured in this work.

\subsubsection{Temperature and pressure}\label{sec:disc_T}

The comparison between the pressure profiles obtained from spectroscopic X-ray and SZ analysis indicates excellent agreement between the results obtained with the two methods, as shown in Fig.~\ref{fig:A2142_p}. This shows that X-ray and SZ observations provide a very consistent picture of the state of the ICM, as already pointed out in detailed comparisons of the pressure profiles derived with the two methods \citep{planck13,sayers13}. Interestingly, we note a slight excess in the X-ray spectroscopic pressure profile around $\sim500$ kpc. Given that the SZ effect is less sensitive to clumping, this excess can be explained by the overestimated gas density in the sloshing region (see above).

Given that the X-ray pressure profile is obtained by the product of the spectroscopic density profile with the spectroscopic temperature profile, this observed excess in the X-ray pressure profile implies that the spectroscopic temperature profile is less affected by clumping than the spectroscopic density profile. Otherwise, the two effects would balance and the X-ray pressure profile would tend to reproduce the features of the SZ pressure profile, which is less sensitive to clumping. Such an effect was also noticed by  \citet{rozo12}  in an analysis of  \emph{Chandra} vs \emph{Planck} data and in \citet{planck_Yx} in an analysis of \emph{XMM-Newton} vs \emph{Planck} data. In both studies the authors studied the scaling relations between $Y_{SZ}$ and $Y_{X}$, where $Y_{SZ}$ is related to the total pressure within the cluster's volume and $Y_{X}=M_{gas}T$, its X-ray analog, and they concluded that $Y_{X}$ was always in excess compared to $Y_{SZ}$.

A similar effect can be seen when comparing the spectroscopic X-ray temperature with the temperature estimated by combining SZ and X-ray imaging data (see Fig. \ref{fig:A2142_T}). Namely, the temperature profile obtained when combining the SZ pressure profile with the azimuthal mean density profile systematically underestimates the spectroscopic X-ray temperature, while the clumping-corrected profile leads to a temperature in agreement with the spectroscopic data. Interestingly, the agreement between spectroscopic X-ray and clumping-corrected X/SZ temperatures is contrary to the expectations of \citet{rasia14}, which predict that inhomogeneities in the density distribution should bias the observed spectroscopic temperature by 10 to 15\%, and is in contradiction with the results of \citet{mazzotta04}, who demonstrated based on numerical hydrodynamical N-body simulations that the projected spectroscopic temperature is lower than the emission-weighted temperature in the presence of inhomogeneities in the ICM. Conversely, our results are consistent with \citet{frank13}, who studied the temperature distribution in a large cluster sample and pointed out that the average spectroscopic temperature even exceeds the median of the temperature distribution.

\subsubsection{Entropy profile}

As shown in Sect.~\ref{sec:S}, the most striking result of our analysis is that the combined X-ray/SZ entropy profile is consistent at 1-$\sigma$ with the self-similar expectation once clumping-corrected gas density profiles are used. This implies that for Abell 2142, the formation history in the outskirts is similar to the one expected from purely gravitational collapse \citep[see e.g.,][]{voit05}. This shows that for this cluster spherical virialisation shocks is the dominant source for building up the entropy level of the ICM. Since the accretion shocks are located at larger radii \citep[$\sim3R_{200}\sim 6.5$~Mpc, ][]{lau15}, we do not expect to observe a turnover in the entropy profile, even in the broad radial range accessible to this study. This conclusion is reinforced by the study performed in \citet{cavaliere11}, where the authors found inverse correlations between the entropy level and the halo concentration, implying that for low halo concentration like A2142, the entropy profile is expected to follow the self-similar expectations and to undergo negligible non-thermal support.
A small contribution of clumps smaller than the resolution of our study ($\sim20$~kpc) could be invoked to bring the two measurements in agreement, although such a contribution is not required from a statistical point of view.

Conversely, when the effect of gas clumping is not taken into account, the entropy profile flattens beyond $R_{500}$ and shows a behavior very similar to most of the studies based on \emph{Suzaku} data \citep[e.g.][]{urban14,akamatsu11,walker13,sato14}. In a study based on a sample of relaxed clusters, \citet{walker12} showed that all clusters exhibit a flattening in their entropy profile beyond $R_{500}$. A2142 is probably the object with the most significant turn-around in the entropy profile in their sample. Such a behavior has been interpreted in the past as evidence for a lack of thermalization in the gas at these radii, because of the presence of a significant non-thermal pressure \citep[e.g.][]{lapi10,kawa10} or non-equilibration between ions and electrons \citep{hoshino10,avestruz15}. For instance, \citet{fusco14} invoked the presence of a non-thermal pressure component to sustain hydrostatic equilibrium in the outskirts of A1795, A2029, A2204 and A133, and concluded that the temperature profile steepening is mostly due to non-thermal effects, while the role of the gas clumping was assumed to be marginal \citep[see also][]{walker12}.
However, our results establish that clumping is the primary mechanism driving the entropy flattening and show that non-thermal effects, if present, should be mild. Interestingly, based on hydrodynamical simulations of galaxy cluster formation, \citet{nelson14} concluded that the non-thermal pressure accounts for only $10-30\%$ of the total pressure support at $R_{200}$, while out-of-equilibrium electrons can cause a drop in temperature by $10\%$ at $R_{200}$ \citep{rudd09,avestruz15}. This implies that neither process seems sufficient to explain the observed entropy drop in the case where the clumping is not taken into account.

\subsection{Hydrostatic mass and gas fraction}

\subsubsection{No hint of hydrostatic bias}

As shown in Fig. \ref{fig:A2142_Mtot} and Table \ref{tab:mass}, all the mass reconstructions presented here agree with the reconstructions based on weak gravitational lensing and galaxy dynamics (and even slightly exceed them). This is somewhat surprising, since the latter measurements do not require any assumption on the state of the gas. Indeed, residual kinetic energy in the form of bulk motions or turbulence should induce an additional pressure term, which should lead to an underestimate of the mass when the energy budget is assumed to be entirely thermalized \citep[e.g.][]{rasia06,nagai07a,burns10}. Simulations consistently predict that non-thermal effects should be proportionally larger beyond $R_{500}$ \citep{lau09,battaglia13}. The lack of difference between hydrostatic-based and weak lensing measurements would therefore imply that the gas in the outskirts of A2142 is relaxed and fully thermalized. This is a surprising result, especially since recent studies have unveiled that A2142 is located at the core of a collapsing supercluster \citep{einasto15,gramann15}.

We note that although weak lensing is insensitive to the dynamical state, it is sensitive to the triaxiality of the observed halo \citep[e.g.][]{limousin13}. Thanks to a large spectroscopic campaign totaling nearly 1,000 redshifts, \citet{owers11} found that A2142 does not show prominent substructures along the line of sight, but it is strongly elongated in the plane of the sky along the NW-SE axis. This could lead to an underestimation of the cluster mass when assuming spherical symmetry, both for weak lensing and galaxy kinematics, which might explain the slightly higher hydrostatic masses observed here all the way out to $R_{200}$. 

Interestingly, we noted in Sect.~\ref{sec:Mtot} that the purely X-ray hydrostatic mass profile does not seem to be affected by the presence of clumps. This can be explained by the facts that (i) the hydrostatic mass depends on the logarithmic clumping factor gradient \citep[see e.g. Eq.~(14) of ][]{roncarelli13}, which is observed to be negligible (see Fig.~\ref{fig:A2142_clumping}), and (ii) the effect of the presence of clumps on the spectroscopic temperature is observed to be small too (see Sect.~\ref{sec:disc_T}).
Therefore, the standard X-ray hydrostatic mass measurement technique is essentially unaffected. This is not the case of the X/SZ method, since in this case the gas density enters directly in the hydrostatic equation (Eq.~(\ref{eq:Mtot})). This explains why the clumping-corrected X/SZ profile agrees with the X-ray-only result, while the mean X/SZ profile returns a mass that is lower by $\sim20\%$.

\subsubsection{Gas fraction}
\label{sec:fgasdisc}

Depending on the adopted method, our measurements of the gas fraction (see Fig. \ref{fig:A2142_fgas}) indicate a rather flat gas fraction which is consistent with the Universal value of the baryon fraction once the stellar fraction ($f_{\star}$) is taken into account: $\Omega_b/\Omega_m-f_{\star}\approx 13-14\%$. Interestingly, we note that the gas fraction only slightly rises from the core to the outskirts, unlike what is usually observed in most clusters \citep[e.g.][]{vikhlinin06}. Indeed, this is expected that the gas fraction should increase with the distance from the cluster center because of entropy injection \citep[see e.g.,][]{pratt10,young11}. 

As pointed out in \citet{simionescu11,vazza13}, the effect of clumping is largest on the reconstructed gas fraction, since it combines a negative bias on the gravitating mass with a positive bias on the gas mass \citep[see the discussion in][]{eckertclumping}. Comparing the gas fraction profiles obtained with and without the correction for the emissivity bias, we found that while the clumping-corrected gas fraction is roughly constant in the range 0.5-3~Mpc and consistent with the cosmic value corrected for the stellar fraction, the X/SZ gas fraction profile uncorrected for clumping increases with radius and exceeds the universal baryon fraction in the outskirts of the cluster. The classical X-ray analysis sits somewhat in between, since the hydrostatic mass measured with this method is relatively unaffected by clumping (see above), while the gas mass is overestimated. 

A similar effect has been observed in \citet{eckert13b}, where the authors measured the average gas fraction in unrelaxed and relaxed clusters using the azimuthal mean density profile from \textit{ROSAT}/PSPC and the pressure profile from \emph{Planck}. They observed that in non-cool-core clusters the gas fraction at $R_{200}$ exceeds the Universal value, while for the relaxed (cool-core) clusters the gas faction is consistent with the expectations. This difference could be explained by a larger amplitude in the inhomogeneities of the gas distribution in unrelaxed clusters than in relaxed clusters, which would lead to a larger clumping factor in the former class. Such a dependence is expected in numerical simulations, in which unrelaxed clusters are characterized by a larger mass accretion rate, and thus by a larger clumping factor. Our results show that in the case of A2142 gas clumping alone can explain the observed excess gas fraction beyond $R_{500}$. This therefore reinforces this interpretation.

\subsection{Reliability of the method in the presence of large substructures}

Abell~2142 is a dynamically active cluster where an infalling galaxy group has been discovered  in the NE region \citep[see caption of Fig.~\ref{fig:annulus_region} and ][]{eckert1a2142}. Given that this accreted substructure is still not in thermodynamic equilibrium with the ICM of the main cluster (with an estimated temperature of 1.3-1.5 keV), the properties of the gas in NE region are not expected to be representative of the cluster ICM. 

In order to quantify the effect of the presence of such a subcluster in our results, we repeated exactly the same procedure as above, but leaving out the NE region from our analysis. The results obtained excluding the NE region from the combined fits in the spectral analysis can be found in Tables  \ref{table:source_indivfit9}, and \ref{table:source_indivfit10}. Interestingly, the results of the combined fit  with and without the NE region are consistent, except for the value of the APEC norm in the 9-12 arcmin annulus (which increases when we remove the NE region from the analyse to the order of 1\%. This effect could be due to statistical fluctuations).

We then applied the method of \citet{ettori10} to these profiles and obtained an hydrostatic mass profile that can be characterized by the quantities  $M_{200}=14.5\pm 0.3$ 10$^{14}$M$_{\sun}$  and $R_{200}=2270 \pm 17$ kpc. The comparison with the results obtained including the NE region (see Table~\ref{tab:M200}) shows  that masking the NE region changes the hydrostatic mass  $M_{200}$ by just 3\%. This shows that our method can be applied even in the presence of substructures as long as their size does not exceed the one of this subcluster \citep[whose mass has been estimated to be of the order of a few $10^{13} $ M$_{\sun}$, see][]{eckert1a2142}.

\section{Conclusion}\label{sec:conclusion}

In this paper, we studied the outskirts of the massive cluster Abell 2142 by combining X-ray (\emph{XMM-Newton}) and SZ (\emph{Planck}) data, which allowed us to trace the state of the intracluster gas out to the virial radius of this system. For the first time, we applied a method insensitive to the presence of gas inhomogeneities, with the aim of disentangling the effects of gas clumping and non-thermal pressure support. Our main findings can be summarized as follows.

\begin{itemize}

\item We found that Abell 2142 is affected by a significant level of clumping in its outskirts, which leads to a mean clumping factor $\sqrt{C}=1.18\pm0.06$ at $R_{200}$. Roughly half of the clumping can be ascribed to the presence of large-scale asymmetries in the gas distribution, while the remaining half should be in the form of accreting clumps (see Fig. \ref{fig:CR}).

\item We recovered the entropy profile of the cluster out to the virial radius by combining the gas density profile from \emph{XMM-Newton} with the pressure profile from \emph{Planck} (Fig.~\ref{fig:A2142_K}). We showed that when gas clumping is taken into account, the entropy profile follows the prediction of purely gravitational collapse \citep{voit05}. Indeed, the flattening of the entropy profile, which is significant when using the azimuthal mean density profile \citep[see also][]{akamatsu11}, disappears when the X-ray analysis is corrected for the clumping bias. Therefore, contrary to \citet{akamatsu11} our data do not require to invoke non-gravitational effects to explain a lack of thermalization of the intracluster gas beyond $R_{500}$. We note however that the analysis performed in \citet{akamatsu11} was limited to the NW direction and that this lack of azimuthal coverage may contribute to their results.

 \item We applied the hydrostatic equilibrium equation to reconstruct the mass profile of the cluster out to its virial radius (Fig.~\ref{fig:A2142_Mtot}). While the hydrostatic mass profile obtained with the azimuthal median is consistent with hydrostatic equilibrium assumption with the thermal gas, the one obtained using the azimuthal mean decreases at $R_{200}$ and beyond, which has been interpreted in several previous studies as evidence for a strong non thermal pressure component to balance gravity. We compared our mass measurements to the results obtained with \textit{Suzaku} \citep{akamatsu11}, to the weak lensing mass estimate from \citet{umetsu09}, and to the galaxy kinematics measurement from \citet{munari14} (see Table \ref{tab:mass}). Our mass estimates are consistent and even slightly exceed the estimates obtained with different methods, which does not require to invoke a hydrostatic bias. Furthermore, we noted that the total mass estimated from the classical spectroscopic X-ray method is only slightly affected by gas clumping (this is a second order effect) and follows the X-ray/SZ mass profile obtained with the azimuthal median. This may indicate that the temperature profile from spectroscopic analysis is mildly affected by the presence of clumps (Fig.~\ref{fig:A2142_T}). 
    
 \item Finally, we combined our hydrostatic mass and gas mass measurements to estimate the radial profile of intracluster gas fraction (Fig.~\ref{fig:A2142_fgas}). Our results show that the profile obtained using a method insensitive to clumping is consistent with $\Omega_b/\Omega_m-f_\star$. Conversely, the gas fraction profile derived using the azimuthal mean increases in the cluster outskirts and exceeds the cosmic value. 
 
 \end{itemize}

 In conclusion, the case of Abell 2142 provides a striking example of the importance of using a method insensitive to outliers in the gas distribution when probing the thermodynamical state of cluster outskirts. When correcting for gas clumping, the radial profiles of entropy, hydrostatic mass and gas faction are consistent with the predictions. Conversely, when using the classical method (azimuthal mean) we observe a strong entropy flattening beyond $R_{500}$ and a gas fraction that exceeds the cosmic values. Neglecting the clumping effect would therefore require to invoke additional effects such as non-equilibration between ions and electrons \citep[e.g.][]{akamatsu11,hoshino10} or non-thermal pressure components to sustain gravity \citep[e.g.][]{fusco14}. 
 
In the near future, the X-COP program will provide a similar data quality for a sizable cluster sample (13 clusters), which will allow us to test if the conclusions drawn here in the case of A2142 can be generalized to the local cluster population.
  
\acknowledgements{We thank F.~Vazza for a careful reading of the manuscript. CT acknowledges the financial support from the Swiss National Science Foundation. EP and GH are grateful for the support of the French Agence Nationale de la Recherche under grant ANR-11-BS56-015. SE, SM, FG acknowledge the financial contribution from contracts ASI-INAF I/009/10/0 and PRIN-INAF 2012 `A unique dataset to address the most compelling open questions about X-Ray Galaxy Clusters'. Based on observations obtained with \emph{XMM-Newton}, as ESA science mission with instruments and contributions directly funded by ESA Member States and NASA. }

 %%%%%%%%%%%%%%%%%%%%%%%%%%%%%%%%%%%%%%%%%%%%%%%%%%%%%%%%%%%%%%%%%%%%%%%%%%%%%%%%%%%%%%%%%%%%%%%%%%%%%%%%%%%%%%%%%%%%%%%%%%%%%%%%%%%%%%%%%%%%%%%%%%%%%%%%%%%%%%%%%%%%%%%%%%%%%%%%%%%%%%%%%%%%%%%%%%%%%%%%%%%%%%%%%%%%%%%%%%%%%%%%%%%%%%%%%%%%%%%%%%%%%%%%%%%%%%%%%%%%%%%%%%%%%%%%%%%%%%%%%%%%%%%%%%%%%%%%%%%%%%%%%%%%%%%%%%%%%%%%%%%%%%%%%%%%%%%%%%%%%%%%%%%%%%%%%%%%%%%%%%%%%%%%%%%%%%%%%%%%%%%%%%%%%%%%%%%%%%%%%%%%%%%%%%%%%%%%%%%%%%%%%%%%%%%%

\appendix
\section{Modeling residual soft protons}
\label{app:sp}
\begin{figure}
\resizebox{0.97\hsize}{!}{\includegraphics{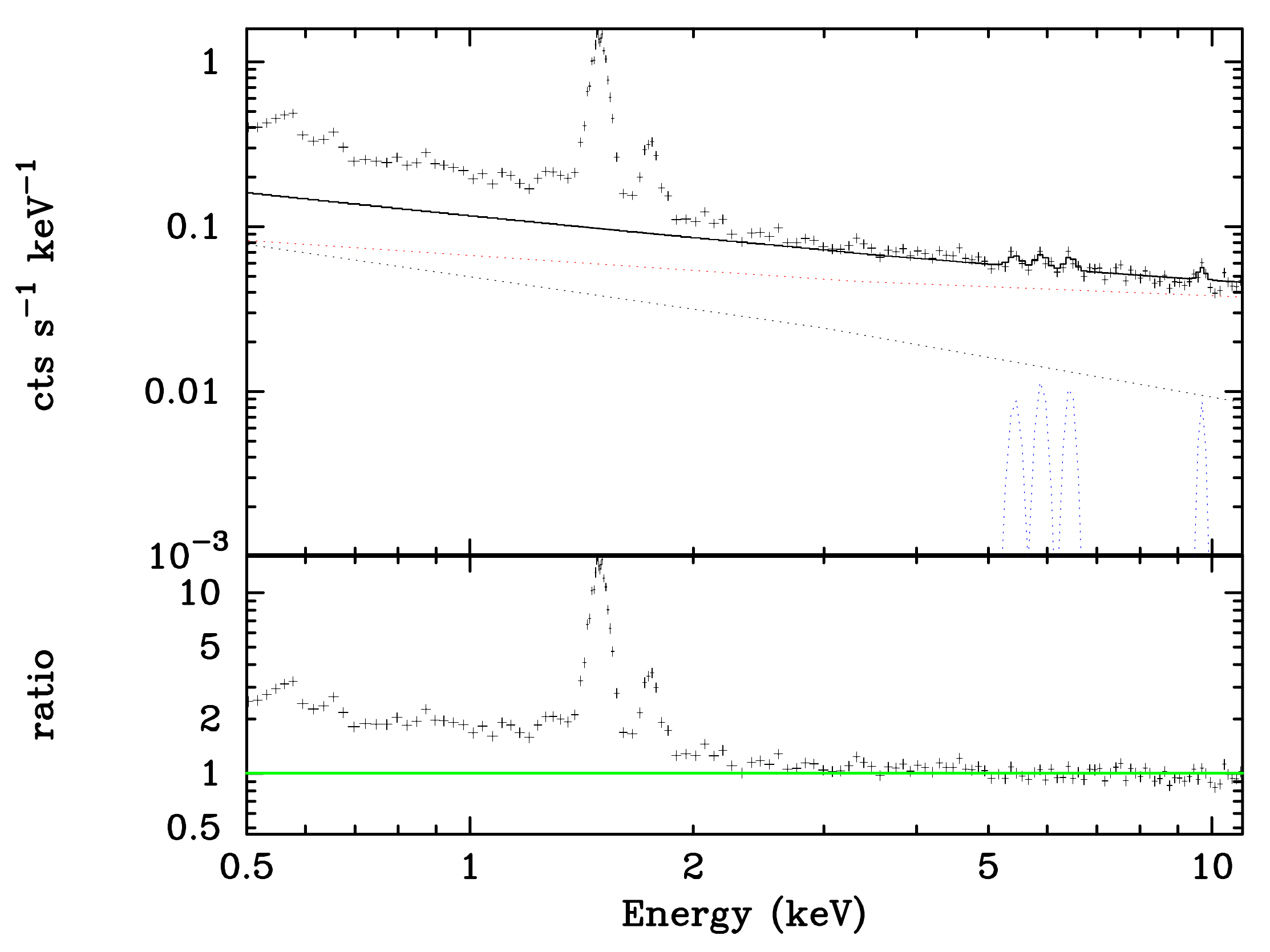}}
\caption{MOS2 Spectrum for observation 0085150101 (see Table \ref{tab:blank}) accumulated over the entire FOV. Top panel: observed spectrum (data points) and spectral model comprising a QPB component (broken power-law, shown as a dashed red line, and  gaussian lines, shown as dashed blue lines) and a QSP component (broken power-law, shown as a dashed black line). The fit is performed in the [4-10] keV energy range and extrapolated to lower energies. Parameters for the QPB component are fixed to the value estimated from filter-wheel-closed data. Note how the contribution of the QSP component increases with respect to that of the QPB as we move down in energy. Bottom panel: Residuals in the form of ratio of data over model, note how in the [0.7-1.2] keV band the data exceeds the model by a factor of 2, indicating that in this range roughly half of the accumulated events come from the X-ray sky and the other half are associated to particles.}
\label{fig:qsp}
\end{figure}

\normalsize
In the considered soft band, soft protons provide a modest but non-negligible contribution, to
account for it we need to follow a complex procedure. 
We start by deriving a spectral model for the quiescent particle background (QPB), over a broad spectral range,
by making use of spectra extracted over the full FOV from the auxiliary background event files.
The model comprises a broken power-law with different parameters for MOS and pn and
several gaussian lines to account for fluorescent emission that is observed in both detectors.
Spectral regions that are polluted by particularly intense lines are excised as they
are not particulary helpful in constraining the shape of the continuum.
We then fit the spectra extracted over the full FOV from the event files of the observation,
with a model comprising a QPB component plus a quiescent soft proton background (QSP) component. For the QPB component the parameters
are fixed to those derived from  the fit of the auxiliary background data.
For the QSP component, which has the form of a broken power-law,  all parameters, with the exception of the normalization,
are fixed to fiducial values \citep[see][for a detailed analysis]{kuntz08,leccardi08}. The fit is
carried out in the hard X-ray band where the QSP contribution is more significant and therefore
more easily gauged. More specifically, for the  MOS we use the  [4.0-11.5] keV band while for the
pn the [4.0-7.1] keV  and [9.2-14.0] keV bands, the region between 7.1 and 9.2 keV is excluded to avoid
the strong fluorescence lines that are found there. We have verified that adopting a somewhat more 
restrictive range, i.e. excluding the [4.0-5.0] keV band, has a negligible effect on the derived parameters and 
on the test presented in Sect.\ref{app:blank-sky}. 
Finally the parameters of the soft proton component are fed to the ESAS task \texttt{proton}
that produces a soft proton image in the 0.7-1.2 keV band for each of the three detectors.
In Fig.\ref{fig:qsp} we provide an example of a fit to the MOS2 Spectrum for observation 0085150101.

\begin{table*}
\caption{\label{tab:blank}Blank field observations used to validate the surface brightness production procedure.}
\begin{center}
{\setstretch{1.3}
\begin{tabular}{lcccccccc}
\hline
Obs.Id. & Obs.Date & $N_H$ & $t_{M1}$ &  $t_{M2}$ & $t_{pn}$ & $\mathcal{R}_{M1}$ & $\mathcal{R}_{M2}$ & $\mathcal{R}_{pn}$\\
\, & [yr/mm/dd] & [$10^{20} \rm cm^{-2}$] & [ks] & [ks] & [ks] &  & \\
\hline
\hline
0085150101 & 2001-10-15 &  2.8 & 37.6 & 38.6 & 31.3 & 1.15 & 1.19 & 1.11 \\
0094310201 & 2002-12-15 &  4.4 & 60.8 & 61.4 & 53.1 & 1.07 & 1.10 & 1.07 \\
0108060701 & 2002-01-14 &  0.7 & 70.7 & 71.1 & 58.2 & 1.22 & 1.27 & 1.18 \\
0108062301 & 2002-01-23 &  0.7 & 71.7 & 72.5 & 57.4 & 1.05 & 1.12 & 1.07 \\
0109661001 & 2001-06-23 &  0.8 & 60.5 & 60.3 & 47.1 & 1.09 & 1.12 & 1.03 \\
0111550401 & 2001-06-01 &  1.0 & 79.5 & 80.6 & 65.0 & 1.08 & 1.16 & 1.05 \\
0112370101 & 2000-07-31 &  2.0 & 34.1 & 36.7 & 23.5 & 1.12 & 1.18 & 1.07 \\
0112370301 & 2000-08-04 &  2.0 & 33.6 & 34.3 & 26.2 & 1.15 & 1.19 & 1.11 \\
0128531601 & 2003-12-12 &  1.8 & 68.6 & 69.3 & 56.2 & 1.08 & 1.17 & 1.09 \\
0147511701 & 2002-12-04 &  0.6 & 88.6 & 88.9 & 73.6 & 1.08 & 1.09 & 1.06 \\
0147511801 & 2002-12-06 &  0.6 & 69.2 & 71.8 & 46.0 & 1.11 & 1.14 & 1.06 \\
0148560501 & 2003-05-22 &  2.6 & 56.3 & 57.7 &  0.0 & 1.10 & 1.13 & 1.00 \\
0148960101 & 2003-05-12 &  3.1 & 38.1 & 39.6 & 32.2 & 1.12 & 1.12 & 1.02 \\
0203362101 & 2004-12-09 &  1.8 & 59.4 & 59.3 & 51.0 & 1.06 & 1.07 & 1.07 \\
0210280101 & 2005-04-09 &  2.5 & 70.1 & 69.5 & 58.1 & 1.07 & 1.10 & 1.04 \\
0302420101 & 2005-07-08 &  3.9 & 72.2 & 72.9 & 53.9 & 1.21 & 1.23 & 1.15 \\
0303260201 & 2005-04-07 &  0.6 & 43.5 & 43.6 & 35.5 & 1.08 & 1.09 & 1.03 \\
0402530201 & 2006-06-04 & 29.3 & 82.5 & 84.1 & 65.0 & 1.04 & 1.06 & 1.00 \\
0500500701 & 2007-05-19 &  6.1 & 49.4 & 68.2 & 35.7 & 1.08 & 1.07 & 1.04 \\
0555780101 & 2008-07-05 &  0.7 & 81.2 & 87.5 & 50.1 & 1.07 & 1.09 & 1.01 \\
0651900201 & 2010-06-11 &  1.4 & 77.2 & 86.1 & 49.9 & 1.09 & 1.10 & 0.87 \\
\hline
\end{tabular}
}
\end{center}
Column description: 1. Observation identifier; 2. Observation date; 3. Equivalent hydrogen column density as estimated from 21 cm maps \citep{kalberla05}; 4. Exposure time for MOS1 detector after flare removal; 5. Exposure time for MOS2 detector after flare removal; 6. Exposure time for pn detector after flare removal; 7. Ratio of counts from observation over counts for auxiliary background event files for MOS1 detector. The ratio is computed over the full FOV and in the [7.5-11.85] keV  energy band; 8. Ratio of counts from observation over counts for auxiliary background event files for the MOS2 detector. The ratio is computed over the full FOV and in the [7.5-11.85] keV  energy band; 9. Ratio of counts from observation over counts for auxiliary background event files for the pn detector. The ratio is computed over the full FOV and in the [9.2-14.0] keV energy band. 
\end{table*}

\section{Assessment of systematic uncertainties}
\label{app:blank-sky}
Given the different radial dependence of the various background components: cosmic background, QPB and QSP,
an indication of the quality of our methodology can be obtained by producing a ``mean'' cosmic background radial
profile by stacking a large number of blank-field observations. The stacking  process allows to: 1) average out any significant gradient in the radial directions associated either with cosmic variance or structure in the galactic foreground; 2) achieve sufficient statistics to address systematic issues at the few percent level.
We have chosen 21 observations from the \emph{XMM-Newton} archive, a subsample of these have been previously analyzed and
discussed elsewhere \citep[]{leccardi08}. The sample covers most of the mission timeline, the
bulk of the observations have equivalent hydrogen column densities  $N_H < 3\times10^{20}$cm$^{-2}$, 
while a few are between $3\times10^{20}$cm$^{-2}$ and a few $10^{21}$cm$^{-2}$. The total observing time 
after flare removal is roughly 1.3 Ms for MOS1 and MOS2 and 1 Ms for pn. Details on the sample can be found in Table \ref{tab:blank}. For each observation we use the ESAS task \texttt{comb} to produce a  MOS1 + MOS2 + pn counts image of the field and an associated exposure map. The latter is obtained by summing the exposure maps for the 3 different detectors using appropriate weights for each detector. The \texttt{comb} task is also used to
produce a combined background image for the 3 detectors. The combined field image, background image and exposure map are then fed to a program that produces a radial profile by taking the combined image counts in a given annulus, subtracting the counts from the same region of the background image and dividing the resulting net counts by the mean exposure in the annulus and the area of the annulus.

\begin{figure}
\begin{center}
  \includegraphics[height=0.7\columnwidth,angle=0]{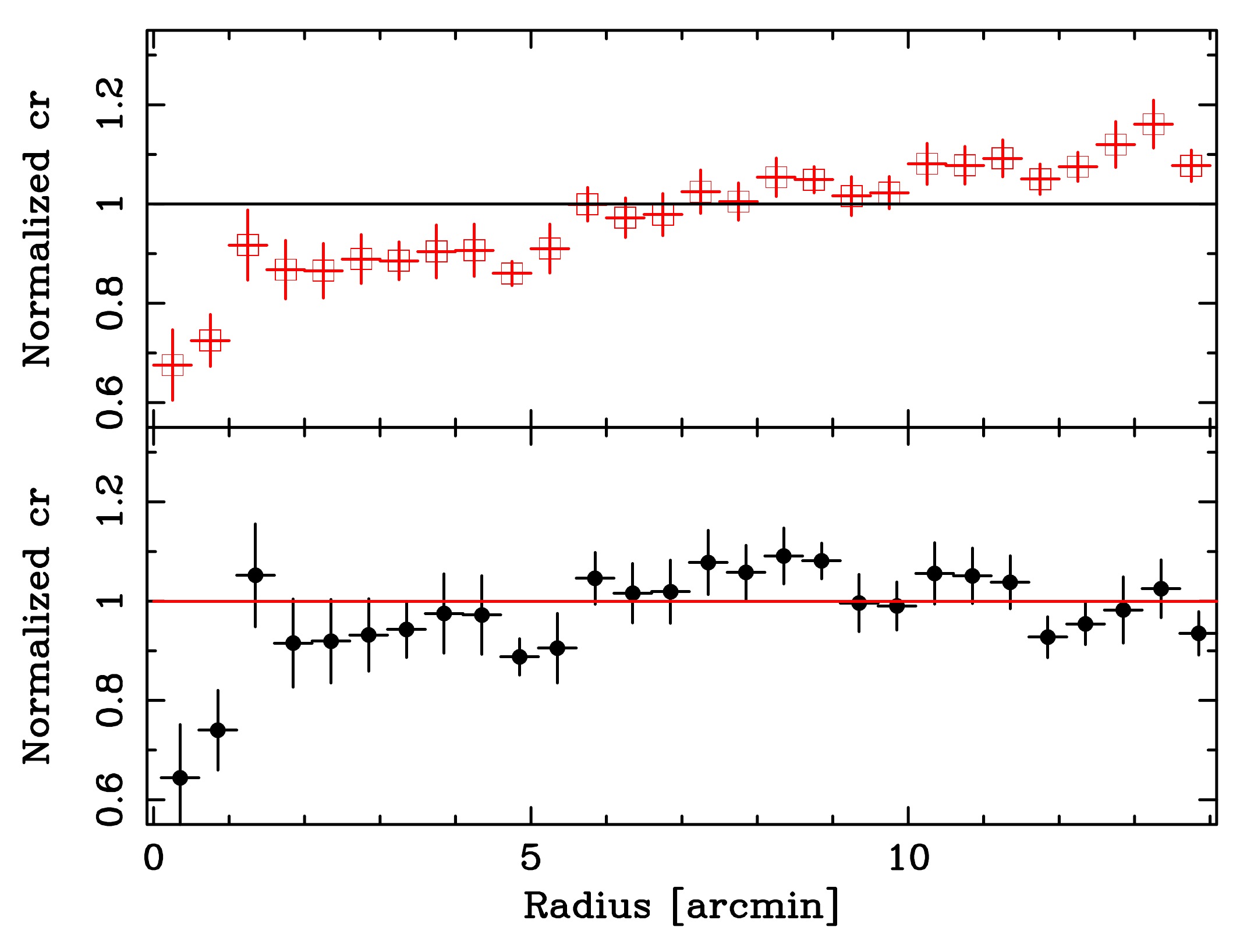}
\caption{Stacked normalized EPIC surface-brightness profile for 21 blank fields (see Table \ref{tab:blank}).
 Top panel: Profile produced without accounting for the QSP component. Note how the surface brightness
increases steadily as we go from the center to outer regions of the detectors.
Bottom panel: Profile produced after subtraction of the QSP component. Note how, beyond the innermost
arcminute, the profile is flat, within deviations contained within 5\%-10\%. }
\label{fig:qsp2}
\end{center}
\end{figure}

Poisson errors from the counts in the field and
background image annuli are propagated to derive statistical errors on the profile. Finally a mean radial profile is derived for the 21 observations by performing an error weighted mean for each annulus. In the bottom panel of Fig.~\ref{fig:qsp2} we show a normalized version of this profile. As we can see, with the exception of the innermost arcminute, the radial profile is flat to within 10\% everywhere, with many data points within the 5\% limit. An important point is that there is no evidence of a large scale trend, i.e. deviations around the mean are scattered over the full radial range and not clustered in a specific region. Regarding the central arcminute, there are a number of possible explanations for the observed drop in surface brightness, it could be related to  poorly removed central sources, insufficiently accurate estimation of the aim point, loss of quantum efficiency in the region of the detectors subject to most intense X-ray irradiation, or to a combination of these. However, what is most important, is that this region is very small, it is less than 1\% of the FOV and therefore not particularly important for our surface brightness estimates. 

A fit to the surface brightness profile with a constant, excluding the central arcminute, yields a $\chi^2 $ of 34.9 for 25 d.o.f. and an associated probability of 9\% for the profile to be consistent with being flat, confirming that systematic deviations must be at the few percent level or smaller. To quantify the presence of residual systematics in our radial profile, we performed an analysis of the scatter. We modeled the intrinsic scatter in the form of a Gaussian. We used a maximum likelihood algorithm \citep{maccacaro88} to fit the mean radial profile and its errors, where the free parameters are the mean and the intrinsic scatter (i.e. the standard deviation of the Gaussian). Our algorithm returns a mean for the profile of $0.99\pm0.01$, which is not surprising, as the profile is normalized. More interestingly, the intrinsic scatter is $0.04\pm0.01$, implying that systematic deviations from flatness are less than 5\% .  

Profiles for the individual detectors, i.e. MOS1, MOS2, and pn are consistent with the one averaged over all instruments. We have also tried cutting our sample in different ways: low and high $N_H$, low and high residual soft
proton component; early and late part of the mission; in no instance  do we detect a substantial
deviation from the mean profile. For comparison, in the top panel of Fig.~\ref{fig:qsp2} we also plot
a radial profile which has been produced without taking the residual soft proton component into account, i.e.
only the QPB image has been subtracted from the observation image. At variance with the black
profile the red profile shows deviations that go up to 15\%. The most important difference
 is that this profile is not flat but rises steadily so that at edge of the FOV it is about 30\% higher
 than at the center (excluding the central arcminute) clearly showing that failure to remove the residual
 soft proton contribution will result in a significant systematic error on extended sources with 
 surface brightness comparable to that of the X-ray background. In conclusion, the method presented here is able to estimate the total background with a precision of 5\% in the [0.7-1.2] keV band. This allows us to extract the surface-brightness profiles of A2142 out to $\sim2R_{500}$, where systematic and statistical uncertainties become comparable. 
 
\section{Comparison between \emph{XMM-Newton} and \emph{ROSAT} results}
\label{app:rosat}
\begin{figure}
\resizebox{\hsize}{!}{\includegraphics{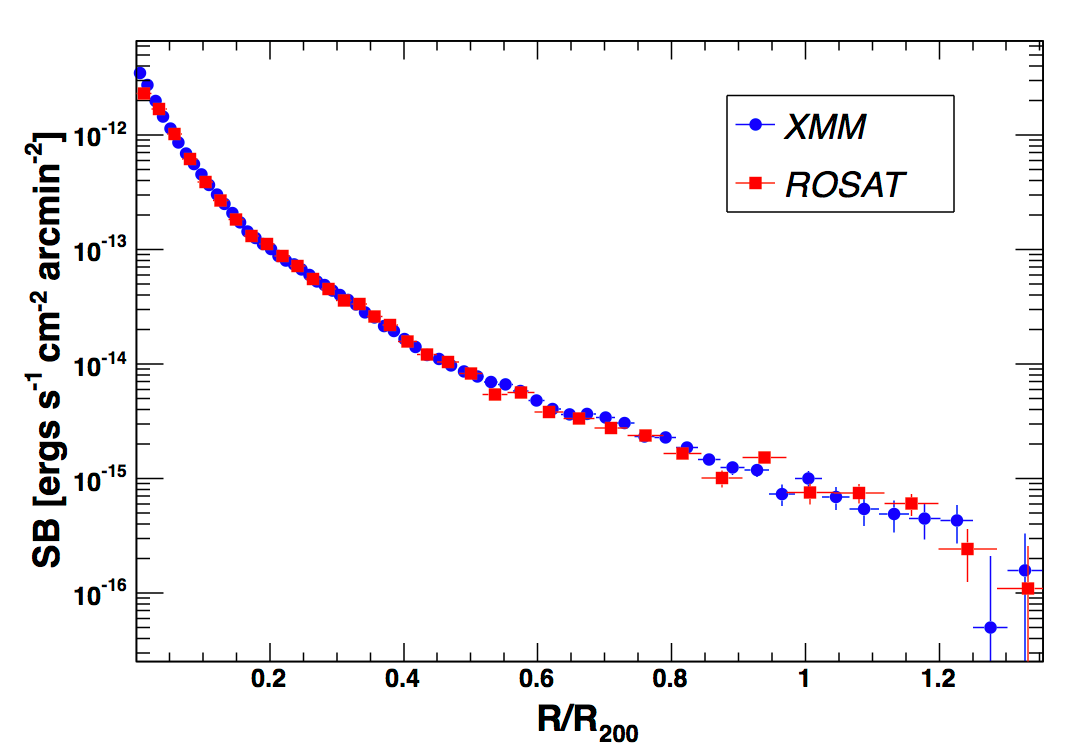}}
\caption{Surface brightness profiles of A2142 obtained with \emph{XMM-Newton} (blue) and \emph{ROSAT}/PSPC (red) converted into energy flux in the [0.5-2] keV band. In the X axis we show the distance to the cluster center divided by $R_{200}\sim2200$ kpc.}
\label{fig:pspc}
\end{figure}

As a final validation of the background subtraction method presented in this paper, we compared the background-subtracted surface brightness profile of A2142 with the surface brightness profile of the same system measured with \emph{ROSAT}/PSPC. Thanks to its very wide FOV ($\sim2$ square degrees) and very low instrumental background, \emph{ROSAT}/PSPC was very well suited to measure low-surface brightness emission. The \emph{ROSAT} data were reduced using the ESAS data reduction scheme \citep{snowden94} following the method described in \citet{eckert12}. The profiles were converted into energy flux by assuming that the spectral shape is reproduced by an absorbed APEC model at a temperature of 8 keV, and folding the spectral model with the response of the two instruments. The resulting profiles are shown in Fig.~\ref{fig:pspc}. A remarkable agreement is observed out to the largest radii probed, which correspond to $1.3R_{200}\sim$ 3,000~kpc.

\section{Tables and Spectra}
\subsection{Spectra of each of the annuli defined in Fig.~\ref{fig:annulus_region}}
%%%%%%%%%%%%%%%%

\begin{figure*}
\resizebox{\hsize}{!}{\vbox{\hbox{
\includegraphics{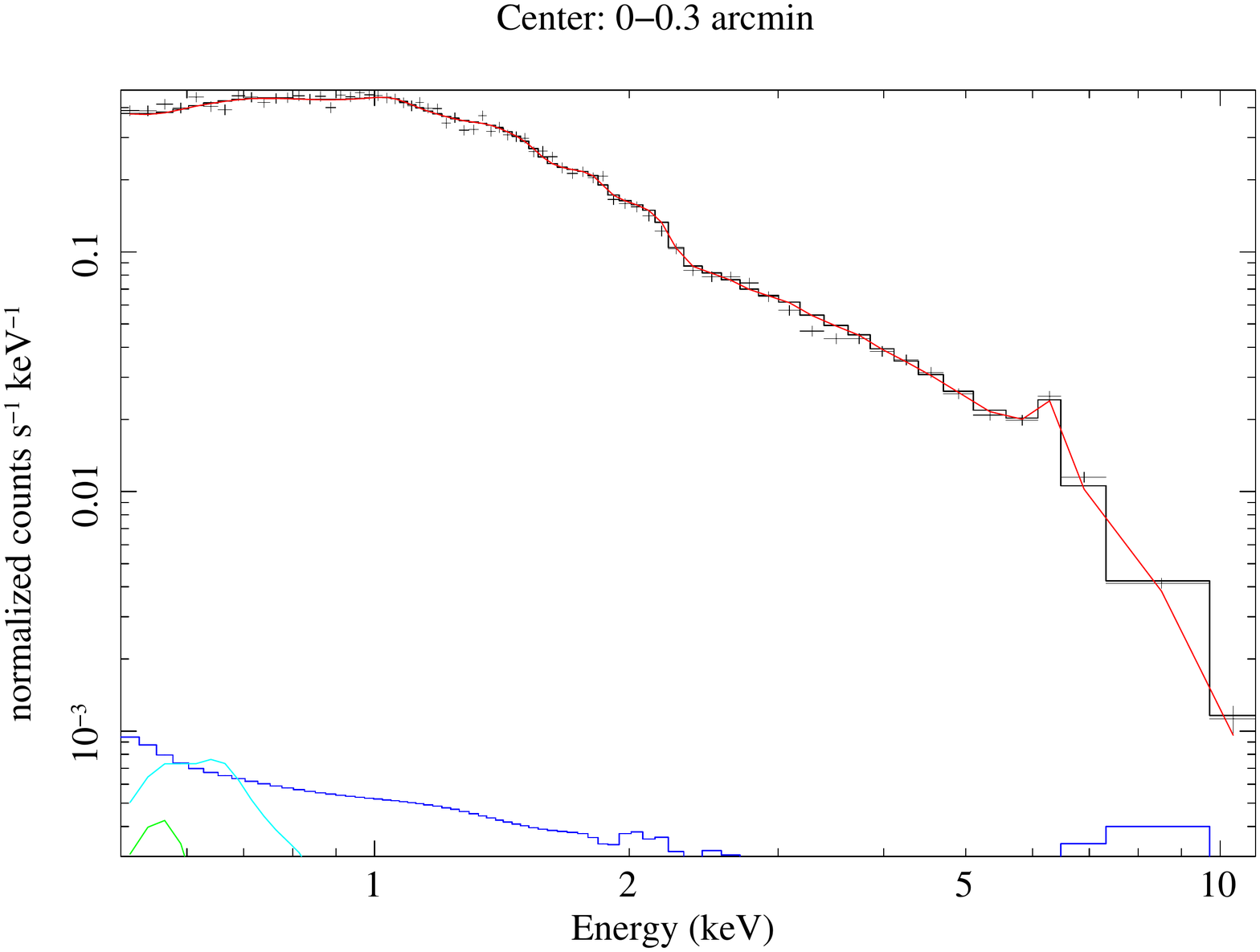}
\includegraphics{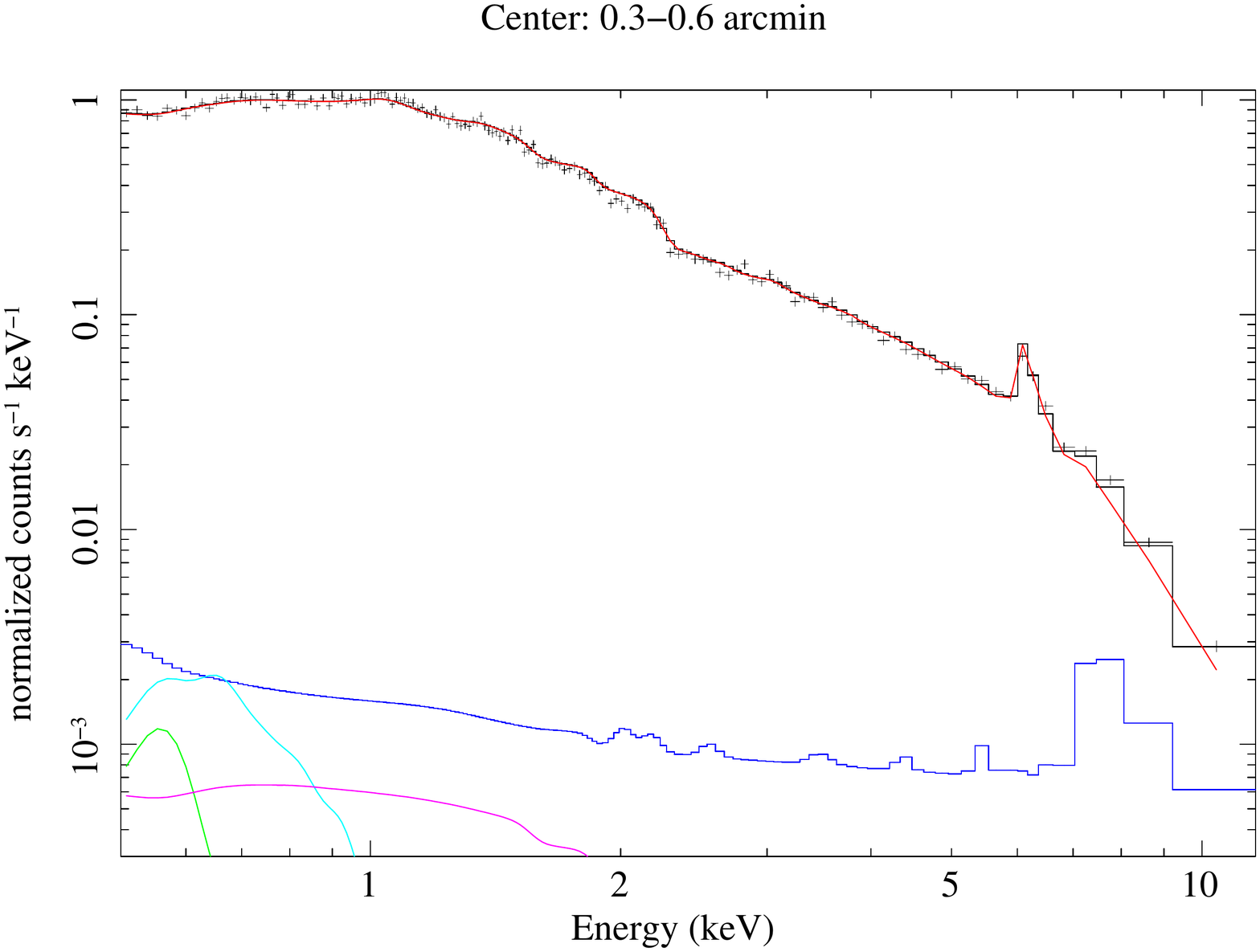}}
\hbox{
\includegraphics{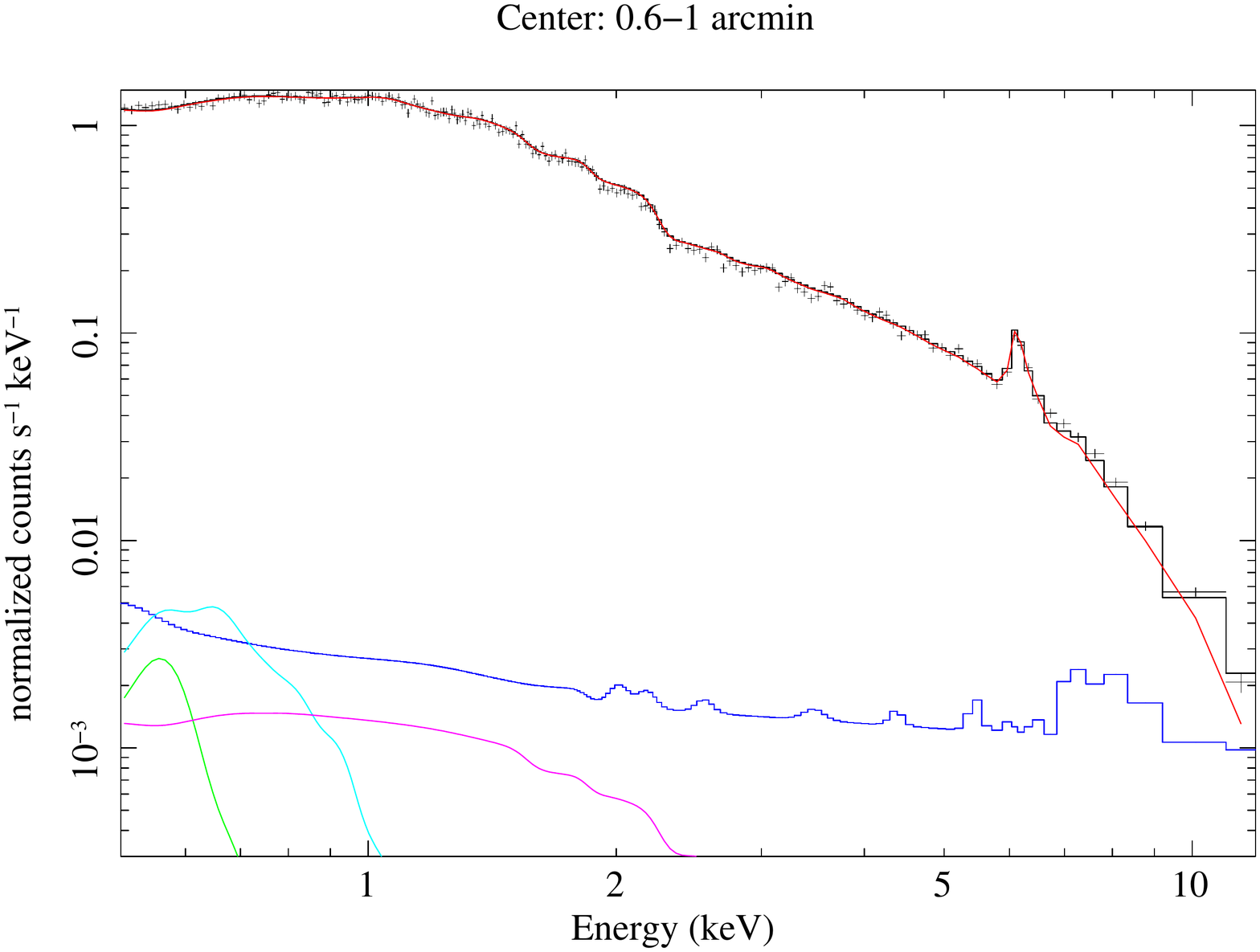}
\includegraphics{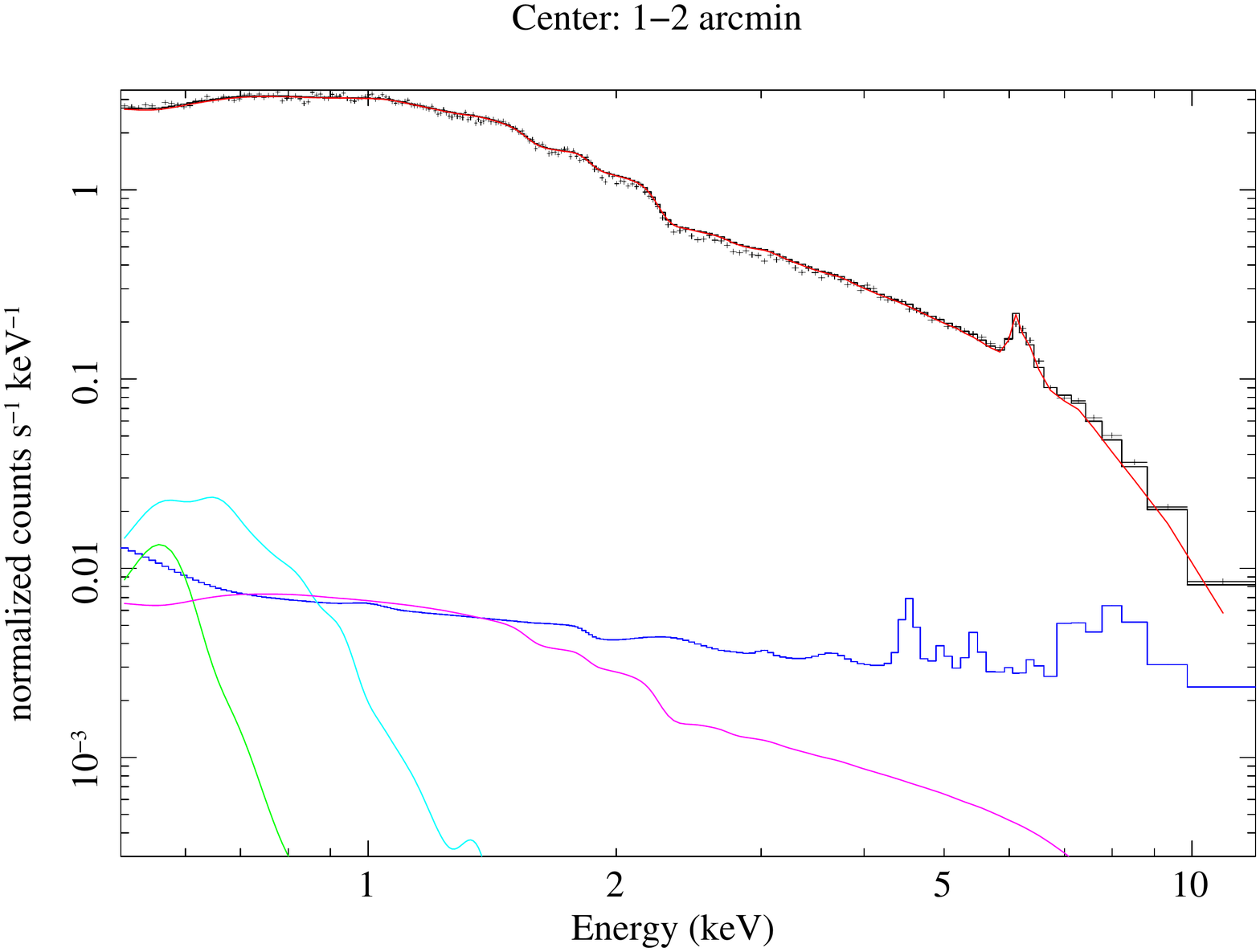}}
\hbox{
\includegraphics{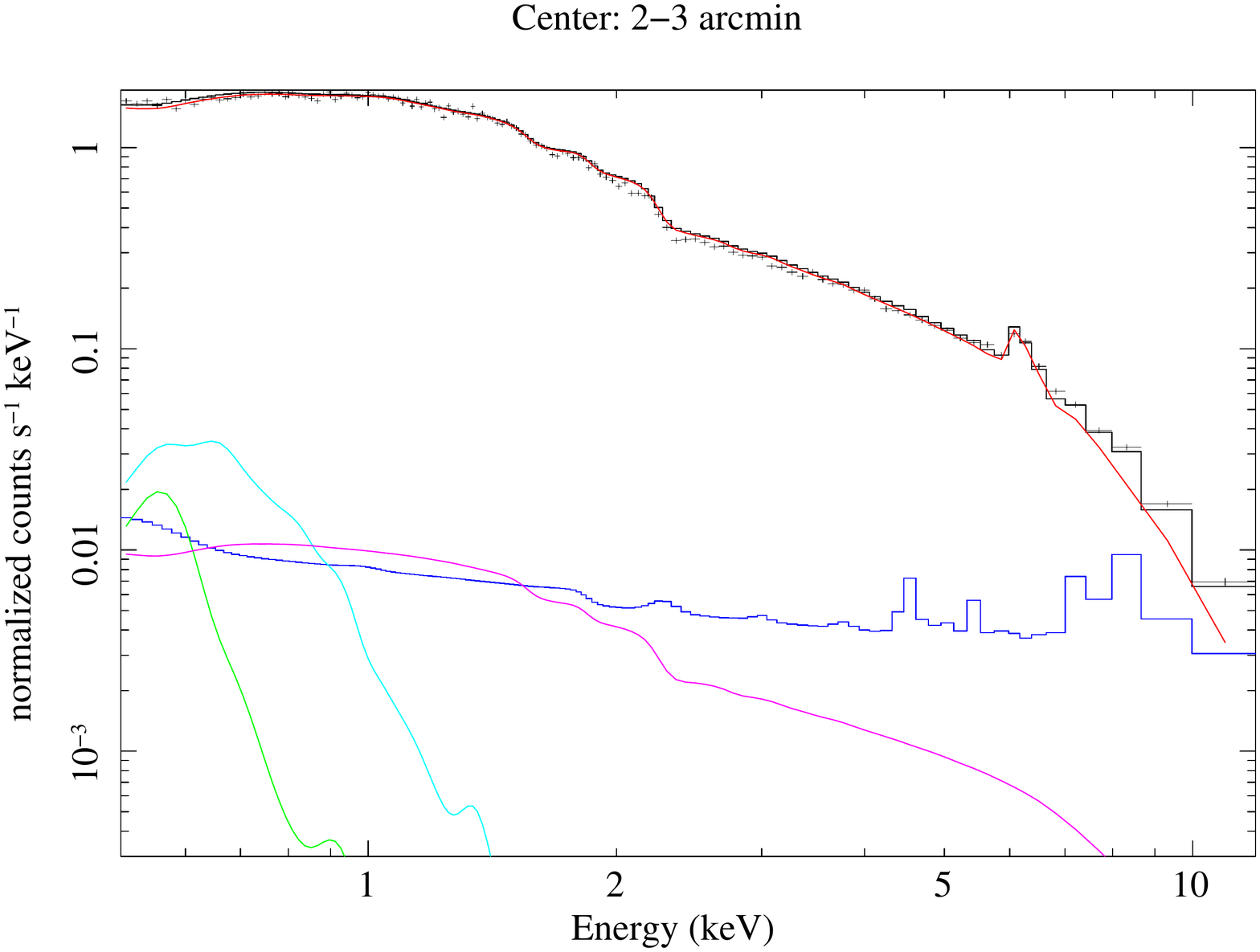}
\includegraphics{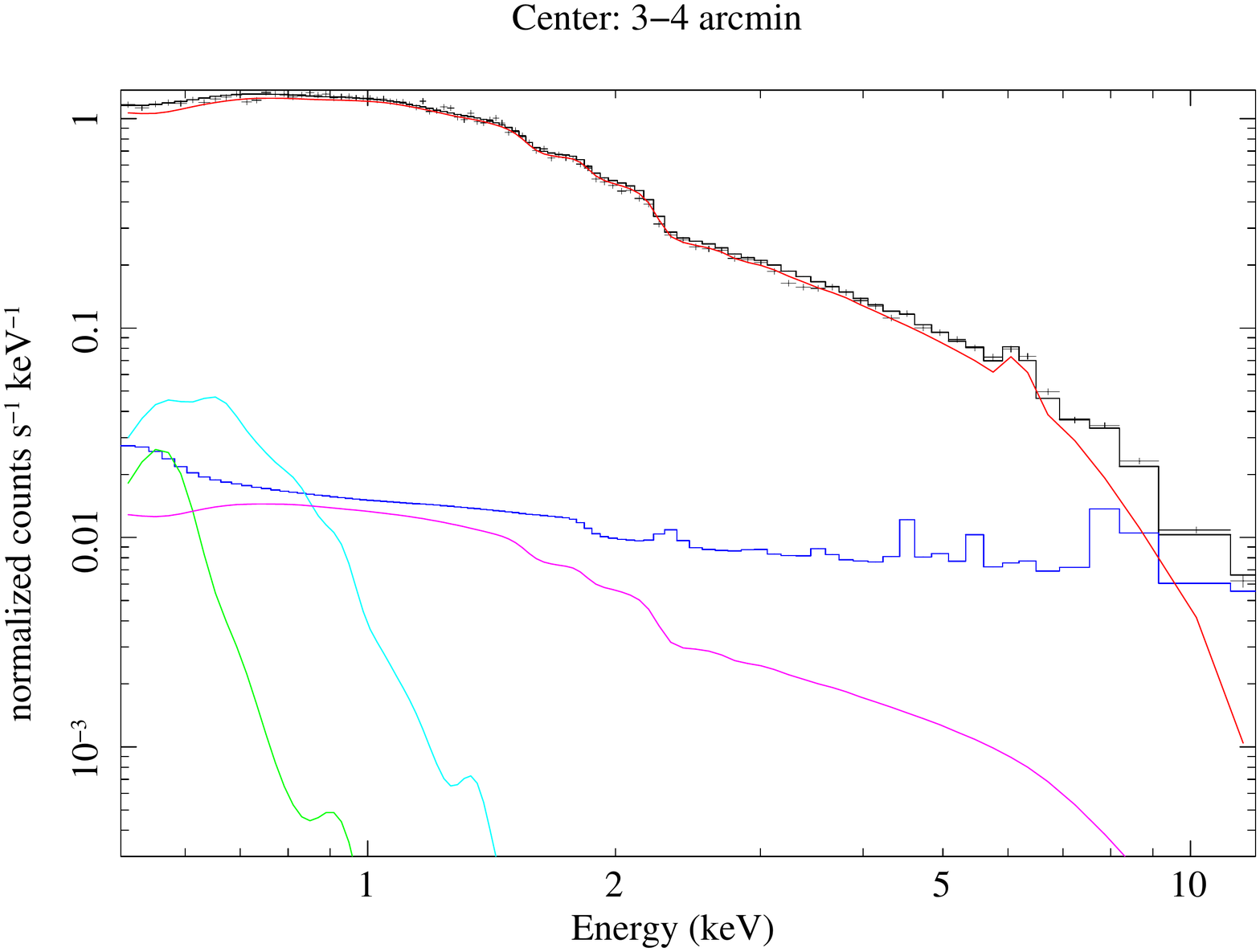}}}}
\caption{EPIC/pn spectra for the annuli defined in Fig.~\ref{fig:annulus_region}. The various components used for the fit are shown in red, for the source; in blue, for the NXB; in magenta, for the CXB; in cyan, for the Galactic halo emission; and in green, for the local hot bubble. Note that the fit was performed jointly on all 3 EPIC instruments, but here only the pn is shown for clarity. The results of the fit are listed in Table~\ref{table:source_fit}.}
\label{fig:allspectra}
\end{figure*}

\begin{figure*}
\resizebox{\hsize}{!}{\vbox{\hbox{
\includegraphics{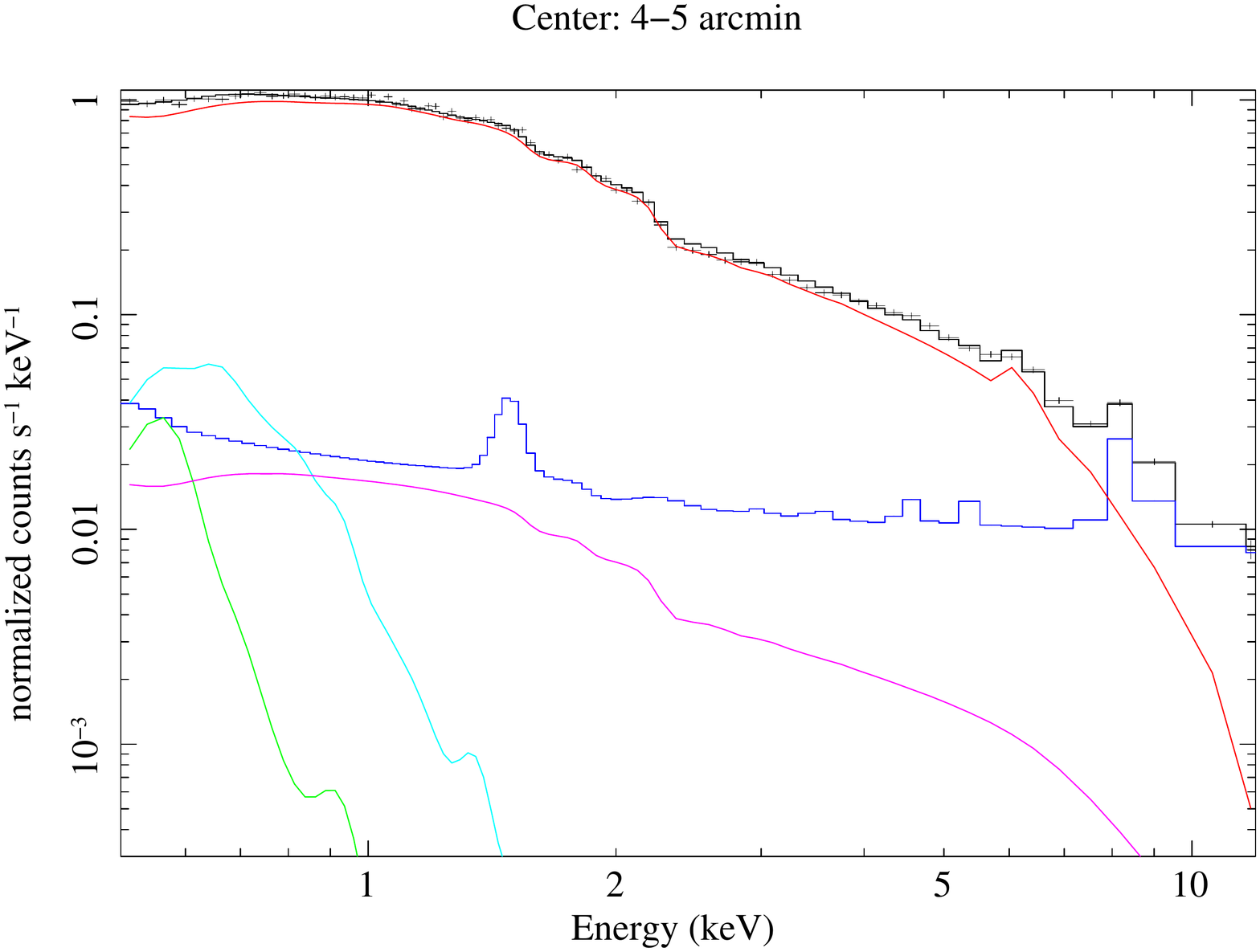}
\includegraphics{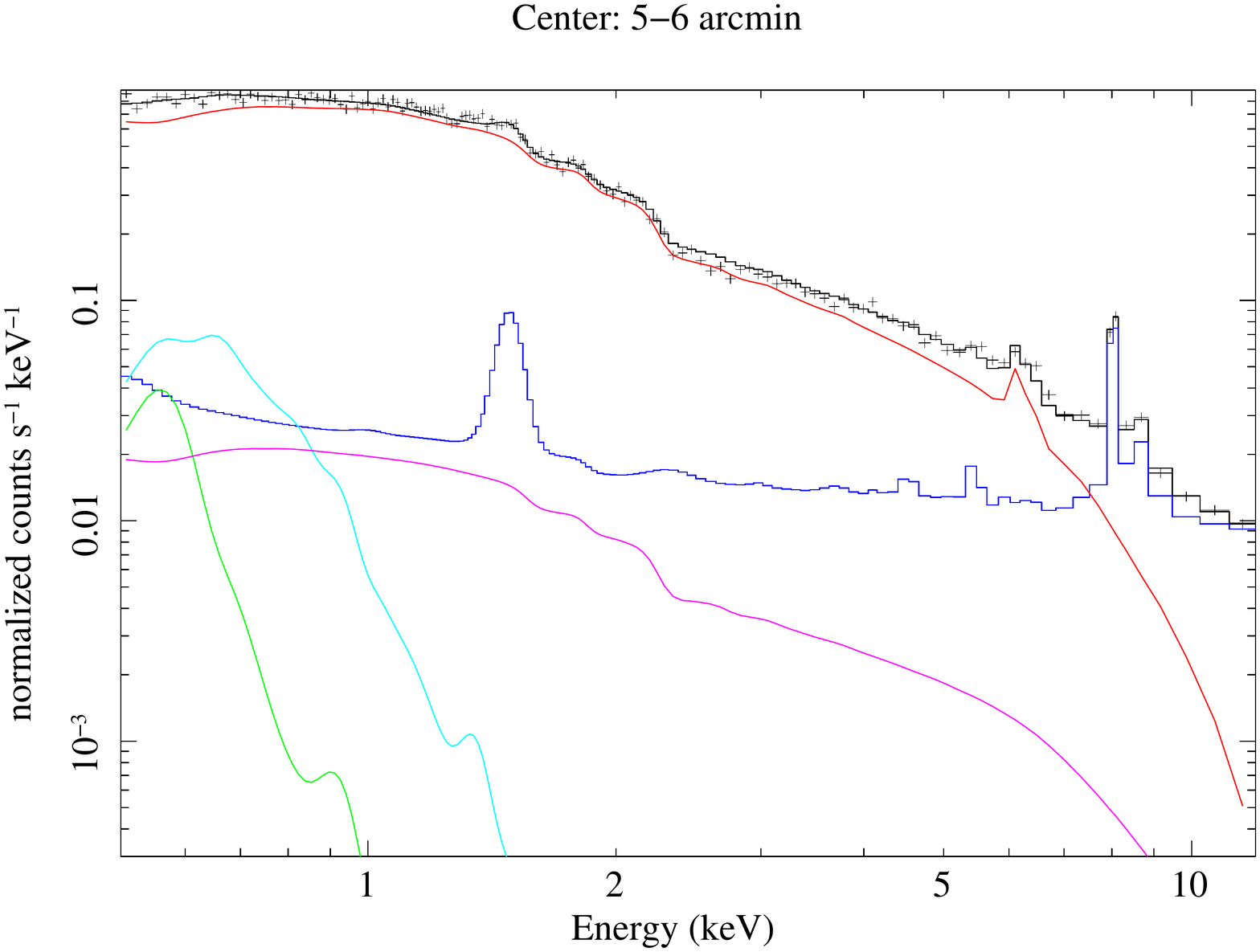}}
\hbox{
\includegraphics{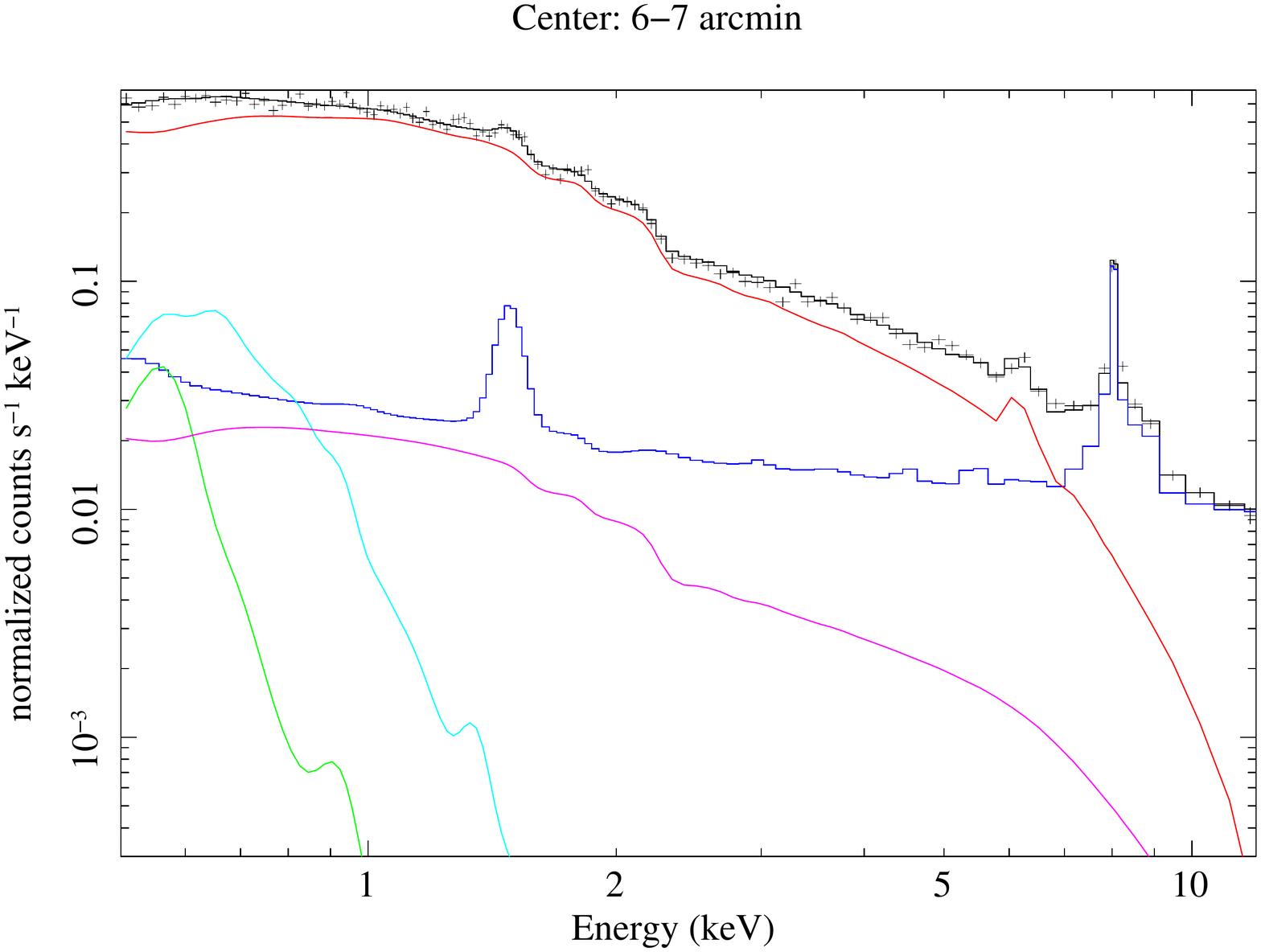}
\includegraphics{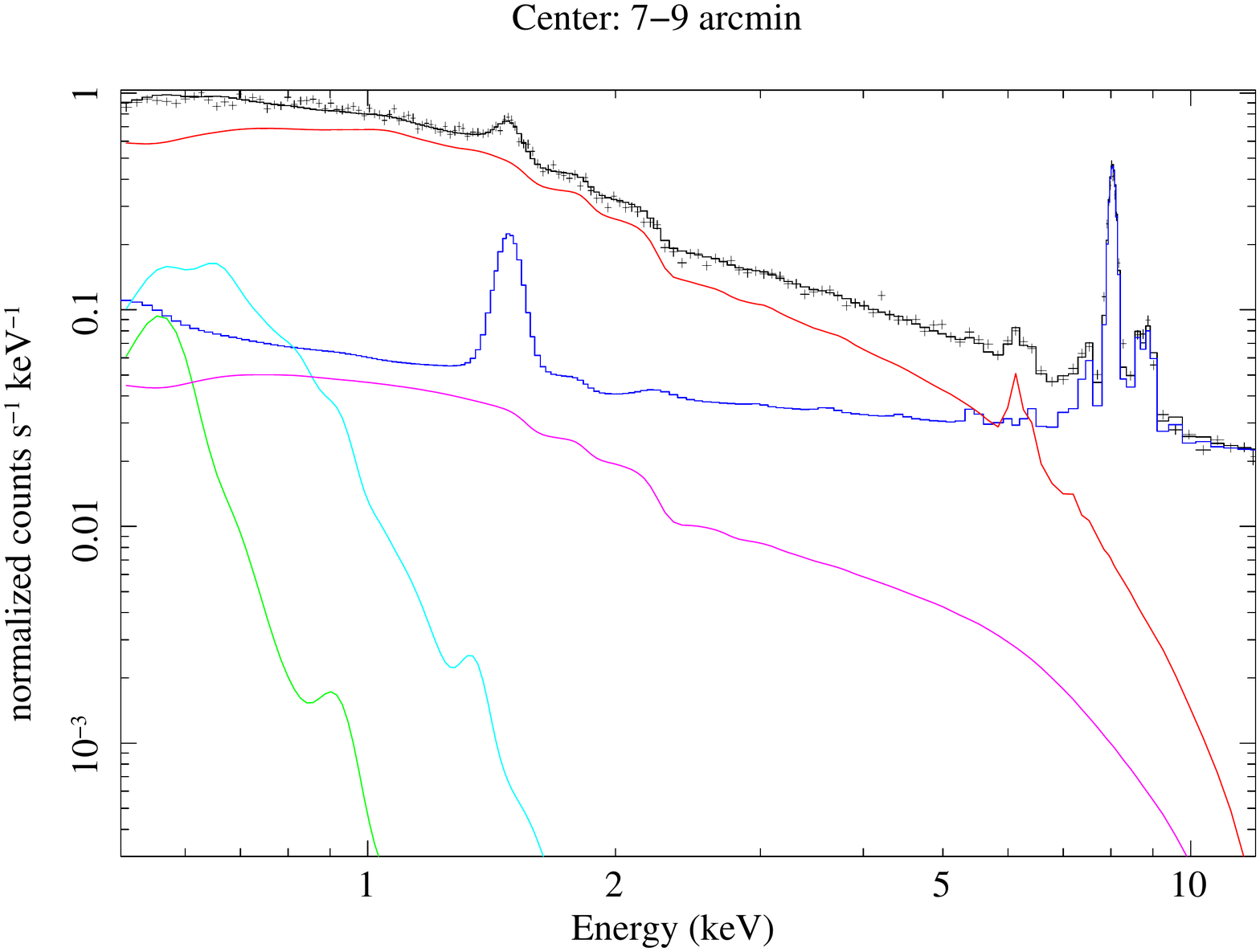}}
\hbox{
\includegraphics{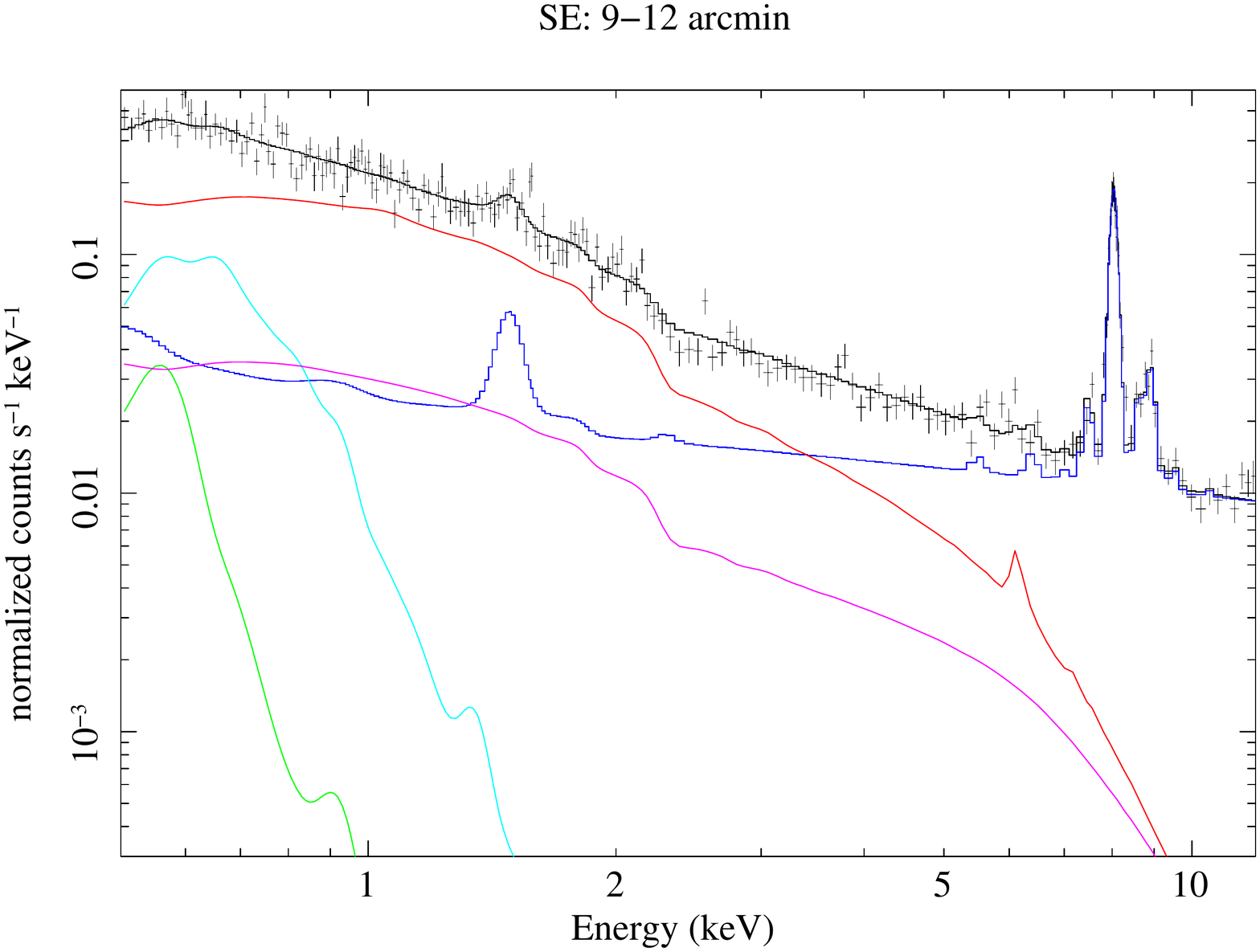}
\includegraphics{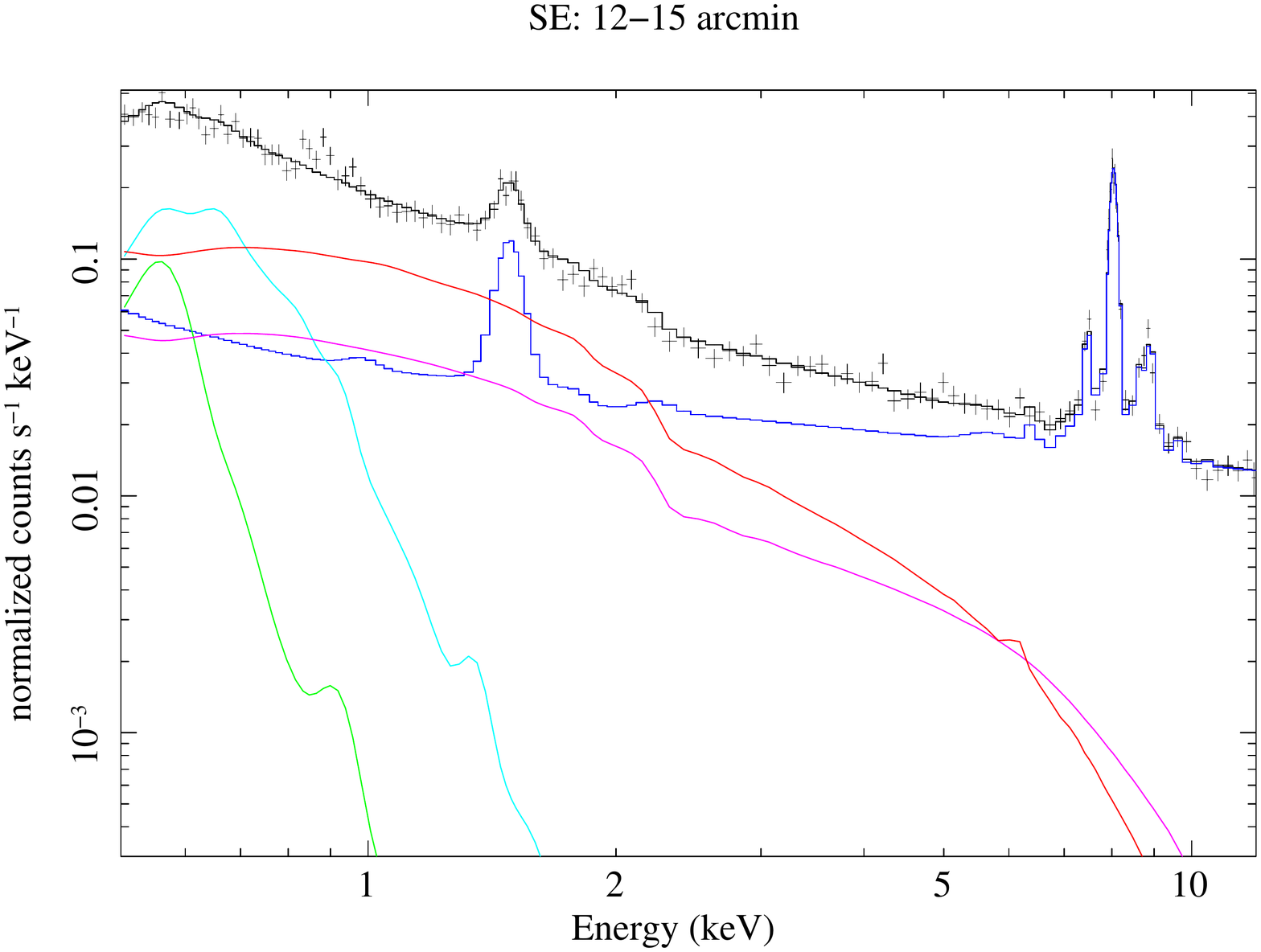}}}}
\caption{Same as Fig. \ref{fig:allspectra} (continued).}
\label{fig:allspectra2}
\end{figure*}

\subsection{Results of the fit of the region in each observation individually }

\begin{table*}
\caption{Result of the fit of the spectrum in the annulus 7-9 arcmin for the individual regions: center, NW, SE and SW separately. The results of the fit of the combined center, NW, SE and SW regions can be found in Table~\ref{table:source_fit}. The temperature is given in keV, the norm in $10^{-3}$cm$^{-5}$ and the abundance in solar metallicity.}
\begin{center}
\begin{tabular}{ccccccc}\hline
Obs &T&$\Delta $T&norm&$\Delta$norm&Z&$\Delta$Z \\\hline\hline
center&7.82&[7.66, 8.06]&0.0574&[0.0569, 0.0579]&0.346&[0.314, 0.387]\\
NW&7.28&[6.59, 8.15]&0.0441&[0.0424, 0.0456]&0.488&[0.319, 0.686]\\
SE&5.69&[5.40, 5.98]&0.104&[0.102, 0.105]&0.169&[0.116, 0.228]\\
SW&6.47&[5.69, 7.11]&0.0347&[0.0334, 0.0365]&0.249&[0.108, 0.423]\\\hline
\label{table:source_indivfit8}
\end{tabular}
\end{center}
\end{table*}

\begin{table*}
\caption{Result of the fit of the spectrum in the annulus 9-12 arcmin for the individual regions: center, NE, NW, SE and SW separately and for the combined fit excluding the NE region (ALL$_{\text{withoutNE}}$). The results of the combined fit including the NE region (center, NE, NW, SE and SW) can be found in Table~\ref{table:source_fit}. The temperature is given in keV, the norm in $10^{-3}$cm$^{-5}$ and the abundance in solar metallicity.}
\begin{center}
\begin{tabular}{ccccccc}\hline
Obs &T&$\Delta $T&norm&$\Delta$norm&Z&$\Delta$Z \\\hline\hline
center&7.25&[6.87, 7.63]&0.0241&[0.0238, 0.0245]&0.197&[0.143, 0.255]\\
NE&5.15&[4.50, 6.27]&0.0155&[0.0147, 0.0162]&0.0590&[0.0, 0.220]\\
NW&7.22&[6.62, 7.82]&0.0213&[0.0208, 0.0221]&0.152&[0.0, 0.342]\\
SE&5.28&[4.95, 5.76]&0.0240&[0.0235, 0.0243]&0.122&[0.0518, 0.200]\\
SW&4.75&[4.20, 5.32]&0.0153&[0.0147, 0.0159]&0.0949&[0.0, 0.219]\\
ALL$_{\text{withoutNE}}$&7.17&[6.81, 7.44]&0.0229&[0.0226, 0.0231]&0.196&[0.153, 0.242]\\\hline
\label{table:source_indivfit9}
\end{tabular}
\end{center}
\end{table*}

\begin{table*}
\caption{Result of the fit of the spectrum in the annulus 12-15 arcmin for the individual regions: NE, NW, SE and SW separately and for the combined fit excluding the NE region (ALL$_{\text{withoutNE}}$). The results of the combined fit including the NE region (NE, NW, SE and SW) can be found in Table~\ref{table:source_fit}. The temperature is given in keV, the norm in $10^{-3}$cm$^{-5}$ and the abundance in solar metallicity.}
\begin{center}
\begin{tabular}{ccccccc}\hline
Obs &T&$\Delta $T&norm&$\Delta$norm&Z&$\Delta$Z \\\hline\hline
NE&2.95&[2.60, 3.32]&0.00747&[0.00696, 0.00788]&0.333&[0.230, 0.477]\\
NW&6.87&[6.01, 8.82]&0.00858&[0.00801, 0.00921]&0.440&[0.139, 0.858]\\
SE&5.89&[5.18, 6.68]&0.00951&[0.00916, 0.00985]&0.117&[0.00914, 0.266]\\
SW&3.16&[2.75, 3.80]&0.00750&[0.00706, 0.00792]&0.0272&[0.0, 0.137]\\
ALL$_{\text{withoutNE}}$&5.20&[4.75, 5.76]&0.00849&[0.00824, 0.00875]&0.140&[0.0530, 0.237]\\\hline
\label{table:source_indivfit10}
\end{tabular}
\end{center}
\end{table*}

\subsection{XMM temperature profile in the NW direction}\label{T_NW}
%%%%%%%%%%%%%%%%
\begin{figure}
\begin{center}
  \includegraphics[height=0.7\columnwidth,angle=0]{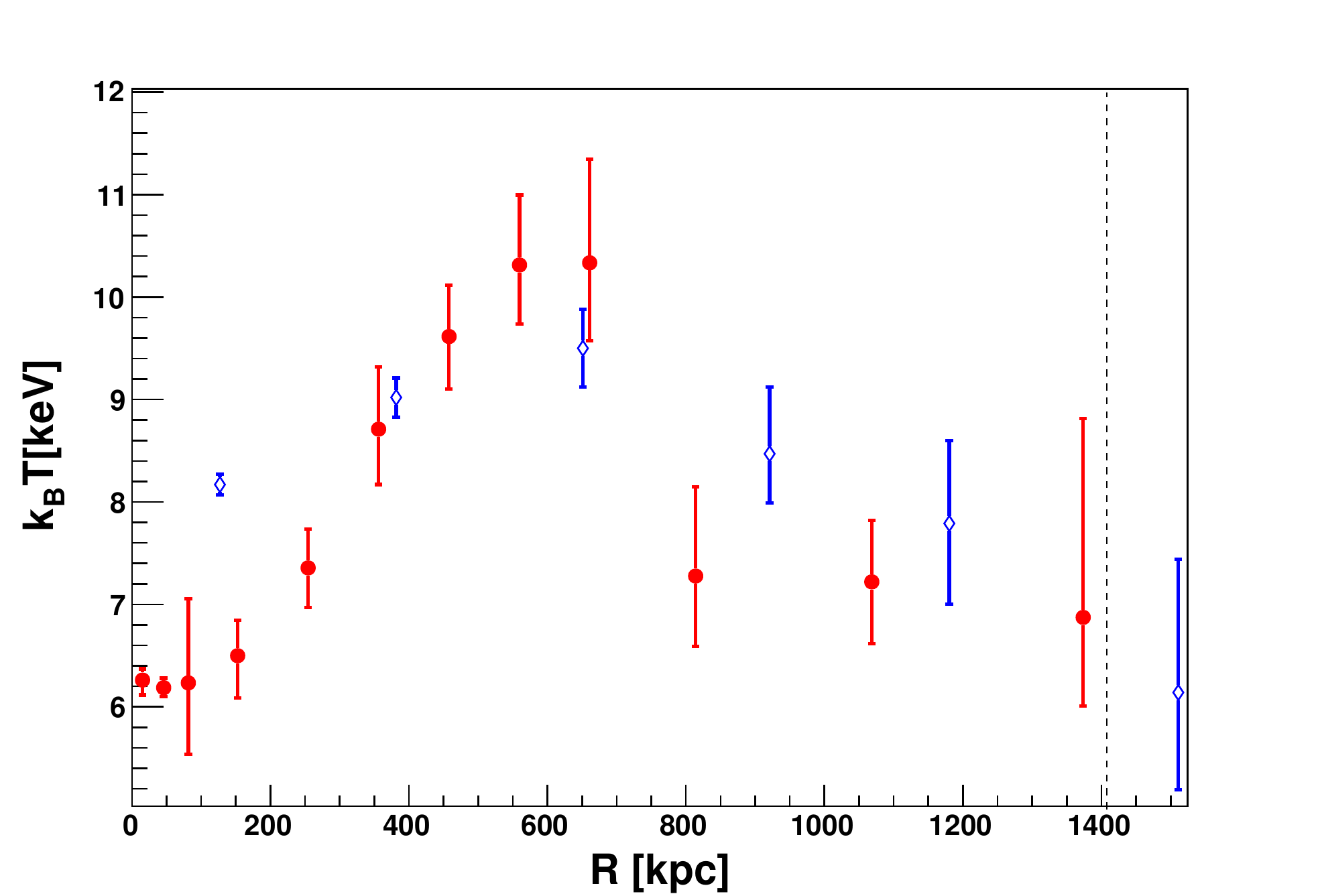}
\caption{Temperature profile in the NW direction. Red: data points using the \textit{APEC} model on \textit{XMM-Newton} observations; In blue: data point from   \citep{akamatsu11} obtained with \textit{Suzaku}. The dashed line represents the value of $R_{500}$.}\label{TemperatureAPEC_NW}
\end{center}
\end{figure}
%%%%%%%%%%%%%%%%


\begin{thebibliography}{4}
\bibitem[Akamatsu et al.(2011)]{akamatsu11} Akamatsu, H., Hoshino, A., Ishisaki, Y., et al.\ 2011, \pasj, 63, 1019
\bibitem[Ameglio et al.(2009)]{ameglio09} Ameglio, S., Borgani, S., Pierpaoli, E., et al.\ 2009, \mnras, 394, 479 
\bibitem[Ameglio et al.(2007)]{ameglio07} Ameglio, S., Borgani, S., Pierpaoli, E., \& Dolag, K.\ 2007, \mnras, 382, 397 
\bibitem[Anders \& Grevesse(1989)]{anders89} Anders, E., \& Grevesse, N.\ 1989, \gca, 53, 197 
\bibitem[Arnaud et al.(2011)]{arnaud10} Arnaud, M., Pratt, G. W., Piffaretti, R., et al. 2010, \aap, 517, A92
\bibitem[Avestruz et al.(2015)]{avestruz15} Avestruz, C., Nagai, D., Lau, E.~T., \& Nelson, K.\ 2015, \apj, 808, 176 
\bibitem[Basu et al.(2010)]{basu10} Basu, K., Zhang, Y.-Y., Sommer, M.~W., et al.\ 2010, \aap, 519, A29
\bibitem[Battaglia et al.(2013)]{battaglia13} Battaglia, N., Bond, J.~R., Pfrommer, C., \& Sievers, J.~L.\ 2013, \apj, 777, 123 
\bibitem[Battaglia et al.(2015)]{battaglia15} Battaglia, N., Bond, J.~R., Pfrommer, C., \& Sievers, J.~L.\ 2015, \apj, 806, 43 
\bibitem[Bonamente et al.(2013)]{bonamente13} Bonamente, M., Landry, D., Maughan, B., et al.\ 2013, \mnras, 428, 2812 
\bibitem[Burns et al.(2010)]{burns10} Burns, J.~O., Skillman, S.~W., \& O'Shea, B.~W.\ 2010, \apj, 721, 1105
\bibitem[Buote(2000)]{buote00} Buote, D.~A.\ 2000, \apj, 539, 172
\bibitem[Carter \& Sembay(2008)]{carter08} Carter, J.~A., \& Sembay, S.\ 2008, \aap, 489, 837 
\bibitem[Carter et al.(2011)]{carter11} Carter, J.~A., Sembay, S., \& Read, A.~M.\ 2011, \aap, 527, A115 
\bibitem[Cavagnolo et al.(2009)]{cavagnolo09} Cavagnolo, K.~W., Donahue, M., Voit, G.~M., \& Sun, M.\ 2009, \apjs, 182, 12 
\bibitem[Cavaliere et al.(2011)]{cavaliere11} Cavaliere, A., Lapi, A., \& Fusco-Femiano, R.\ 2011, \apj, 742, 19  
\bibitem[Croston et al.(2006)]{croston06} Croston, J.~H., Arnaud, M., Pointecouteau, E., \& Pratt, G.~W.\ 2006, \aap, 459, 1007
\bibitem[De Luca \& Molendi(2004)]{deluca04} De Luca, A., \& Molendi, S.\ 2004, \aap, 419, 837
\bibitem[Eckert et al.(2015a)]{eckertclumping} Eckert, D., Roncarelli, M., Ettori, S., et al.\ 2015a, \mnras, 447, 2198
\bibitem[Eckert et al.(2015b)]{XXL} Eckert, D., Ettori, S., Coupon, J., et al.\ 2015b, arXiv:1512.03814
\bibitem[Eckert et al.(2014)]{eckert1a2142} Eckert, D., Molendi, S., Owers, M., et al.\ 2014, \aap, 570, A119
\bibitem[Eckert et al.(2013a)]{eckert13a} Eckert, D., Molendi, S., Vazza, F., Ettori, S., \& Paltani, S.\ 2013a, \aap, 551, A22
\bibitem[Eckert et al.(2013b)]{eckert13b} Eckert, D., Ettori, S., Molendi, S., Vazza, F., \& Paltani, S.\ 2013b, \aap, 551, A23
\bibitem[Eckert et al.(2012)]{eckert12} Eckert, D., Vazza, F., Ettori, S., et al.\ 2012, \aap, 541, A57 
\bibitem[Eckert et al.(2011)]{eckert11} Eckert, D., Molendi, S., \& Paltani, S.\ 2011, \aap, 526, A79 
\bibitem[Einasto et al.(2015)]{einasto15} Einasto, M., Gramann, M., Saar, E., et al.\ 2015, \aap, 580, A69 
\bibitem[Ettori et al.(2013)]{ettorimgas} Ettori, S., Donnarumma, A., Pointecouteau, E., et al.\ 2013, \ssr, 177, 119
\bibitem[Ettori \& Molendi(2011)]{ettori11} Ettori, S., \& Molendi, S.\ 2011, Memorie della Societa Astronomica Italiana Supplementi, 17, 47 
\bibitem[Ettori et al.(2010)]{ettori10} Ettori, S., Gastaldello, F., Leccardi, A., et al.\ 2010, \aap, 524, A68 
\bibitem[Ettori et al.(2002)]{ettorideproj} Ettori, S., De Grandi, S., \& Molendi, S.\ 2002, \aap, 391, 841
\bibitem[Fabian et al.(1981)]{fabian81} Fabian, A.~C., Hu, E.~M., Cowie, L.~L., \& Grindlay, J.\ 1981, \apj, 248, 47
\bibitem[Farnsworth et al.(2013)]{farnsworth13} Farnsworth, D., Rudnick, L., Brown, S., \& Brunetti, G.\ 2013, \apj, 779, 189 
\bibitem[Frank et al.(2013)]{frank13} Frank, K.~A., Peterson, J.~R., Andersson, K., Fabian, A.~C., \& Sanders, J.~S.\ 2013, \apj, 764, 46
\bibitem[Foreman-Mackey et al.(2013)]{foreman13} Foreman-Mackey, D., Hogg, D.~W., Lang, D., \& Goodman, J.\ 2013, \pasp, 125, 306
\bibitem[Fusco-Femiano \& Lapi(2014)]{fusco14} Fusco-Femiano, R., \& Lapi, A.\ 2014, \apj, 783, 76
\bibitem[Gonzalez et al.(2007)]{gonzales07} Gonzalez, A.~H., Zaritsky, D., \& Zabludoff, A.~I.\ 2007, \apj, 666, 147
\bibitem[G\'orski et al.(2005)]{gorski05} G\'orski, K. M., Hivon, E., Banday, A. J., et al. 2005, \apj, 622, 759
\bibitem[Gramann et al.(2015)]{gramann15} Gramann, M., Einasto, M., Hein{\"a}m{\"a}ki, P., et al.\ 2015, arXiv:1506.05252 
\bibitem[Hoshino et al.(2010)]{hoshino10} Hoshino, A., Henry,J.~P., Sato, K., et al.\ 2010, \pasj, 62, 371
\bibitem[Hurier et al.(2013)]{hurier13} Hurier, G., Mac\`ias-P\'erez, J.~F., \& Hildebrandt, S.\ 2013, \aap, 558, 118
\bibitem[Jarosik et al.(2011)]{jarosik11} Jarosik, N., Bennett, C.~L., Dunkley, J., et al.\ 2011, \apjs, 192, 14
\bibitem[Kalberla et al.(2005)]{kalberla05} Kalberla, P.~M.~W., Burton, W.~B., Hartmann, D., et al.\ 2005, \aap, 440, 775 
\bibitem[Kawaharada et al.(2010)]{kawa10} Kawaharada, M., Okabe, N., Umetsu, K., et al.\ 2010, \apj, 714, 423 
\bibitem[Khedekar et al.(2013)]{khedekar13} Khedekar, S., Churazov, E., Kravtsov, A., et al.\ 2013, \mnras, 431, 954
\bibitem[Kriss et al.(1983)]{kriss83} Kriss, G.~A., Cioffi, D.~F., \& Canizares, C.~R.\ 1983, \apj, 272, 439
\bibitem[Kuntz \& Snowden(2008)]{kuntz08} Kuntz, K.~D., \& Snowden, S.~L.\ 2008, \aap, 478, 575 
\bibitem[Lamarre et al. (2010)]{lamarre10} Lamarre, J.-M., Puget, J.-L., Ade, P.~A.~R., et al.\ 2010, \aap, 520, 9
\bibitem[Lapi et al.(2010)]{lapi10} Lapi, A., Fusco-Femiano, R., \& Cavaliere, A.\ 2010, \aap, 516, A34 
\bibitem[Lau et al.(2009)]{lau09} Lau, E.~T., Kravtsov, A.~V., \& Nagai, D.\ 2009, \apj, 705, 1129
\bibitem[Lau et al.(2015)]{lau15} Lau, E.~T., Nagai, D., Avestruz, C., Nelson, K., \& Vikhlinin, A.\ 2015, \apj, 806, 68 
\bibitem[Leccardi \& Molendi(2008)]{leccardi08} Leccardi, A., \& Molendi, S.\ 2008, \aap, 486, 359 
\bibitem[Limousin et al.(2013)]{limousin13} Limousin, M., Morandi, A., Sereno, M., et al.\ 2013, \ssr, 177, 155 
\bibitem[Maccacaro et al.(1988)]{maccacaro88} Maccacaro, T., Gioia, I.~M., Wolter, A., Zamorani, G., \& Stocke, J.~T.\ 1988, \apj, 326, 680 
\bibitem[Mathiesen et al.(1999)]{mathiesen99} Mathiesen, B., Evrard, A.~E., \& Mohr, J.~J.\ 1999, \apjl, 520, L21
\bibitem[Markevitch et al.(2000)]{markevitch00} Markevitch, M., Ponman, T.~J., Nulsen, P.~E.~J., et al.\ 2000, \apj, 541, 542
\bibitem[Mazzotta et al.(2004)]{mazzotta04} Mazzotta, P., Rasia, E., Moscardini, L., \& Tormen, G.\ 2004, \mnras, 354, 10  
\bibitem[McCammon et al.(2002)]{mccammon02} McCammon, D., Almy, R., Apodaca, E., et al.\ 2002, \apj, 576, 188
\bibitem[McLaughlin(1999)]{mclaughlin99} McLaughlin, D.~E.\ 1999, \aj, 117, 2398 
\bibitem[Mitsuda et al.(2007)]{mitsuda07} Mitsuda, K., Bautz, M., Inoue, H., et al.\ 2007, \pasj, 59, 1 
\bibitem[Morandi et al.(2013)]{morandi13} Morandi, A., Nagai, D., \& Cui, W.\ 2013, \mnras, 436, 1123
\bibitem[Morandi \& Cui(2014)]{morandi14} Morandi, A., \& Cui, W.\ 2014, \mnras, 437, 1909  
\bibitem[Moretti et al.(2003)]{moretti03} Moretti, A., Campana, S., Lazzati, D., \& Tagliaferri, G.\ 2003, \apj, 588, 696
\bibitem[Munari et al.(2014)]{munari14} Munari, E., Biviano, A., \& Mamon, G.~A.\ 2014, \aap, 566, A68 
\bibitem[Nagai \& Lau(2011)]{nagai11} Nagai, D., \& Lau, E.~T.\ 2011, \apjl, 731, L10
\bibitem[Nagai et al.(2007a)]{nagai07a} Nagai, D., Kravtsov, A.~V., \& Vikhlinin, A.\ 2007, \apj, 668, 1 
\bibitem[Nagai et al.(2007b)]{nagai07b}Nagai, D., Vikhlinin, A., \& Kravtsov, A. V. \ 2007, \apj, 655, 98
\bibitem[Navarro et al.(1997)]{NFW}Navarro J.F., Frenk C.S., White S.D.M., 1997, \apj, 490, 493
\bibitem[Nelson et al.(2014)]{nelson14} Nelson, K., Lau, E.~T., \& Nagai, D.\ 2014, \apj, 792, 25
\bibitem[Nevalainen et al.(2010)]{ne10} Nevalainen, J., David, L., \& Guainazzi, M.\ 2010, \aap, 523, A22
\bibitem[Nord et al.(2009)]{nord09} Nord, M., Basu, K., Pacaud, F., et al.\ 2009, \aap, 506, 623 
\bibitem[Okabe et al.(2014)]{okabe14} Okabe, N., Umetsu, K., Tamura, T., et al.\ 2014, \pasj, 66, 99
\bibitem[Owers et al.(2009)]{owers09} Owers, M.~S., Nulsen, P.~E.~J., Couch, W.~J., \& Markevitch, M.\ 2009, \apj, 704, 1349 
\bibitem[Owers et al.(2011)]{owers11} Owers, M.~S., Nulsen, P.~E.~J., \& Couch, W.~J.\ 2011, \apj, 741, 122 ., et al.\ 2013, \aap, 550, A131
\bibitem[Piffaretti et al.(2011)]{mcxc} Piffaretti, R., Arnaud, M., Pratt, G.~W., Pointecouteau, E., \& Melin, J.-B.\ 2011, \aap, 534, A109
\bibitem[Planck HFI Core Team (2011)]{planckHFI} Planck HFI Core Team\ 2011, \aap, 536, 4
\bibitem[Planck Collaboration et al.(2012)]{planck_Yx} Planck Collaboration, Aghanim, N., Arnaud, M., et al.\ 2012, \aap, 543, A102 
\bibitem[Planck Collaboration (2013)]{planck13} Planck Collaboration\ 2013, \aap, 550, A131
\bibitem[Planck Collaboration (2014)]{planck14Ymap} Planck Collaboration\ 2014, \aap, 571, A21 
\bibitem[Planck Collaboration (2015)]{planckdr2015} Planck Collaboration, \ 2015, arXiv:1502.01582
\bibitem[Planck Collaboration (2015b)]{planck15Ymap} Planck Collaboration\ 2015, arXiv:1502.01596
\bibitem[Planck Collaboration (2015c)]{planck15cosmo} Planck Collaboration\ 2015, arXiv:1502.01589 
\bibitem[Planck Collaboration (2015d)]{planckSZ2013} Planck Collaboration\ 2015, \aap, 581,14 
\bibitem[Planck Collaboration (2015e)]{planckSZ2015} Planck Collaboration\ 2015, arXiv:1502.01598
\bibitem[Pratt et al.(2010)]{pratt10} Pratt, G.~W., Arnaud, M., Piffaretti, R., et al.\ 2010, \aap, 511, A85 
\bibitem[Rasia et al.(2006)]{rasia06} Rasia, E., Ettori, S., Moscardini, L., et al.\ 2006, \mnras, 369, 2013 
\bibitem[Rasia et al.(2014)]{rasia14} Rasia, E., Lau, E.~T., Borgani, S., et al.\ 2014, \apj, 791, 96
\bibitem[Reiprich et al.(2013)]{reiprich13} Reiprich, T.~H., Basu, K., Ettori, S., et al.\ 2013, \ssr, 177, 195
\bibitem[Reiprich et al.(2009)]{reiprich09} Reiprich, T.~H., Hudson, D.~S., Zhang, Y.-Y., et al.\ 2009, \aap, 501, 899 
\bibitem[Roncarelli et al.(2013)]{roncarelli13} Roncarelli, M., Ettori, S., Borgani, S., et al.\ 2013, \mnras, 432, 3030
\bibitem[Rossetti et al.(2013)]{rossetti13} Rossetti, M., Eckert, D., De Grandi, S., et al.\ 2013, \aap, 556, A44
\bibitem[Rozo et al.(2012)]{rozo12} Rozo, E., Vikhlinin, A., \& More, S.\ 2012, \apj, 760, 67 
\bibitem[Rudd \& Nagai(2009)]{rudd09} Rudd, D.~H., \& Nagai, D.\ 2009, \apjl, 701, L16 
\bibitem[Sato et al.(2014)]{sato14} Sato, K., Matsushita, K., Yamasaki, N.~Y., Sasaki, S., \& Ohashi, T.\ 2014, \pasj, 66, 85 
\bibitem[Sayers et al.(2013)]{sayers13} Sayers, J., Czakon, N.~G., Mantz, A., et al.\ 2013, \apj, 768, 177 
\bibitem[Simionescu et al.(2011)]{simionescu11} Simionescu, A., Allen, S.~W., Mantz, A., et al.\ 2011, Science, 331, 1576
\bibitem[Snowden et al.(2008)]{snowden08} Snowden, S.~L., Mushotzky, R.~F., Kuntz, K.~D., \& Davis, D.~S.\ 2008, \aap, 478, 615
\bibitem[Snowden et al.(1994)]{snowden94} Snowden, S.~L., McCammon, D., Burrows, D.~N., \& Mendenhall, J.~A.\ 1994, \apj, 424, 714
\bibitem[Smith et al.(2001)]{smith01} Smith, R.~K., Brickhouse, N.~S., Liedahl, D.~A., \& Raymond, J.~C.\ 2001, \apjl, 556, L91 
\bibitem[Sunyaev \& Zeldovich(1972)]{sunyaev72} Sunyaev, R.~A., \& Zeldovich, Y.~B.\ 1972, Comments on Astrophysics and Space Physics, 4, 173 
\bibitem[Tauber et al. (2010)]{tauber10} 	Tauber, J.~A., Mandolesi, N., Puget, J.-L., et al.\ 2010,\aap, 520, 1
\bibitem[Umetsu et al.(2009)]{umetsu09} Umetsu, K., Birkinshaw, M., Liu, G.-C., et al.\ 2009, \apj, 694, 1643 
\bibitem[Urban et al.(2014)]{urban14} Urban, O., Simionescu, A., Werner, N., et al.\ 2014, \mnras, 437, 3939 
\bibitem[Vazza et al.(2013)]{vazza13} Vazza, F., Eckert, D., Simionescu, A., Br{\"u}ggen, M., \& Ettori, S.\ 2013, \mnras, 429, 799
\bibitem[Vazza et al.(2011)]{vazza11} Vazza, F., Roncarelli, M., Ettori, S., \& Dolag, K.\ 2011, \mnras, 413, 2305 
\bibitem[Vazza et al.(2009)]{vazza09} Vazza, F., Brunetti, G., Kritsuk, A., et al.\ 2009, \aap, 504, 33 
\bibitem[Vikhlinin et al.(2006)]{vikhlinin06} Vikhlinin, A., Kravtsov, A., Forman, W., et al.\ 2006, \apj, 640, 691 
\bibitem[Voges et al.(1999)]{voges99} Voges, W., Aschenbach, B., Boller, T., et al.\ 1999, \aap, 349, 389 
\bibitem[Voit et al.(2005)]{voit05} Voit, G.~M., Kay, S.~T., \& Bryan, G.~L.\ 2005, \mnras, 364, 909
\bibitem[Walker et al.(2013)]{walker13} Walker, S.~A., Fabian, A.~C., Sanders, J.~S., Simionescu, A., \& Tawara, Y.\ 2013, \mnras, 432, 554
\bibitem[Walker et al.(2012)]{walker12} Walker, S.~A., Fabian, A.~C., Sanders, J.~S., \& George, M.~R.\ 2012, \mnras, 427, L45
\bibitem[Young et al.(2011)]{young11} Young, O.~E., Thomas, P.~A., Short, C.~J., \& Pearce, F.\ 2011, \mnras, 413, 691
\bibitem[Zhuravleva et al.(2013)]{zhu13} Zhuravleva, I., Churazov, E., Kravtsov, A., et al.\ 2013, \mnras, 428, 3274 
\end{thebibliography}
\end{document}